\documentclass[aps,nofootinbib,twocolumn,longbibliography]{revtex4-1}
\usepackage{graphicx,amsmath,amssymb,amsbsy,latexsym,verbatim}
\usepackage{graphicx}

\usepackage{hyperref}
\usepackage{multirow} \usepackage{rotating}

\newcommand{\be}{\nopagebreak[3]\begin{equation}}
\newcommand{\ee}{\end{equation}}
\newcommand{\ba}{\nopagebreak[3]\begin{eqnarray}}
\newcommand{\ea}{\end{eqnarray}}

\newcommand{\bc}{\begin{comment}}
\newcommand{\ec}{\end{comment}}
\newcommand{\Co}{\mathbb{C}}

\newcommand{\eq}[1]{(\ref{#1})}

\newcommand{\bra}[1]{\ensuremath{\langle#1|}}
\newcommand{\ket}[1]{\ensuremath{|#1\rangle}}
\newcommand{\bk}[2]{{\langle#1\,|\,#2\rangle}}
\newcommand{\bek}[3]{{\langle#1\,|\,#2\,|\,#3\rangle}}
\def\vol{{\mathtt{v}}}

\newcommand{\insertion}[1]{{\footnotesize \vskip.3cm
 #1

}\vspace{1em}}

\newcommand{\problem}[1]{{\footnotesize\em 
\vskip.3cm
\noindent\underline{Problem:}  #1

}\vspace{1em}}

\newcommand{\problems}[1]{{\footnotesize\em 
\vskip.3cm
\noindent\underline{Problems:}  #1

}\vspace{1em}}

\newcommand{\C}{\mathbb{C}}
\def\id{1\hspace{-.9mm}\mathrm{l}}
\newcommand{\slc}{$SL(2,\mathbb{C})$ }

\begin{document}
\title{ \Large Zakopane lectures on loop gravity}
     \author{Carlo Rovelli}
     \affiliation{Centre de Physique Th\'eorique de Luminy\footnote{Unit\'e mixte de recherche (UMR 6207) du CNRS et des Universit\'es de Provence (Aix-Marseille I), de la M\'editerran\'ee (Aix-Marseille II) et du Sud (Toulon-Var); affili\'e \`a la FRUMAM (FR 2291).}, Case 907, F-13288 Marseille, EU}
\date{\small  \today}
\begin{abstract} \noindent These are introductory lectures on loop quantum gravity. The theory is presented in self-contained form, without emphasis on its derivation from classical general relativity. Dynamics is given in the covariant form. Some applications are described.
\end{abstract}

\maketitle

\section{Where are we in quantum gravity?} 

Our current knowledge on the elementary structure of nature  is summed up in three theories: quantum theory, the particle-physics standard model (with neutrino mass) and general relativity (with cosmological constant).  With the notable exception of the  ``dark matter" phenomenology, these theories appear to be in accord with virtually {all} present observations.   But there are physical situations where these theories lack predictive power: We do not know the gravitational scattering amplitude for two particles, if the center-of-mass energy is of the order of the impact parameter in Planck units; we miss a reliable framework for very early cosmology, for predicting what happens at the end of the evaporation of a black hole, or describing the quantum structure of spacetime at very small scale. This is because the standard model is based on flat-space quantum field theory (QFT), which disregards the general relativistic aspects of spacetime, and GR disregards quantum theory.  

There are \emph{two} problems raised by this situation. The first is to {complete} the picture and make it {consistent}.  This is called the problem of \emph{quantum gravity}, since what is missing are the quantum properties of gravity.  A second, distinct, problem, is \emph{unification}, namely the hope of reducing the full phenomenology to the manifestation of a single entity.  (Maxwell theory unifies electricity and magnetism, while QCD consistently completes the standard model, but is not unified with electroweak theory.)

Loop quantum gravity (LQG), or loop gravity, is a tentative solution to the  \emph{first} of these problems, and not the second.\footnote{It is sometime said that quantum gravity requires unification. All arguments to this effect rely on assumptions of conventional local QFT which are violated in loop gravity, because of the quantum properties of spacetime itself.}  Its aim is to provide predictions for quantum gravitational phenomena, and a coherent framework for GR and QFT, consistent with the standard model.   LQG is not yet complete, but is  a mature theory, where physical calculations can be performed.

The theory defines a version of QFT that does not disregard the lesson of GR and --the other way around--a theoretical account of space, time and gravitation that does not disregard quantum theory.  The idea that underlies the theory is to take seriously the import of quantum theory as well as that of  GR.  GR has proven spectacularly effective for describing relativistic gravitation. It has achieved this result by modifying in depth the way we describe of space and time. LQG merges the general-relativistic understanding of space and time into QFT.   

GR and quantum theory (adapted to GR's temporal evolution) are therefore the well-established physical ground of LQG. Assuming this ground to remain valid all the way to the Planck scale is a substantial extrapolation. But extrapolation is the most effective tool in science. Maxwell equations, found in a lab, work from nuclear to galactic scale. Up to contrary empirical indications --always possible--, a good bet is that what we have learned so far may well continue to hold.  Some Planck-scale observations are becoming possible today, and their results support the confidence in such extrapolations (see e.g. \cite{Laurent:2011he}). 

Full direct empirical access to quantum gravitational phenomena, on the other hand, is not easy. This is a nuisance. But it is not an obstacle, because the current problem is not to select among different theories of quantum gravity: it is to find at least one complete and consistent theory. LQG aims at providing one such a complete and consistent quantum theory of gravity.

In these lectures I give a self-contained presentation of the theory. I focus on the technical construction of the covariant theory.\footnote{For the alternative, canonical, formulation, see the recent papers  \cite{Sahlmann:2010zf,Domagala:2010bm} and reference therein, or the classic textbook \cite{ThiemannBook}.}  For a wider presentation of the many aspects of LQC and its applications, see \cite{Rovelli:2010bf}. I start by sketching the structure of the theory, below.  Then, Section \ref{Hs} defines states and operators.  Section  \ref{Ta} the transition amplitudes.  Section  \ref{Ap}  some applications.   A list of problems is given in Appendix  \ref{problems}.  Mathematical review, advanced comments, and pointers to alternative formulations are in {\footnotesize smaller characters}.

\subsection{The structure of the theory}\label{intro}
 
LQG utilizes the Ashtekar's formulation of GR \cite{Ashtekar86} and its variants, and can be ``derived" in different ways.  The three major ones are: canonical quantization of GR, covariant quantization of GR on a lattice, and a formal quantization of geometrical ``shapes".  Surprisingly,  these very different techniques and philosophies converge towards the same formalism. The convergence supports the idea that LQG is a natural formalism for general relativistic QFT.  

I sketch these derivations in Section  \ref{derivation}. But in the main part of these lectures I do not follow any of them. Rather, I give directly the definition of the quantum theory, as if I introduced QED by giving the definition of the Hilbert space of photons and electrons, and the Feymnan rules defining the transition amplitudes.  In the rest of this Section, I anticipate a brief non-technical overview of the structure of the theory. 

\subsubsection{States and operators}\label{so}

The gravitational analog of QED's photons and electrons Hilbert space  is defined in Section  \ref{Hs}. The quanta of LQG differ from those of QED, because the Maxwell and Dirac fields live over a fixed spacetime metric space, while the gravitational field forms itself the spacetime metric space. It follows that the quanta of gravity are also ``quanta of space". They do not live in space, but give rise to space themselves.  

\begin{figure}[h]
\centerline{\includegraphics[scale=0.3]{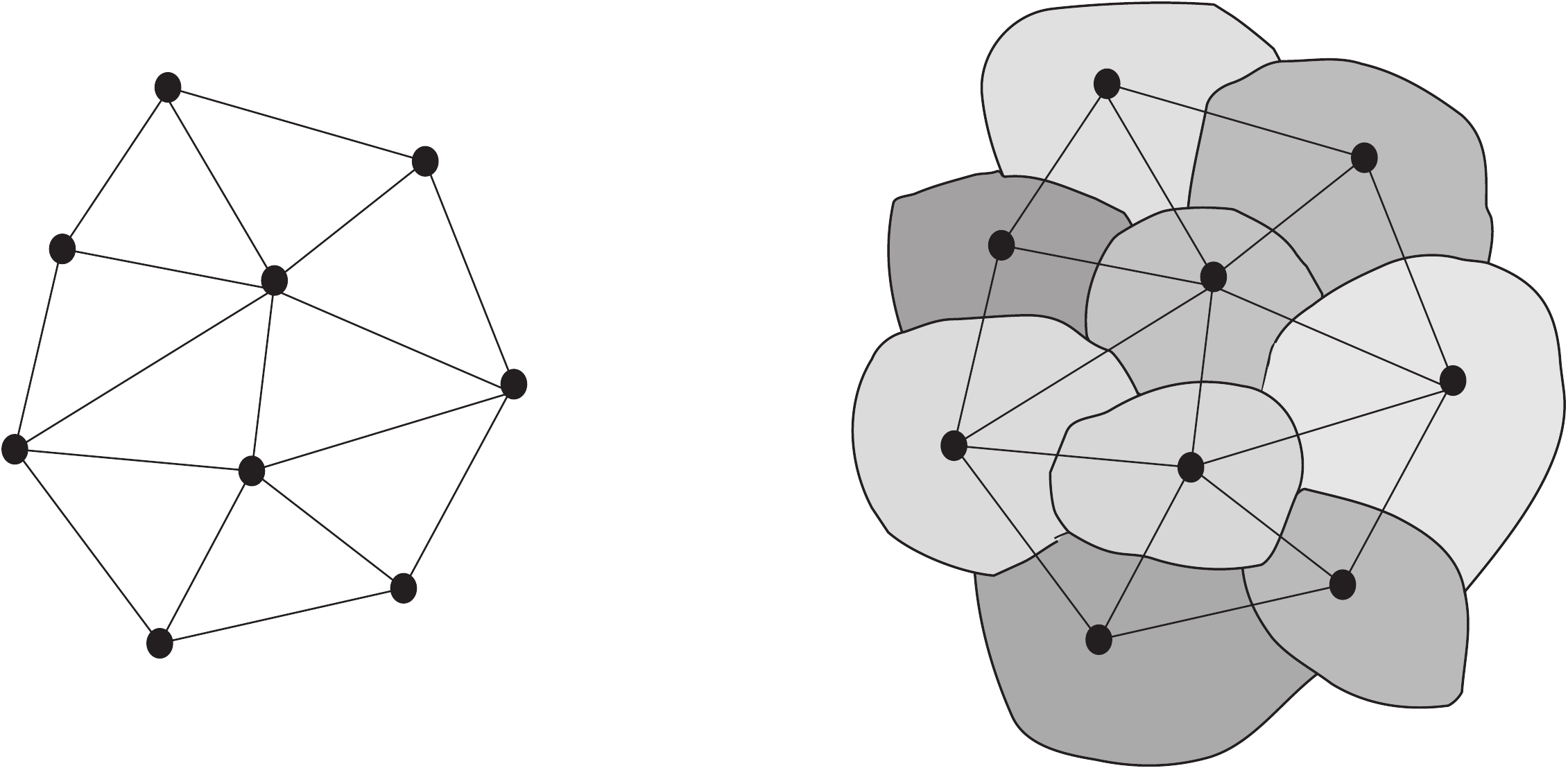}}
$\hspace{-15em}|\Gamma, j_l,v_n\rangle$
\caption{A spin network and the ``quanta of space" it describes.} 
\label{ch1}
\end{figure}

The mathematics needed to describe such quanta of space is provided by the theory of spin networks (essentially graphs colored with spins, see Figure \ref{ch1}), first developed by Roger Penrose, and then independently rediscovered as the result of a textbook canonical quantization of GR in Ashtekar variables. 

The spin-networks Hilbert space is not an exotic object.  It is essentially nothing else than the conventional Hilbert space of $SU(2)$ lattice Yang-Mills theory.  Ashtekar has shown that the kinematics of GR can be neatly cast into the same form as the kinematics of $SU(2)$ Yang-Mills theory.  The other way around, the Hilbert space of  $SU(2)$ Yang-Mills lattice theory admits an interpretation as a description of quantized geometries, formed by quanta of space, as we shall see in a moment.  This  interpretation  forms the content of the ``spin-geometry" theorem by Roger Penrose, and an earlier related theorem by Hermann Minkowski. These two theorems ground the kinematics of LQG.

Alternatively, this same Hilbert space can be seen as the quantization of the moduli space of the flat $SU(2)$ connections on the topologically non-trivial manifold ${\cal M}^*$ obtained removing the one skeleton of a Regge triangulation from the 3d space.  In a Regge geometry, the connection is flat on ${\cal M}^*$, because curvature is concentrated on the Regge bones.  Therefore the moduli space describes Regge geometries, which, in turn, can approximate Riemanian geometries arbitrarily well. 

The resulting description of quantum space is a solid part of LQG.  It provides a clear mathematical and intuitive picture of quantum space, and it is used in all versions of the theory.  

Its most remarkable feature is the discreteness of the geometry at the Planck scale, which appears in this context as a rather conventional  quantization effect:  In GR, the gravitational field determines lengths, areas and volumes. Since the gravitational field is a quantum operator, these quantities are given by quantum operators. Planck scale discreteness follows from the spectral analysis of such operators. 

To avoid a common misunderstanding, I emphasize that the discreteness is not given by the fact that the grains of space in Figure \ref{ch1} are discrete objects. Rather, it is given by the fact that the  \emph{size} of each grain is quantized in discrete steps\footnote{The quantum aspect of a photon is not the discreteness of the Fourier modes of the electromagnetic field in a box: it is the fact that the energy of each mode is quantized in multiples of $h\nu$.}, with minimum non-vanishing size at the Planck scale. This is the key result of the theory, which becomes later responsible of the UV finiteness of the transition amplitudes.

\subsubsection{Transition amplitudes}

The covariant formulation of LQG is based on a concrete definition of the formal ``sum over 4-geometries" of the exponential of the GR action \cite{Misner:1957fk}
\be
   Z \sim \int Dg \ \ e^{\frac{i}{\hbar}\! \int \! \!R\sqrt{g}\,d^4\!x}.
   \label{ZZZ}
\ee 
In the much simpler context of three Euclidean spacetime dimensions, a beautiful and surprising definition of this ``sum over geometries" was found by Ponzano and Regge in 1968 \cite{Ponzano:1968uq} and made rigorous by Turaev and Viro in 1992 \cite{Turaev:1992hq}. The Ponzano-Regge theory fixes a triangulation $\Delta$ of spacetime, assigns half integers, or spins, $j_f$ to each bone (segment) $f$ of $\Delta$ and is defined by the partition function
\be
   Z=\sum_{j_f} \prod_f (2j_f+1)\ \prod_v \{6j\}
   \label{ZPR}
\ee 
where $v$ labels the tetrahedra of $\Delta$ and $\{6j\}$ is the Wigner 6-$j$ symbol (the natural invariant object of $SU(2)$ representation theory constructed with six spins).  Interpret the spins as assigning (discrete) lengths to the bones of a piecewise flat geometry on $\Delta$.  Then Ponzano and Regge showed that, for large spins, $\{6j\}$ is essentially the exponent of the Regge action, which is turns approximates the action of GR. In other words, \eqref{ZPR} provides a simple geometrical way to discretize and define \eqref{ZZZ}. 

The relation between Ponzano-Regge theory and LQG was recognized early, with the realization that the Ponzano-Regge \emph{assumption} that the length of the edges are discrete, is nothing else than the LQG \emph{result} of the discretization of the geometry, in its 3d version \cite{Rovelli:1993kc}. But it is only in the last years that the full power of similarity has emerged, with the discovery of a four dimensional version of the Ponzano-Regge amplitude  \eqref{ZPR}.  This is indeed the 4d amplitude that defines the covariant dynamics of LQG:
\be
   Z=\sum_{j_f,i_e} \prod_f (2j_f+1)\ \prod_v A_v(j_e,i_v),
   \label{ZEPRL}
\ee 
where the spins are now associated to the faces of a cellular decomposition $\Delta$ of spacetime  (or ``foam"), $i_e$ are other $SU(2)$ quantum numbers associated to 3-cells, called intertwiners (defined in the next section); $v$ labels the 4-cells and $A_v(j_e,i_v)$ is a simple generalization of the $\{6j\}$ symbol, which  involves both $SU(2)$ and $SL(2,C)$, which I define in detail in Section \ref{Ta}.  

The main properties of  \eqref{ZEPRL} are the following. \begin{enumerate}
\item[i.] In a suitable semiclassical limit, \eqref{ZEPRL} approaches  \eqref{ZZZ}.   $A_v(j_e,i_v)$ approaches the exponential of the Regge action, which in turns approaches the action of general relativity. Therefore  \eqref{ZEPRL} is a discretization of the path integral for quantum gravity.
\item[iii.]  \eqref{ZEPRL} is ultraviolet finite, a property strictly connected to the Planck discreteness of the spin networks. It admits a quantum-deformed version \cite{Fairbairn:2010cp,Han:2011vn} that describes the cosmological constant coupling \cite{Ding:2011hp} and is IR finite.\footnote{In 3d, this gives the Turaev-Viro theory \cite{Turaev:1992hq}.}  In this version, a theorem assures that  \eqref{ZEPRL} is finite.

\item[iv.] The amplitude   \eqref{ZEPRL}  is for pure gravity, but it can be coupled to fermions and Yang Mills fields  \cite{Bianchi:2010bn}. The finiteness result continues to hold.
 \end{enumerate}

The expression \eqref{ZEPRL}  was found independently and developed during the last few years by a number of research groups \cite{Barrett:1997gw,Engle:2007uq,Livine:2007vk,Engle:2007qf,Freidel:2007py,Engle:2007wy,Kaminski:2009fm}, using different path and different formalisms (and a variety of notations). Different definitions have  later been recognized to be equivalent.  The resulting theory is variously denoted as ``EPRL model",  ``EPRL-FK model",  ``EPRL-FK-KKL model",  ``new BC model"... in the literature.    I call it here simply  the partition function of LQG.  The presentation I  give below does not follow any of the original derivations.   

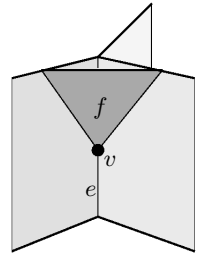
\begin{figure}[h]
\centerline{
\includegraphics[scale=0.2]{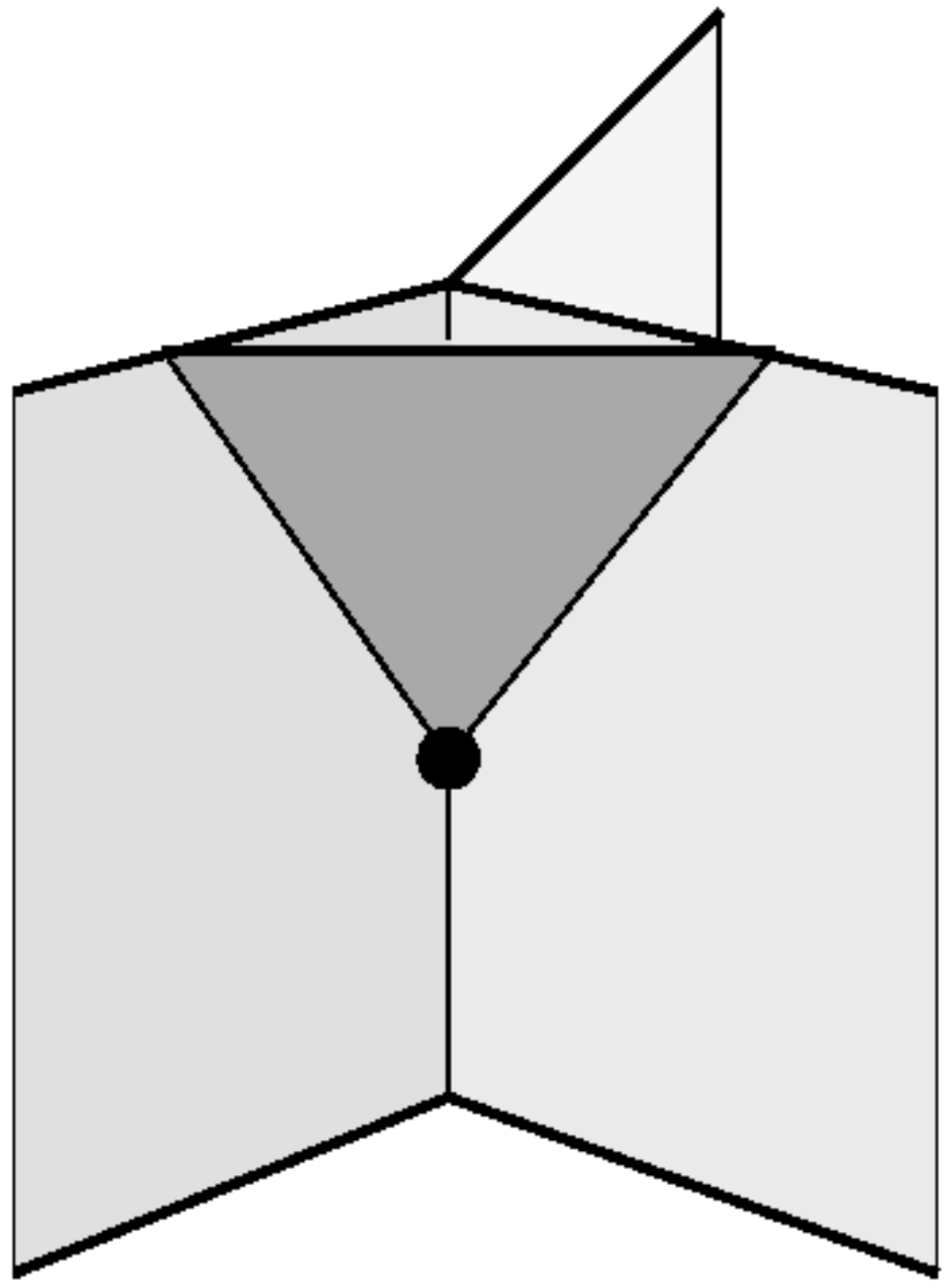}
$\begin{picture}(20,25)
\put(-38,33){\small $v$}
\put(-42,53){\small $f$}
\put(-45,22){\small $e$}
\end{picture}$}
\caption{A simple two-complex with one internal vertex.}
\label{vrtx22}
\end{figure}

It is convenient to view \eqref{ZEPRL} as defined on the dual of the cellular decomposition, or, more precisely, on the two-complex $\cal C$ formed by the two-skeleton of the dual, and extend its definition to two-complexes that do not come from a triangulation.  In $\cal C$, a 4-cell $v$ becomes vertex, a 3-cell $e$ becomes an edge and a face $f$ becomes a dual face; see Figure \ref{vrtx22}. Such a two-complex, colored with spins $j_f$ and intertwiners $i_e$ is called a ``spinfoam". Accordingly, \eqref{ZEPRL} (which I shall denote $Z_{\cal C}$ below to emphasize the dependence on the two-compex) is also called a ``spinfoam sum", or a ``spinfoam model". 

A remarkable aspect of the definition  \eqref{ZEPRL} of the dynamics of LQG is its extreme simplicity.  The amplitude  \eqref{ZEPRL} can also be derived by requiring a certain number of general properties such as locality, linearity and Lorentz invariance to hold \cite{Bianchi:2010Nice}.  It is quite surprising that this simple algebraic definition leads to the Einstein action. 

\subsubsection{Physical amplitudes and the continuum limit}

Because of  diff-invariance, it is not easy to extract physical information from \eqref{ZEPRL} inserting bulk operator, as usually done in QFT. This is a well-known general difficulty of quantum gravity.  But there is an alternative technique that works in quantum gravity, which is to compute \eqref{ZEPRL} on a foam \emph{with boundaries}, as a function of the boundary state \cite{Oeckl:2003vu}. The boundary $\partial{\cal C}$ of a two complex is a (not necessarily connected) graph  $\Gamma$, and the boundary data for \eqref{ZEPRL} are spins on the links and intertwiners on the nodes. That is, they are spin network states, as described above in the subsection \ref{so}:
\be
   W_{\cal C}(j_l,i_n)=\sum_{j_f,i_e} \prod_f (2j_f+1)\ \prod_v A_v(j_e,i_v)\in{\cal H}_{\partial{\cal C}}
   \label{ZEPRL2}
\ee 
where $j_l$ and $i_n$ are the quantum numbers of the boundary spin networks.\footnote{Indeed, a spinfoam can also be viewed as a history of an evolving spin network.  The evolution is non-trivial only at the vertices, where the nodes of the spin network branch.  The branching of the nodes is precisely the form of the evolution generated by the hamiltonian dynamics, in the canonical formulation of LQG.  Indeed, the  result that prompted the interest in LQG in the eighties was precisely the discovery that in the loop representation of quantum general relativity the Wheeler-deWitt operator is trivial everywhere except at the nodes \cite{Rovelli:1987df,Rovelli:1989za}. Historically, spinfoams first appeared in LQG in this form, as histories of spin networks \cite{Reisenberger:1996pu,Reisenberger:2000zc,Reisenberger:2000fy,Baez99,Baez:1997zt}.}

When the boundary graph is formed by several disconnected components, the expression \eqref{ZEPRL2} defines transition amplitudes between spin network states, and standard techniques can then be used to  derive various  other physically meaningful  quantities, for instance quantum cosmology amplitudes or $n$-point functions for the gravitational field over a background. The information about the backgound over which the  $n$-point functions are computed is taken into account in the choice of the boundary states themselves. I illustrate this in Section \ref{Ap}.  

Exact transition amplitudes are defined by a refinement limit, namely by a foam with a large number of vertices. Indeed,  \eqref{ZEPRL} is akin to the lattice definition of QCD, where the continuum theory requires a refinement limit to be taken.  But diff-invariance leads to a fundamental difference in the way the continuum limit is achieved. QCD requires a parameter in the action to be taken to a critical value, while here the continuum limit is defined uniquely by the refinement of the foam \cite{Rovelli:2011fk}. 

More importantly, in a suitable regime of  boundary values, the number vertices can provide a good expansion parameter. The expansion in \emph{small} foams, can be an effective perturbation expansion in an appropriate regime.\footnote{This opens up another interpretation of the spinfoams: on a given foam, \eqref{ZEPRL2} can be seen as a Feynman graph amplitude, a term in an expansion, describing a physical process. \\ In fact, the amplitude \eqref{ZEPRL2} can be literally obtained as the Feynman graph of an auxiliary field theory. This is a very interesting way of formulating the theory, called the ``group field theory" formulation, which I will not explicitly cover here, although I implicitly use techniques derived from it. See \cite{Geloun:2010vj,Krajewski:2010yq,Oriti06}. \\  The double interpretation of \eqref{ZEPRL2} --as Feynman-graphs, as in QED (or high energy QCD), and as a lattice, as in nonperturbative QCD-- is puzzling at first. But a moment of reflection shows that it is natural: a Feynman graph is a history of quanta. The lattice is a collection of small regions of spacetime.  But in quantum gravity  regions of space are quanta of the gravitational field, and therefore the lattice is itself a ``history of quanta of the field", namely a Feynman graph.  Such convergence between the QED perturbative picture and the QCD lattice one is an intriguing feature of the theory.}
   I discuss this technique in Section \ref{Ap}.  

The reason why this may happen is that a refinement of the foam brings each vertex amplitude closer to the flat regime, and in this regime the theory approaches $BF$ theory, which is a topologically invariant theory, namely invariant under a refinement of the lattice. In other words, there is a regime where quantum gravity can be studied as a perturbation of a topological quantum field theory.  The topological QFT plays a role similar to that of the free theory is the conventional perturbation expansion:  a non physical QFT sufficiently well understood to define non trivial theories by a perturbation expansion around it.  

\vskip0.5cm

After these general introductory notes, it is time to start the real work.

\section{States and operators}\label{Hs}

\subsection{Elementary math: $SU(2)$}\label{math1}

\hfill  \begin{minipage}{6.5cm}\small 
\em `` It is the mark of the educated man to look for precision in each class of things just so far as the nature of the subject admits."\\  Aristotle,  Nicomachean Ethics, I,3.
\end{minipage}\\
\vspace{1em}

{\footnotesize 

\noindent LQG uses heavily the group $SU(2)$, its representation theory and the Hilbert spaces of  the square-integrable functions over the group.  Here is a reminder of some elementary facts about these.

$SU(2)$ is the group of $2\times2$ unitary matrices $h$ with unit determinant. A basis in its algebra is provided by the three matrices $\tau_i=-\frac i2 \sigma_i,\ i=1,2,3$, where $\sigma_i= {\tiny \left\{ \left( \begin{array}{cc}   & 1  \\  1 &  \end{array}\right) , \left( \begin{array}{cc}   & -i  \\  i &  \end{array}\right) , \left( \begin{array}{cc} 1  &   \\   &-1  \end{array}\right)\right\}}$ are the Pauli matrices.  Every $h\in SU(2)$ can be written as
\be
h=e^{\alpha^i\tau_i}=\cos\!{\left(\frac\alpha2\right)}\; 1\!\!1+i \sin\!\left(\frac\alpha2\right)\frac{\alpha^i}{\alpha}\,\sigma_i.
\ee
where $\alpha\equiv\sqrt{\alpha_i\alpha_i}<2\pi$ is the rotation angle of the $SO(3)$ rotation corresponding to the $SU(2)$ element $h$.  The group manifold of $SU(2)$ is the three sphere $(x^0)^2+(x^1)^2+(x^2)^2+(x^3)^2=1$, where $x^0=\cos\!\left(\frac\alpha2\right)$ and $x^i =\sin\!{\left(\frac\alpha2\right)}\frac{\alpha^i}{\alpha}$. The standard Haar measure on $SU(2)$ can be written as the invariant  measure on this sphere: $dh=d^4x \ \delta(|x|^2-1)$.

The irreducible unitary representations of $SU(2)$ are labelled by a half-integer $j=0,\frac12, 1,\frac32,  ...$ called ``spin".  The representation space ${\cal H}_j$ has dimension $d_j=2j+1$. The standard basis that diagonalizes $\tau_3$ is denoted $v^m,\ m=-j,...,+j$. The representation matrices are the Wigner matrices  $D^j(h)^m{}_n$. The spin-$j$ character is defined as 
$\chi^j(h)=tr[D^j(h)]$. A key property of these matrices is to be orthogonal in the Haar measure
\be
\int_{SU(2)}\hspace{-1.5em} dh \  \ \overline{D^{j'}(h)^{m'}{}_{n'}}\  \ D^j\!(h)^m{}_n=\frac1d_j \delta^{jj'}
 \delta^{mm'} \delta_{nn'}.
 \label{ortho}
\ee
Using the fact that the Wigner matrices are unitary, this can also be written in the more useful form
\be
\int_{SU(2)}\hspace{-1.5em} dh \  \  {D^{j'}\!(h^{\scriptscriptstyle-1})_{n'}{}^{m'}}\  \  D^j\!(h)^m{}_n= 
\frac1d_j \delta^{jj'}
 \delta^{mm'} \delta_{nn'}.
 \label{wigner}
\ee
which admits the simple graphical representation:
\be
\int_{SU(2)} \!\!\! dh \ \ \ 
\begin{picture}(20,25)
\put(8,-15){\tiny n'}
\put(8,22){\tiny m'}
\put(10,5){\circle*{5}}
\put(10,18){\vector(0,-1){28}}
\hspace{1em}
\begin{picture}(20,30)
\put(8,-15){\tiny n}
\put(8,22){\tiny m}
\put(10,5){\circle*{5}}
\put(10,-10){\vector(0,1){28}}
\end{picture}
\end{picture}
\hspace{1em}
= 
\hspace{.5em}
\frac1{d_j} \ \ 
\begin{picture}(20,25)
\put(0,-15){\tiny n'}
\put(0,22){\tiny m'}
\put(10,-15){\tiny n}
\put(10,22){\tiny m}
\qbezier(3,17)(6.5,7)(10,17)
\qbezier(3,-8)(6.5,3)(10,-8)
\end{picture}.
\ee
\vskip3mm

Spaces of functions on the group play an important role in LQG: in particular, the space  $L_2[SU(2)]$ of the functions square-integrable in the Haar measure. Because of the orthogonality \eqref{ortho}, the Wigner matrices form a basis in this space. Writing, in Dirac notation, $D^j(h)^m{}_n=\langle h|j,m,n\rangle$ and $\langle \psi|\phi \rangle = \int dh \overline{\psi(h)}\phi(h)$, equation \eqref{ortho} reads 
\be
\langle j',m',n'|j,m,n\rangle= \delta^{jj'}
 \delta^{mm'} \delta_{nn'}\ \frac{1}{2j+1}.
\ee
This is the content of the Peter-Weyl theorem, which plays a major role below.  This can equally be expressed as follows.  Since $D^j:{\cal H}_j\to {\cal H}_j$, we can write $D^j\in({\cal H}_j^* \otimes {\cal H}_j)$ and the Peter-Weyl theorem can be expressed in the useful notation
\be
L_2[SU(2)]=\oplus_j \left({\cal H}_j^*\otimes {\cal H}_j\right).
\ee

Some operators are naturally defined on $L_2[SU(2)]$. The (matrix elements of the) $SU(2)$ group element $h$ act as multiplicative operators.  The (hermitian) left and right invariant vector fields $\vec L=\{L_i\}$ and $\vec R=\{R_i\}$ are defined by
\be
L_i\psi(h)\equiv \left.i\frac{d}{dt}\, \psi(he^{t \tau_i})\right|_{t=0},\label{left}
\ \ \ \ 
R_i\psi(h)\equiv \left.i\frac{d}{dt}\, \psi(e^{t \tau_i}h)\right|_{t=0}.
\ee
Acting on the Wigner matrices, they give 
\be
\vec L\ D^j(h)=i\ D^j(h)\; \vec J^j, \ \ \ \ \   
\vec R\ D^j(h)=i\ \vec J^j\; D^j(h),
\ee
where $\vec J^j$ are the (anti-hermitian) generators in the representation $j$.  The Casimir operator $L^2:=L_iL_i$ acts on the individual ${\cal H}_j$  in the Peter-Weyl decomposition (because it acts only on the $m$ indices and not on the $j$ indices) and is diagonal in the spins 
\be
L^2\, D^j(h)=j(j+1)\, D^j(h).
\ee

Given $k$ spins $j_1, ..., j_k$, the tensor product ${\cal H}_{j_1}\otimes ...
\otimes {\cal H}_{j_k}$ is the space of the tensors $i^{m_1...m_k}$ with indices in different representations. This tensor product can be decomposed into irreducibles, as in standard angular momentum theory. In particular, its invariant subspace is formed by the invariant tensors, satisfying $D^{j_1}(h)^{m_1}{}_{n_1} ... D^{j_k}(h)^{m_k}{}_{n_k}  i^{n_1...n_k}= i^{m_1...m_k}$. These are called \emph{intertwiners} and the linear space they span 
\be
{\cal K}_{j_1...j_k}={\rm Inv}[{\cal H}_{j_1}\otimes ...\otimes {\cal H}_{j_k}]
\ee
is called ``intertwiner space". Examples of invariant tensors are the fully antisymmetric tensor $i_{ijk}=\epsilon_{ijk}$ in  ${\cal K}_{111}={\cal H}_{1}\otimes {\cal H}_{1}\otimes {\cal H}_{1}$, the tensor $i_i{}^A{}_B=\sigma_i{}^A{}_B$ formed by the components of the Pauli matrices  in ${\cal K}_{1\frac12\frac12}={\cal H}_{1}\otimes {\cal H}_{\frac12}\otimes {\cal H}^*_{\frac12}$, and the three tensors $i_{ijkl}=\delta_{ij}\delta_{kl}$,  $i'_{ijkl}=\delta_{ik}\delta_{jl}$ and  $i'''_{ijkl}=\delta_{il}\delta_{jk}$ in ${\cal K}_{1111}={\cal H}_{1}\otimes {\cal H}_{1}\otimes {\cal H}_{1}\otimes {\cal H}_{1}$. Since an $SU(2)$ representation appears at most once in the tensor product of two others, it is easy to see that ${\cal K}_{j_1j_2j_3}$ is always 1-dimensional, and therefore $i_{ijk}$ and $i_i{}^A{}_B$ are unique (up to scaling); while  ${\cal K}_{1111}$ is 3-dimensional.

\problems{(i) Compute the normalization and the scalar products between the three intertwiners in   ${\cal K}_{1111}$ mentioned above. Find an orthonormal basis.  (ii) Find the dimension of ${\cal K}_{\frac12\frac12\frac12\frac12}$. (iii) Find an orthonormal basis.}

\subsection{Elementary math: graphs}\label{math1}

Graphs play a role in the following. Roughly, the adjacency relations (who is next to who) between the elementary quanta of space is described by graphs.  

A graph $\Gamma$ is a combinatorial object.  It is defined as a triple $\Gamma=({\cal L},{\cal N},\partial)$, where $\cal L$ is a finite set of $L$ elements $l$, which we call ``links", $\cal N$ is a finite set of $N$ elements $n$, which we call ``nodes", and the boundary relation $\partial=(s,t)$ is an (ordered) couple of functions $s:{\cal L}\to{\cal N}$ called ``source" and $t:{\cal L}\to{\cal N}$, called ``target". 

The simplest way of visualizing a graph is of course to imagine the nodes as points and the links as (oriented) lines that join these points. Each link goes from its source to its target.

\begin{figure}[h]
\centerline{\includegraphics[scale=0.15]{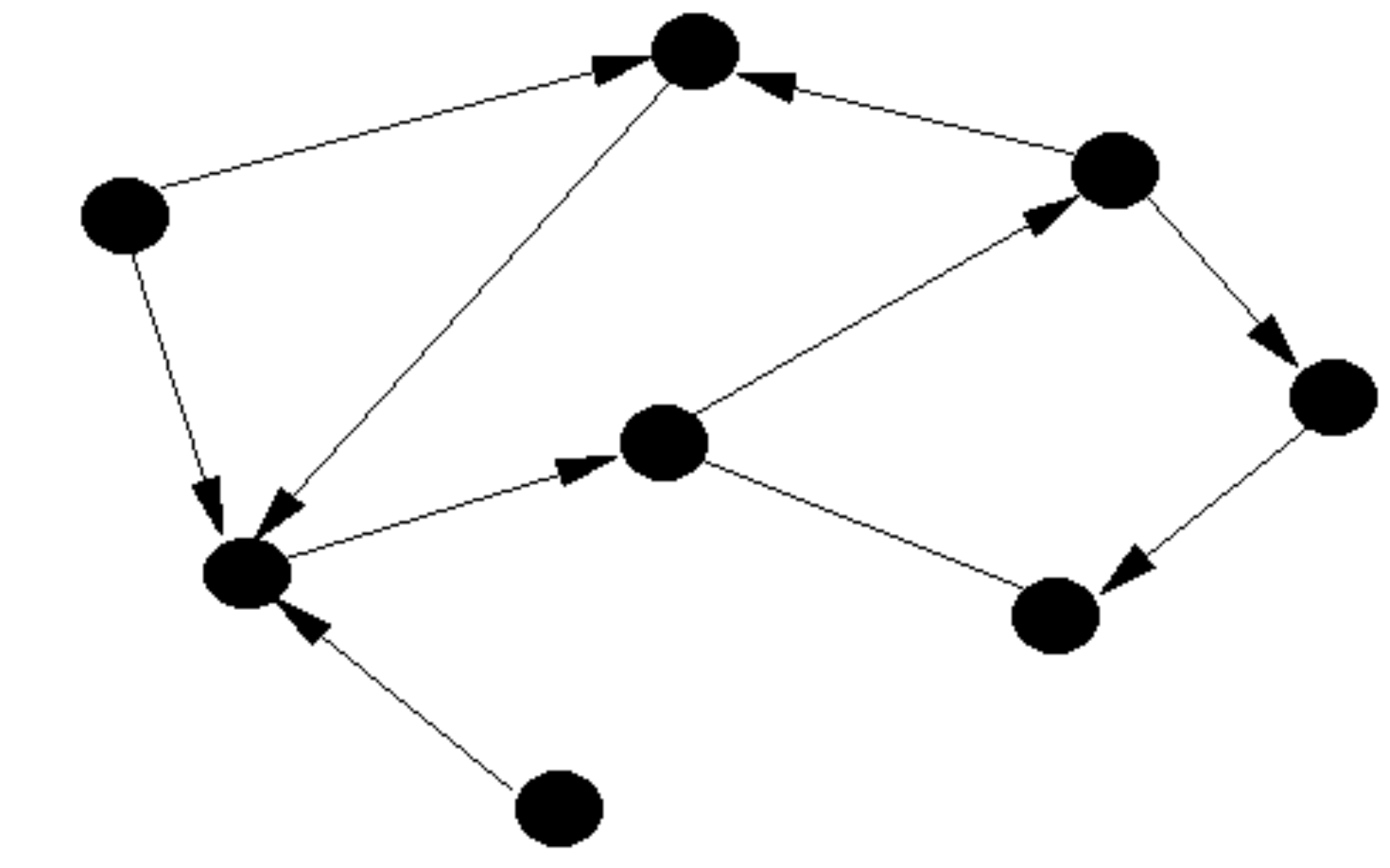}}
%\vspace{-6mm}
\caption{{Picturing of an graph with $N=8$ and $L=10$.}} 
\end{figure}

It is convenient to define also a ``reversed" link $l^{-1}$ for each $l$, where $s(l)=t(l^{-1})$ and $t(l)=s(l^{-1})$. I use the notation $l\in n$ to indicate that $l$ is a link or a reversed link with $s(l)=n$.  Thus the set of $l\in n$ is the set of the oriented links bounded by the node $n$, all considered with outgoing orientation.

An automorphism of a graph is a map from $\Gamma$ to itself which preserves $\partial$. Given a graph, its automorphisms form a discrete group.

We say that a graph $\Gamma'$ is a subgraph of a graph $\Gamma$, we write $\Gamma'\le\Gamma$ and we say that  $\Gamma$ ``contains" $\Gamma'$,  if there exists a map sending the links and the nodes of $\Gamma'$ into the links and the nodes of $\Gamma$, preserving the boundary relations.  Of course if $\Gamma$ contains $\Gamma'$ there may be more than one of the maps.  That is, it might be possible to ``place" $\Gamma'$ into $\Gamma$ in different manners.

The relation $\le$ equips the set of all graphs with a partial order. This order admits an upper bound, namely given any two graphs $\Gamma$ and $\Gamma'$ we can always find a third graph $\Gamma''$ which contains  both  $\Gamma$ and $\Gamma'$.  This implies that we can define the limit $\Gamma\to\infty$ of any quantity $f_{\Gamma}$ that depends on graphs. If it exists, we say that 
\be
        f_{\infty}= \lim_{\Gamma\to\infty} f_{\Gamma}
\ee
if for any $\epsilon$ there is a $\Gamma'$ such that $|f_{\infty}-f_{\Gamma}|<\epsilon$ for all $\Gamma>\Gamma'$.

 }
 \problem{What is $ \lim_{\Gamma\to\infty} 1/L$?
 }
 
 \vspace{.5em}
 
With these preliminaries, we are ready to define states and observables of quantum gravity. The kinematics of any quantum theory is given by a Hilbert space $\cal H$ carrying an algebra of operators $A$ that have a physical interpretation in terms of observables quantities of the system considered. Let us start with $\cal H$.

\subsection{Hilbert space}\label{Hilbert}

Let me start by recalling the structure of the Hilbert spaces used in QED and in QCD. A key step in constructing any interactive QFT is always a \emph{finite} truncation of the dynamical degrees of freedom. In weakly coupled theories, such as low-energy QED or high-energy QCD, the truncation is provided by the particle structure of the free field, which allows us to consider virtual processes involving \emph{finitely many} particles, described by Feynman diagrams. 
In strongly coupled theories, such as QCD, we can resort to a non-perturbative truncation, such as a \emph{finite} lattice approximation.\footnote{In either case, the relevant effect of the remaining degrees of freedom can be subsumed under a dressing of the coupling constants, if a criterion of renormalizability or criticality is met.} The full theory is then formally obtained as a limit where all degrees of freedom are recovered.  

In the first case, we can start by defining the single-particle Hilbert space $ {\cal H}_1$. For a massive scalar theory, for instance, this can be taken to be the space ${\cal H}_1=L_2[M]$ of the square-integrable functions on the Lorentz hyperboloid $M$.  The $n$-particle Hilbert space is
\be
{\cal H}_n=L_2[M^n]/\sim
\ee
 where the factorization is by the equivalence relation determined by the action of the permutation group, which symmetrizes the states.  The Hilbert space
\be
 {\cal H}_N= \bigoplus_{n=0}^N{\ \cal H}_n 
\ee
contains all states up to $N$ particles, and is actually sufficient for all calculations in perturbation theory.  ${\cal H}_N$ is naturally a subspace of ${\cal H}_{N'}$ for $N<N'$, and the full Fock space is the limit 
\be
 {\cal H}_{\rm Fock}=\lim_{N\to\infty}{\cal H}_N. 
\ee

In the second case, namely in lattice gauge theory, the canonical theory is defined on a lattice $\Gamma$ with, say, $L$ links $l$ and $N$ nodes $n$. The variables are group elements $U_l\in G$, where $G$ is the gauge group, associated to the links, and the non-gauge-invariant Hilbert space is  \cite{KoguthSusskind}
\be
\tilde{\cal H}_{\Gamma}=L_2[G^L].
\ee
Gauge transformations act on the states $\psi(h_l)\in \tilde{\cal H}_{\Gamma}$ at the nodes as
\be
\psi(h_l) \to \psi(g_{\,s(l)}\,h_l\,g_{t(l)}^{-1}),\hspace{2em}Êg_n\in G
\label{gauge}
\ee
and the space of gauge invariant states is the physical Hilbert space 
\be
{\cal H}_\Gamma= L_2[G^L/G^{N}]
\label{gilgt}
\ee
formed by the states invariant under \eqref{gauge}.  Again the full theory is obtained by appropriately taking $L$ and $N$ to infinity.  

The Hilbert space of LQG has aspects in common with both these constructions. Let me now define it in three steps:
\begin{enumerate}
\item[(i)]
For each graph $\Gamma$, consider a ``graph space"
\be
{\cal H}_\Gamma= L_2[SU(2)^L/SU(2)^{N}]
\label{n-1}
\ee
which is precisely the Hilbert space \eqref{gilgt} of an $SU(2)$ lattice gauge theory, over a graph which is not necessarily a cubic lattice. As we shall see, the local $SU(2)$ gauge is related to the freedom of rotating a 3d reference frame in space.

\item[(ii)]
If $\Gamma$ is a subgraph of $\Gamma'$ then ${\cal H}_{\Gamma}$ can be naturally identified with a subspace of  ${\cal H}_{\Gamma'}$ (the subspace formed by the states $\psi(h_l)\in {\cal H}_{\Gamma'}$ depending on $h_l$ only if $l$ is in the subgraph $\Gamma$). 
Define an equivalence relation $\sim$ as follows: two states are equivalent if they can be related (possibly indirectly) by this identification, or if they are mapped into each other by the group of the automorphisms of $\Gamma$.  Let 
\be
\tilde{\cal H}_\Gamma= {\cal H}_\Gamma/\sim.
\ee 
\item[(iii)]
The full Hilbert space of quantum gravity is finally defined as 
\be
    {\cal H}= \lim_{\Gamma\to\infty} \tilde{\cal H}_\Gamma.
    \label{graphspaces}
\ee\\[-3mm]
It is separable.\end{enumerate}
  This completes the construction of the Hilbert space of the theory. 

${\cal H}$ has aspects in common with Fock space as well as with the state space of lattice gauge theory.  As we shall see, states in ${\cal H}_\Gamma$ can be viewed as formed by $N$ quanta, where $N$ is the number of nodes of the graph.  Thus, each node of the graph is like a particle in QED, namely a quantum of electromagnetic field. Here, each node represents a quantum of gravitational field.   

But there is a key difference. QED Fock quanta carry quantum numbers coding where they are located in the background space-manifold. Here, since in general relativity the gravitational field is also physical space, individual quanta of gravity are also quanta of space. Therefore they do not carry information about their localization in space, but only information about the \emph{relative} location with respect to one another. This information is coded by the graph structure.  Thus, the quanta of gravity form themselves the texture of of physical space.  Therefore the graphs in \eqref{graphspaces} can \emph{also} be seen as a generalization of the lattices of lattice QCD.  

This convergence between the perturbative-QED picture and the lattice-QCD picture follows directly from the key physics of general relativity:  the fact that the gravitational field is physical  space itself. Indeed, the lattice sites of lattice QCD are small regions of space; according to general relativity, these are excitations of the gravitational field, therefore they are themselves quanta of a (generally covariant) quantum field theory.  An $N$-quanta state of gravity has therefore the same structure as a Yang-Mills state on a lattice with $N$ sites. This convergence between the perturbative-QED and the lattice-QCD pictures is a beautiful feature of loop gravity. 

Another similarity appears in the factorization by graph automorphisms, which is analogous to the symmetrization of  individual particle states defining the Fock $n$-particle states.\\[-2.mm]

{\footnotesize \emph{Comments.}  This is the ``combinatorial $\cal H$". An alternative studied in the literature is to consider embedded graphs in a fixed three-manifold $\Sigma$ --namely collections of lines $l$ embedded in $\Sigma$ that meet only at their end points $n$-- and to define $\Gamma$ as an equivalence class of such embedded graphs under diffeomorphisms of $\Sigma$. This choice defines the ``Diff $\cal H$". A third alternative is to do the same but using \emph{extended} diffeomorphisms \cite{Fairbairn:2004qe}. This choice defines the ``Extended-Diff $\cal H$". With these definitions a graph is characterized also by its knotting and linking. (If $\Sigma$ is chosen with non-trivial topology, by the homotopy class of the graph as well). In addition, with the first of these alternatives graphs are characterized by moduli parameters at the nodes as well (extended diffeos factor away these moduli  \cite{Fairbairn:2004qe}). The space Diff $\cal H$ is non-separable, leading to a number of complications in the construction of the theory.  The combinatorial $\cal H$ considered here and the extended-Diff $\cal H$ are separable. 

Neither knotting or linking, nor the moduli, have found a physical meaning so far, hence I tentatively prefer the combinatorial definition.  But there are also ideas and interesting attempts to interpret knotting and linking as matter degrees of freedom  \cite{BilsonThompson:2006yc,BilsonThompson:2009fh,Gurau:2010nd}. If it worked it would be very remarkable success, but it is a long shot.

Another option, which I found particularly interesting, and I would instinctively favor, is to restrict the graphs to those which are dual to a cellular decomposition of three space. 

More restrictive is to only consider graphs that are dual to triangulations, namely to restrict the theory to graphs $\Gamma$ where all nodes are four valent. (The valence of a node $n$ is the number of links for which $n$ is the source plus the number of links for which it is the target.)  I do not take this option here, although several results in the literature refer to the theory restricted in this manner.

\subsection{GR as a topological theory, I}\label{tqft1}

There is another very interesting way of interpreting the Hilbert space ${\cal H}_\Gamma$, pointed out by Eugenio Bianchi \cite{Bianchi:2009tj}. Consider a Regge geometry in three (euclidean) dimensions. That is, consider a triangulation (or, more in general, a cellular decomposition) of a 3d manifold $\cal M$, where every cell is flat and curvature, determined by the deficit angles, is concentrated on the bones.  Let $\Delta_1$ be one-skeleton of the cellular decomposition, namely the union of all the bones.  

Notice that the spin connection of the Regge metric is flat everywhere except on $\Delta_1$. Consider the space ${\cal M}^*={\cal M}-\Delta_1$ obtained removing all the bones from $\cal M$. Let $\cal A$ be the moduli space of the flat connections on ${\cal M}^*$ modulo gauge trasformations.  

A moment of reflection will convince the reader that this is precisely the configuration space $[SU(2)^L/SU(2)^N]$ considered above, determined by the graph $\Gamma$ which is dual to the cellular decomposition. This is the graph obtained by representing each cell by a node and connecting any two nodes by a link if the corresponding cells are adjacent.  It is the graph capturing the fundamental group of ${\cal M}^*$.

Therefore the Hilbert space ${\cal H}_\Gamma$ is naturally a quantization of a 3d Regge geometry. Since Regge geometries can approximate Riemanian geometries arbitrarily well, this can be see as a way to capture quantum states of 3d geometries.

The precise relation between these variables and geometry becomes more clear in light of the Ashtekar formulation of GR. Ashtekar has shown that GR can be formulated using the kinematics of an SU(2) YM theory.  The canonical variable is an $SU(2)$ connection and the corresponding conjugate momentum is the triad field. Accordingly, we might expect that the quantum derivative operators on the wave functions on ${\cal H}_\Gamma$ represent the triad, namely metric information. We'll  see below that this in indeed the case.  

A word of caveat: in the Ashtekar formalism, the $SU(2)$ connection is not the spin connection $\Gamma$ of the triad: it is a linear combination of $\Gamma$ and the extrinsic curvature.  Therefore the momentum conjugate the connection will code information about the metric, while the information about the conjugate variable, namely the extrinsic curvature, is included in the connection itself, or, in the discretization, in the group elements $h_l$.

}

\subsection{Operators}

The fact that ${\cal H}$ can be interpreted as a space describing quanta of space follows from its structure, as revealed by a crucial theorem due to Roger Penrose.   Indeed, each Hilbert space ${\cal H}_\Gamma$ has a natural interpretation as a space of quantum metrics, early recognized by Penrose. Let's see how this happens.  

The momentum operator on the Hilbert space of a particle $L_2[R]$ is the derivative operator $\vec p=-i\nabla=-i\frac{d}{d\vec x}$. The corresponding natural ``momentum" operator on $L_2[SU2]$ is the derivative operator \eqref{left}. There is one of these for each link, call it $\vec L_l$. \\[-2.mm]

\insertion{As in lattice gauge theory, operators are defined on the individual spaces ${\cal H}_\Gamma$, not on $\cal H$.  Later I explain how these operators are used in computing observable quantities.}

Because of the gauge invariance \eqref{gauge}, we have
\be
      C_n=\sum_{l\in n} \vec L_l = 0.
\label{closure}
\ee
at each node $n$.       
The operator $\vec L_l$ is not gauge invariant, namely it is not defined on gauge invariant functions. But is easy to write a gauge invariant operator:
\be
G_{ll'}= \vec L_l\cdot \vec L_{l'}
\ee
where $s(l)=s(l')=n$. For reasons that will be clear in a moment, call this operator the ``metric operator".  In particular, denote the diagonal entries of $G_{ll'}$ as
\be
A^2_l=G_{ll}.
\ee

The operator $G_{ll'}$ coincides with Penrose's metric operator \cite{Major:1999mc}. 
Penrose spin-geometry theorem states that the operator $G_{ll'}$ can be interpreted as defining angles in three dimensional  space, at each node \cite{Penrose2,Moussouris:1983uq,Major:1999mc}. The theorem states that these angles obey the dependency relations expected of angles in three dimensional  space.  

I give here in more detail a sharper version of Penrose's original spin-geometry theorem, based on a result by Minkowski.  Consider the classical limit of the Hilbert space ${\cal H}_\Gamma$, that is, consider classical quantities $\vec L_l$ satisfying \eqref{closure}. Minkowski's  theorem \cite{Minkowski:1897uq} states that whenever there are $F$ non-coplanar 3-vectors $\vec L_l$ satisfying the condition  \eqref{closure}, there exists a convex polyhedron in $\mathbb{R}^3$, whose faces have outward normals parallel to $\vec L_l$ and areas $A_l$. 

\problem{Consider a solid polyhedron immersed in a fluid with constant pressure $p$. What is the force on one face due to the pressure? What is the total force on the polyhedron due to pressure? Derive \eqref{closure} from this. (This is the proof of  \eqref{closure} Roger Penrose immediately came up with, when I mentioned him Minkowski theorem.) }
 
\begin{figure}[h]
\centerline{\includegraphics[scale=0.3]{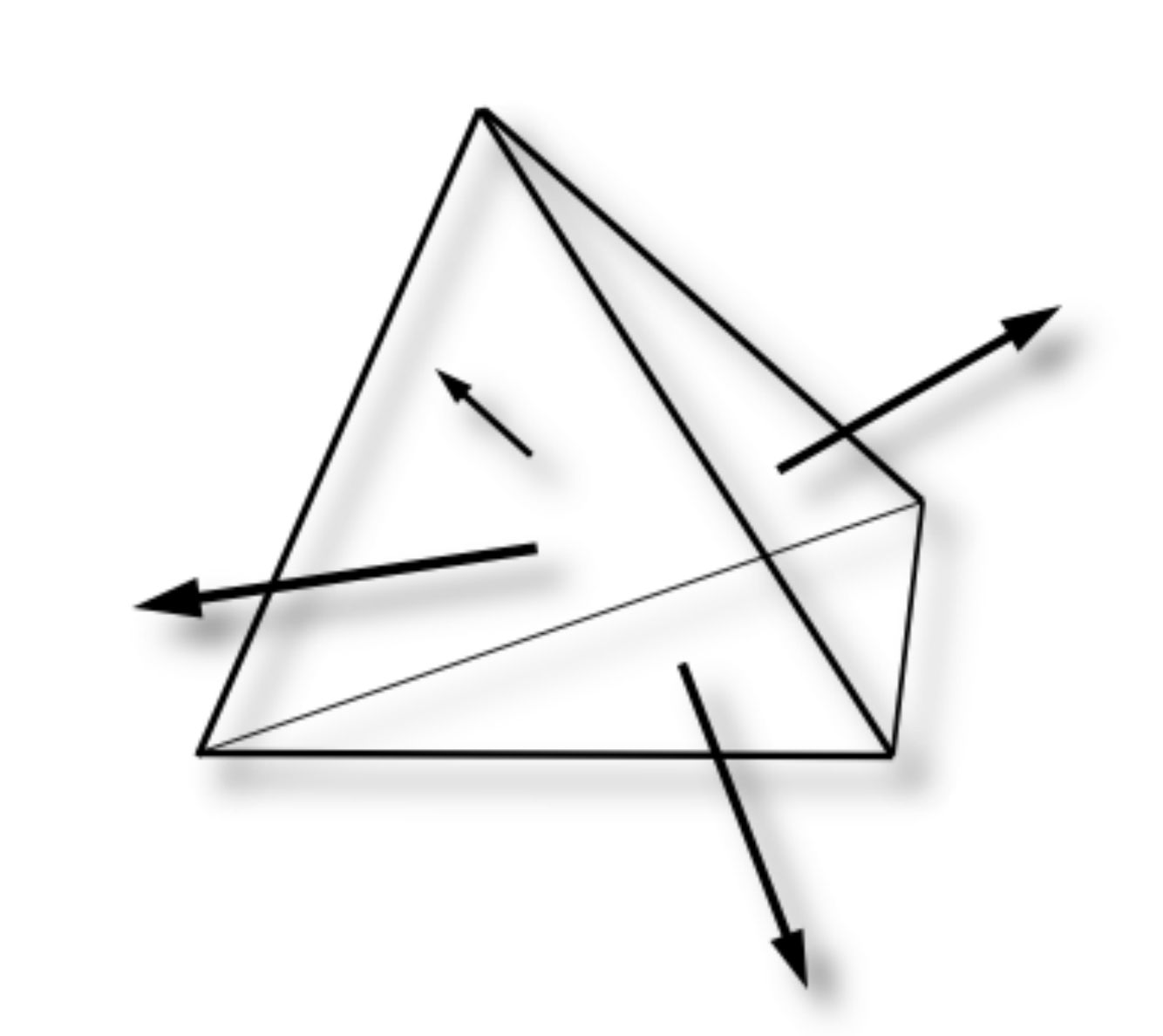}}
\caption{Normals (here arrows)  to the faces, and proportional to the area of the face, satisfy \eqref{closure} and uniquely determine the polyhedron (here a tetrahedron).} 
\label{normals}
\end{figure}

The resulting polyhedron is unique, up to rotation and translation. It follows that if we write
$G_{ll'}=A_lA_{l'}\cos(\theta_{ll'})$, then the quantities $\theta_{ll'}$ satisfy all the relations satisfied by the angles normal to the faces of the polyhedra. The operators $G_{ll'}$ fully capture the ``shape" of the polyhedron, namely its (flat) metric geometry, up to rotations.  In other words, in the classical limit the states in the Hilbert space ${\cal H}_\Gamma$ describe a collection of flat polyhedra with different shapes, one per each node of $\Gamma$. The quantum operators $A_l$ can be interpreted as giving the areas of these faces and the quantum operators $G_{ll'}$ as the (cosine of the) angles between two faces (multiplied by the areas). See Figure \ref{normals}.

What has a collection of polyhedra to do with the gravitational field, which is classically described by a continuous metric?  The answer is suggested by Regge gravity: a collection of flat polyhedra glued to one another defines a (non-differentiable) metric, where the curvature is concentrated on the edges of the polyhedra.   Thus, a collection of glued polyhedra provides a discretized geometry, and therefore a gravitational field up to a finite truncation of its degrees of freedom. 

Thus, the Hilbert space ${\cal H}_\Gamma$ describes a truncation of the degrees of freedom of GR, like the $N$-particle Hilbert space of QED, or the Hilbert space of lattice QCD, describe  truncations of the degrees of freedom of a Yang-Mills field.

Using standard geometrical relations, we can write the volume of these polyhedra in terms of the $\vec L_l$ operators.  For instance, for a 4-valent node $n$, bounding the links $l_1,...,l_4$ the volume operator $V_{n}$ is given by the expression for the volume of a tetrahedron
\be 
V_{n}=\frac{\sqrt{2}}3\ \sqrt{|\vec L_{l_1}\cdot(\vec L_{l_2}\times \vec L_{l_3})|};
\ee
gauge invariance \eqref{gauge} at the node ensures that this definition does not depend on which triple of links is chosen.\\ 

\insertion{As pointed out by Thomas Thiemann and Cecilia Flori \cite{Flori:2008nw}, the definition of the vertex operator for higher valent nodes given in the literature, is not entirely satisfactory. A good definition of the operator for general $n$-valent nodes (which reduces to the one on $(n-1)$-valent nodes when one of the links has zero spin) is still missing (see \cite{Bianchi:2010gc} for an interesting track). This does not affect what follows.}

\problem{Derive the $\frac{\sqrt{2}}3$ factor: Hint: take a tetrahedron having three sides equal to the three orthogonal basis vectors in $R^3$.  It has three faces orthogonal to one another and with area $\frac12$. Therefore the triple product gives  $\left(\frac12\right)^3$.}

Notice that the volume operator $V_n$ acts precisely on the node space ${\cal K}_n$, which, I recall is the space of the intertwiners between the representations associated to the node $n$. It is therefore convenient to choose at each node $n$ a basis of intertwiners $\vol_n$ that diagonalizes the volume operator, and label it with the corresponding eigenvalue $\vol_n$. I use the same notation  $\vol_n$ for the intertwiner and for its eigenvalue. 

\problem{(Important!) Find the basis that diagonalizes the volume in  ${\cal K}_{\frac12\frac12\frac12\frac12}$ and in  ${\cal K}_{1111}$. }

Finally, the \emph{holonomy} operator is the multiplicative operator $h_l$ associated to each link $l$. The operators $\vec L_l$ and $h_l$ form a closed algebra and are the basic operators in terms of which all other observables are built, like the creation and annihilation operators in QFT.

\subsection{Spin network basis}

While the full set of $SU(2)$ invariant operators $G_{ll'}$ do not commute, the Area and Volume operators  $A_l$ and $V_n$ commute. In fact, they form a complete set of commuting observables in ${\cal H}_\Gamma$, in the sense of Dirac (up to possible accidental degeneracies in the spectrum of $V_n$).  We call the orthonormal basis that diagonalizes these operators the \emph{spin-network} basis. 

This basis has a well defined physical and geometrical interpretation \cite{Rovelli:1995ac,Baez95a,Baez95aa}.  The basis can be obtained via the Peter-Weyl theorem and it is defined by 
\be
\psi_{\Gamma, j_l,\vol_n}(h_l)=\big\langle \otimes_l\,d_{j_l}\, D^{j_l}(h_l)\; \big |\; \otimes_n \vol_n\; \big\rangle {}_\Gamma
\label{sn}
\ee
where $D^{j_l}(h_l)$ is the Wigner matrix in the spin-$j$ representation and $\langle \cdot |\cdot  \rangle_\Gamma$ indicates the pattern of index contraction between the indices of the matrix elements and those of the intertwiners given by the structure of the graph.

%\footnote{Both tensor products live in  \be H_\Gamma\subset L_2[(SU2)^L]=\bigoplus_{j_l} \bigotimes_l\, {\cal V}_{j_l}\otimes{\cal V}_{j_l}=\bigoplus_{j_l}  \bigotimes_n\bigotimes_{e\in \partial n}{\cal V}_{j_l}. \ee where ${\cal V}_j$ is the $SU2$ spin-$j$ representation space, here identified with its dual.}  

Since the Area is the $SU(2)$ Casimir, while the volume is diagonal in the intertwiner spaces, the spin $j_l$ is easily recognized as the Area quantum number and $\vol_n$ as the Volume quantum number. \\

\insertion{More in detail, the Peter-Weyl theorem states that $L_2[SU(2)^L]$ can be decomposed into irreducible representations
\be
L_2[SU(2)^L]= \bigoplus_{j_l}\ \bigotimes_l  ({\cal H}_{j_l}^*\otimes {\cal H}_{j_l}). \label{PW}
\ee
Here ${\cal H}_j$ is the Hilbert space of the spin-$j$ representation of $SU(2)$, namely a $2j+1$ dimensional space, with a basis $|j,m\rangle, m=-j,...,j$ that diagonalizes $L^3$. 
The star indicates the adjoint representation, but since the representations of $SU(2)$ are equivalent to their adjoint, we can forget about the star.\footnote{The star does not regard the Hilbert space itself: it specifies the way it transforms under $SU(2)$.} For each link $l$, the two factors in the r.h.s.~of \eq{PW} are naturally associated to the two nodes $s(l)$ and $t(l)$ that bound $l$, because under (\ref{gauge}) they transform under the action of $g_{s(l)}$ and  $g_{t(l)}$, respectively. 
We can hence rewrite the last equation as 
\be
L_2[SU(2)^L]= \bigoplus_{j_l}\ \bigotimes_n {\cal H}_n
\label{sumrep}
\ee
where the node Hilbert space ${\cal H}_n$ associated to a node $n$ includes all the irreducible ${\cal H}_j$ that transform with $g_n$ under (\ref{gauge}), that 
is\footnote{More precisely, ${\cal H}_n = (\bigotimes _{s(l)=n} {\cal H}^*_l)\otimes (\bigotimes _{t(l)=n} {\cal H}_l)$.}
\be
{\cal H}_n = \bigotimes _{l\in n} {\cal H}_{j_l}.
\label{calHn}
\ee
The $SU(2)$ invariant part of this space
\be
{\cal K}_n = {\rm Inv}_{SU(2)}[{\cal H}_{n}].
\label{Hn}
\ee
under the diagonal action of $SU(2)$ is the intertwiner space of the node $n$. The volume operator $V_n$ acts on this space.

The Hilbert space ${\cal K}_\Gamma$ is the subspace of ${\cal H}_\Gamma$ formed by the gauge invariant states. Thus clearly
\be
{\cal H}_\Gamma=L_2[SU(2)^L/SU(2)^N]= \bigoplus_{j_l}\ \bigotimes_n {\cal K}_n. 
\ee
I denote $P_{\scriptscriptstyle SU(2)}:{\cal H}_\Gamma\to {\cal K}_\Gamma$ the orthogonal projector on the gauge invariant states. It can be written explicitly in the form
\be
P_{\scriptscriptstyle SU(2)}\psi(h_{l})=\int_{SU(2)^N} dg_n \ \psi(g_{s({l})}h_{l} g_{t({l})}^{-1}). 
\label{orthpro}
\ee}

Notice also that the states where $j_l=0$ for some $l$ are precisely the states that belong also to the Hilbert space ${\cal H}_{\Gamma'}$ where $\Gamma'$ is the subgraph of $\Gamma$ obtained erasing those $l$'s. It is therefore convenient to define the subspace $\hat{\cal H}_{\Gamma}$ of ${\cal H}_{\Gamma}$, spanned by the spin network states with all $j_l$ nonvanishing. By doing so, we can rewrite 
\eqref{graphspaces} as 
\be
   {\cal H}= \bigoplus_\Gamma\ \hat{\cal H}_\Gamma
    \label{graphspaces2}
\ee\\[-3mm]
without then having to bother to factor the equivalence between spaces with different graph. 

Concluding, a basis in $\cal H$ is labelled by three sets of ``quantum numbers": an abstract graph $\Gamma$; a coloring $j_l$ of the links of the graph with irreducible representations of $SU(2)$ different from the trivial one ($j=\frac12, 1, \frac32, ...$); and a coloring of each node of $\Gamma$ with an element $\vol_n$ in an orthonormal basis in the intertwiner space ${\cal H}_n$. The states $\ket{\Gamma, j_l, \vol_n}$ labelled by these quantum numbers are called ``spin network states" \cite{Rovelli:1995ac}. 

\problem{(Immediate) Use the above to prove that the Hilbert space of Loop Quantum Gravity is separable.}

\problem{Consider the state $\ket{\Gamma_{\Theta}, 1\frac12 \frac12}$ where $\Gamma_{\Theta}$ is the graph formed by two nodes connected by three links. Write this state as a sum of products of ``loops", where a loop is the trace of a product of a sequence of $h's$ along a closed cycle on the graph.  Any spin network state can be written in this way. This is the historical origin of the denomination ``loops quantum gravity" for the theory. See \cite{Rovelli:2010bf}.}

\subsection{Physical picture} \label{interpr}

Spin network states are eigenstates of the area and volume operators.  A spin network state has a simple geometrical interpretation.  It represents a ``granular" space where each node $n$ represents a ``grain" or ``quantum" of space  \cite{Rovelli:1994ge}.   These quanta of space do not have a precise shape because the operators that decide their geometry do not commute.  Classically, each node represents a polyhedron, thanks to Minkowski's theorem, but the polyhedra picture holds only in the classical limit and cannot be taken literally in the quantum theory. In the quantum regime, the operators $G_{ll'}$ do not commute among themselves, and therefore there is no sharp polyhedral geometry at the quantum level.  In other words, these are ``polyhedra" in the same sense in which a particle with spin is a ``rotating body".  

The spectrum of $A_l$ is easy to find out, since $A^2_l$ is simply the Casimir of one of the $SU(2)$ groups. Therefore
the area eigenvalues are 
\be
          a_j=\sqrt{j(j+1)}
\ee
where $j\in I\!\!{N}/2$. Notice that the spectrum is discrete and it has a minimum step between zero and the lowest non-vanishing eigenvalue
\be
a_{\frac12}=\sqrt{\frac12\left(\frac12+1\right)}=\frac{\sqrt 3}2.    \label{areagap}
\ee

The volume of each grain $n$ is $\vol_n$. Volume eigenvalues are not as easy to compute as area eigenvalues.  They can be computed numerically for arbitrary intertwiner spaces (the problem is just to diagonalize a matrix) and there are elegant semiclassical techniques that give excellent results.

\problem{(Important) Find the eigenvalues of the volume in ${\cal K}_{\scriptscriptstyle\frac12\frac12\frac12\frac12}$ and ${\cal K}_{\scriptscriptstyle 1111}$. $\left( Answer\ to\ the \ first: v_1=v_2=\sqrt{\frac{1}{6\sqrt{3}}}.\right)$}

Two grains $n$ and $n'$ are adjacent if there is a link $l$ connecting the two, and in this case the area of the elementary surface separating the two grains is determined by the spin of the link joining $n$ and $n'$. Physical space is ``weaved up"  \cite{Ashtekar:1992tm} by this net of atoms of space.  \\

\begin{figure}[h]
\centerline{\includegraphics[scale=0.3]{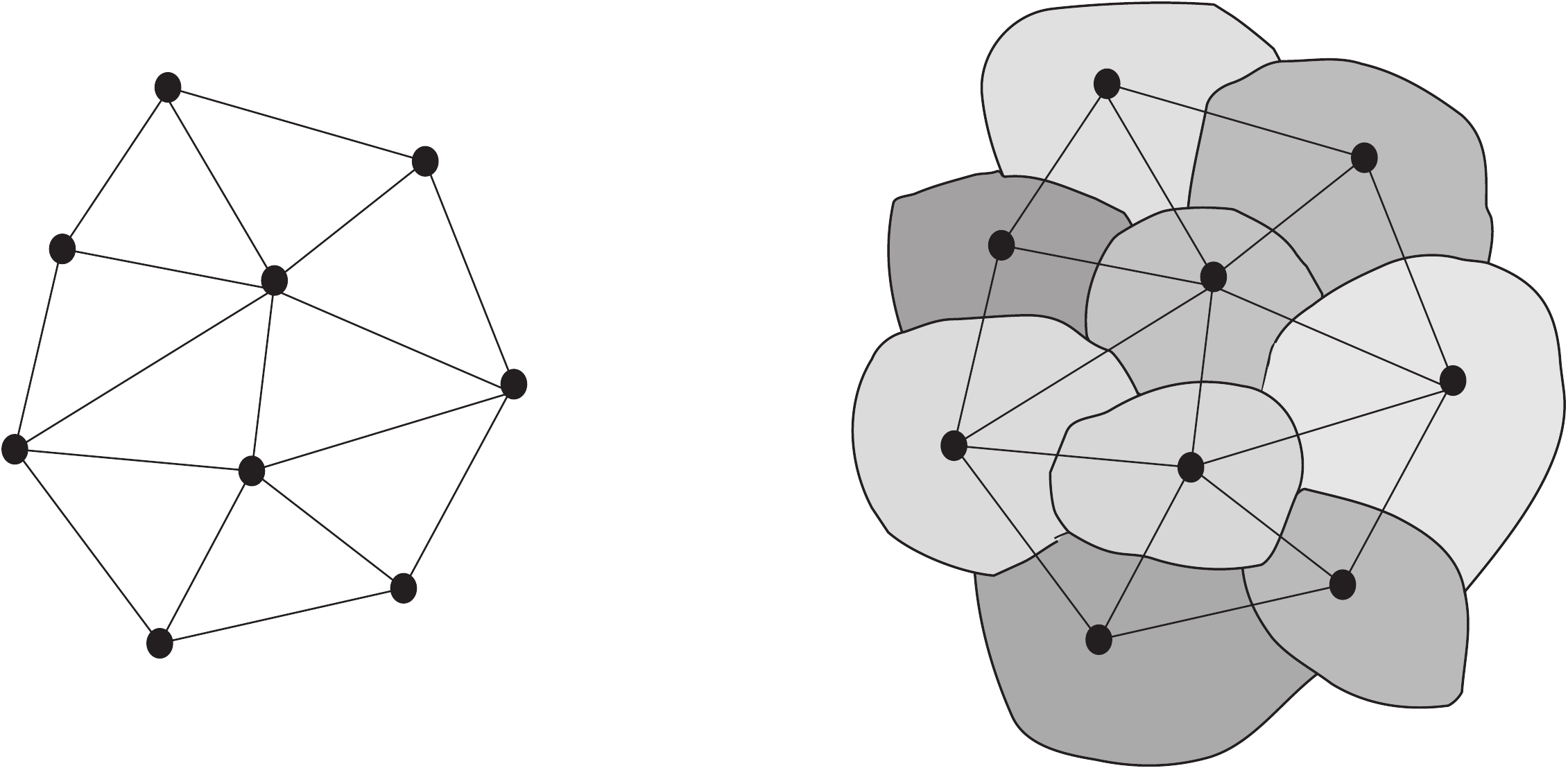}}
$\hspace{-15em}|\Gamma, j_l,v_n\rangle$
\caption{{``Granular" space}. A node $n$ determines a ``grain" or ``chunk" of space.} 
\end{figure}

The geometry represented by a state  $|{\Gamma, j_l, \vol_n}\rangle$ is a \emph{quantum} geometry for three distinct reasons. 
\begin{enumerate}
\item[i.] It is discrete. The relevant quantum discreteness is not the fact that the continuous geometry has been discretized ---this is just a truncation of the degrees of freedom of the theory. It is the fact that area and volume are quantized and their spectrum turns out to be discrete. It is the same for the electromagnetic field. The relevant quantum discreteness is not that there are discrete modes for the field in a box: it is that the energy of these modes is quantized.
\item[ii.] The components of the Penrose metric operator do not commute. Therefore the spin network basis diagonalizes only a subset of the geometrical observables, precisely like the $|j,m\rangle$ basis of a particle with spin. Angles between the faces of the polyhedra are quantum spread in this basis.  
\item[iii.] A generic state of the geometry is not a spin network state: it is a linear superposition of spin networks. In particular, the \emph{extrinsic} curvature of the 3-geometry\footnote{In canonical general relativity the extrinsic curvature of a spacelike surface is the quantity canonically conjugate to the intrinsic geometry of the surface.}, which, as we shall see later on, is captured by the group elements $h_l$, is completely quantum spread in the spin network basis. It is possible to construct coherent states in $H_\Gamma$ that are peaked on a given intrinsic as well as extrinsic geometry, and minimize the quantum spread of both.  A technology for defining these semiclassical states in $H_\Gamma$ has been developed by a number of authors, yielding beautiful mathematical developments  \cite{Thiemann:2002vj,Livine:2007mr,Freidel:2010bw,Bianchi:2009ky,Bianchi:2010gc}.  I sketch some basic ideas below in Section \ref{coherent}.
\end{enumerate}

\subsection{The Planck scale}

So far, I have not mentioned units and physical dimensions.  The gravitational field $g_{\mu\nu}$ has the dimensions of an area.\footnote{This follows from $ds^2=g_{\mu\nu}dx^\mu dx^\nu$ and the fact that it is rather unreasonable to assign dimensions to the coordinates of a general covariant theory: coordinates are functions on spacetime, that can be arbitrarily \emph{nonlinearly} transformed. For instance, they are often angles.}
The dimension of the Ashtekar's electric field $E$ (the densitized inverse triad), is also an area.  The geometrical interpretation described above depends on a unit of length $L_{\rm loop}$, which characterizes the theory. For instance, in, say, centimeters, the minimum area eigenvalue $a_{\frac12}$ will have the value
\be
a_{\frac12}=\frac{\sqrt 3}2\ L^2_{\rm loop}.    \label{areagap2}
\ee 
and the metric operator will be defined by  
\be
G_{ll'}= L^4_{\rm loop}\ \vec L_l\cdot \vec L_{l'}
\ee 
What is the value of $L_{\rm loop}$? The Hilbert space and the operator algebra described here can be derived from a canonical quantization of GR. In this case $\vec L_l$ is easily identified with the flux of the Ashtekar electric field, or the densitized triad, across the polyhedra faces, and canonical quantization fixes the multiplicative factor to be 
\be 
             L^2_{\rm loop}= 8\pi \gamma\ \hbar G
             \label{units}
\ee
where $\gamma$, the Immirzi-Barbero parameter is a positive real number, $G$ is the Newton constant.  This relation may be affected by radiative corrections (the Newton constant may run between Planck scale and the infrared), therefore it is more prudent to keep $L^2_{\rm loop}$ as a free parameter in the theory for the moment. It is the parameter that fixes the scale at which geometry is quantized. 

\problem{Using \eqref{units} and assuming $\gamma\sim1$, compute how many four-valent  quanta of space are needed to fill the volume  of a proton $V_p\sim 1 fm^3 =(10^{-15}m)^3$, if no spin $j_l$ is larger than $\frac12$. Can a single quantum of space have volume $V_p$?} 

\subsection{Boundary states}

\begin{figure}[h]
\centerline{\includegraphics[scale=0.4]{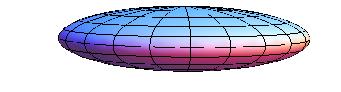}}
\caption{The state described by a spin network can be taken to give the geometry of the three dimensional hypersurface surrounding a finite 4d spacetime region.} \label{boundary}
\end{figure}

The states in $\cal H$ can be viewed as describing quantum space at some given coordinate time. A more useful interpretation, however, and the one I adopt here, is to take them to describe the quantum space {\em surrounding a given 4-dimensional finite region $\cal R$ of spacetime}.  
This second interpretation is more covariant and will be used below to define the dynamics. That is, a state in $\cal H$ is not interpreted as ``state at some time", but rather as a ``boundary state".  See Figure \ref{boundary}.
\\

In the non-general-relativistic limit, therefore, $\cal H$ must be identified with the tensor product ${\cal H}^*_{fin} \otimes {\cal H}_{init} $  of the initial and final state spaces of conventional quantum theory. 

\problem{Consider a single harmonic oscillator in its first excited state. Write explicitly its boundary state for the region $t\in[0,T]$.}

This is the quantum geometry at the basis of loop gravity. Let me now move to the transition amplitudes between quantum states of geometry. 

\section{Transition amplitudes}\label{Ta}

\subsection{Elementary math: $SL(2,\C)$}\label{math2}

{\footnotesize 
I start with a few notions about $SL(2,\C)$, the (double cover of the) Lorentz group  $SO(3,1)$. $SL(2,\C)$ is a six dimensional group.  I denote by $\psi$ the spinors of the fundamental representation defined on $\Co^2$ by the $2\times 2$ complex matrices with unit determinant. By $v$ the vectors in the 4d real representation defined on Minkowski space by the Lorentz transformations. And by $J$ the antisymmetric tensors in the adjoint representation (as the electromagnetic field).

It is convenient to study $SL(2,\C)$ by choosing a ``rotation" subgroup $H=SU(2)\subset SL(2,\C)$. Choosing an $SU(2)$ subgroup in $SL(2,\C)$ is like choosing a Lorentz frame in special relativity.  In the vector representation $H$ leaves a timelike vector $t$ invariant, and we can choose Minkowski coordinates where, say $t=(1,0,0,0)$. Then we can distinguish the time space components of any vector $v=v^0t+\vec v$, where $\vec v=(0,v^i), i=1,2,3$ is orthogonal to $t$.   In the fundamental representation a choice of $H$ is equivalent to the choice of a scalar product.  $H$ is given by the matrices unitary with respect to this scalar product. A change of Lorentz frame is equivalent to rotation of the scalar product in $\Co^2$. Given a scalar product $\langle \psi |\phi\rangle=g_{AB}\overline{\psi^A}\phi^B$, we can choose a basis in $\Co^2$ where $g=\id$. The relation between the choice of basis in $\Co^2$ and in Minkowski space is given by the Clebsch-Gordan map $v\to v^0 \id+v^i\sigma_i$. 

In the adjoint representation, a basis in the $SL(2,\C)$ algebra is formed by the generators $L_i$ of the $SU(2)$ rotations and the corresponding boost generators $K_i$.   Any group element can be written in the form
\be
g=e^{\alpha^i\tau_i+i\beta^i\tau_i}
\ee
The left invariant vector fields are then given by 
\be
L_i\psi(h)\equiv \left.i\frac{d}{dt}\, \psi(he^{t \tau_i})\right|_{t=0},\label{left2}
\ \ \ \ 
K_i\psi(h)\equiv \left.i\frac{d}{dt}\, \psi(he^{it \tau_i})\right|_{t=0}.
\ee
The two Casimirs of the group are $\vec L\cdot \vec K$ and $|\vec L|^2-|\vec K|^2$. 

The finite-dimensional representations of $SL(2,\C)$ are non-unitary. For instance the Minkowski ``scalar product" $x\cdot y=x^0y^0-x^1y^1-x^2y^2-x^3y^3$ is not a scalar product, because it is not positive definite.  LQG uses instead \emph{unitary} representations of $SL(2,\C)$, which are infinite dimensional.  These can be studied for instance in \cite{Ruhl:1970fk} or  \cite{Gelfand:1966uq}. Roberto Pereira's thesis \cite{Pereira:2010} is also very useful for this.  Here I give the essential information about these representations.

Unitary representations of $SL(2,\C)$  are labeled by a spin $k$ and a positive real number $p$. The representation spaces are denoted ${\cal H}_{p,k}$. The two Casimirs take the values $2pk$ and $p^2-k^2$, respectively, on  ${\cal H}_{p,k}$. Each  ${\cal H}_{p,k}$ can be decomposed into irreducibles of the $SU(2)$ subgroup as follows
\be
 {\cal H}_{p,k} = \oplus_{j=k}^\infty\  {\cal H}_{p,k}^{j}
 \ee
where ${\cal H}_{p,k}^{j}$ is a (2j+1) dimensional space that carries the spin-$j$  representation of $SU(2)$.  A useful basis in  ${\cal H}_{p,k}$ is obtained diagonalizing the total spin and the third component of the spin of the $SU(2)$ subgroup. States in this basis can be written as $ |p,k; j,m\rangle$. The representation matrices $D^{pk}(g)$ in this basis have the form $D^{pk}(g)^{jm}{}_{j'm'}$. 
The representation matrices $D^{pk}(g)$ span a linear space of functions on $SL(2,C)$. As elements of this space, they can be written in Dirac notation as \\
\be
\langle g | p,k;(j,m),(j'm')\rangle = D^{pk}(g)^{jm}{}_{j'm'}.
\ee\\
$\chi^{p,k}(g)=tr[D^{pk}(g)]$ is the $SL(2,\C)$ character in the $(p,k)$ unitary representation. (This is  generally a distribution on the group, since the representation spaces are infinite dimensional.) 

Of particular importance in LQG is a subspace of this space of functions. This is the space 
spanned by the subspaces  ${\cal H}_{p,k}^{j}$ where
\be
p=\gamma j, \hspace{2em} k=j;
\label{pk}
\ee
that is, the space 
\be
{\cal H}_\gamma = \oplus_j \left(({\cal H}_{\gamma j,j}^{j})^* \otimes {\cal H}_{\gamma j,j}^{j}\right).
\ee
In other words, this is the space of functions on $SL(2,\C)$ of the form
\be
\psi(g)= \sum_{j,mn} c_{j,mm'}\ D^{\gamma j,j}(g)^{jm}{}_{jm'}.
\ee
equivalently 
\be
 |\psi\rangle= \sum_{j,mn} c_{j,mm'}\  | \gamma j,j;(j,m),(jm')\rangle.
\ee

The space ${\cal H}_\gamma$  has two remarkable properties
\begin{itemize}\itemsep=.1mm
\item[-] It is naturally isomorphic to $L_2[SU(2)]$. The map 
\be
Y_\gamma: |j,m,n\rangle \mapsto |\gamma j,j; (j,m),(j,n)\rangle 
\label{Y}
\ee
sends $L_2[SU(2)]$ onto ${\cal H}_\gamma$. This map is clearly $SU(2)$ covariant. 

\item[-] For all $\psi,\phi\in H_\gamma$, we have \cite{Ding:2009jq,Ding:2010ye,Ding:2010fw}
\be
\langle\psi |\vec K +\gamma \vec L | \phi\rangle\sim 0.
\label{quantumsimpl}
\ee
where $\sim$ means that the relation holds in the large $j$ limit (which, as we shall see later on is the semiclassical limit of the quantum theory).
The relation
\be
\vec K +\gamma \vec L = 0.
\label{clsimpl}
\ee
 has an important meaning in quantum gravity, because, as we shall see in 
Section \ref{derivationdynamics} it is precisely the ``simplicity condition" that reduce BF theory to GR --- see equation \eqref{simplicity}.
\end{itemize}

\problem{(Important) Compute the value of the two Casimirs if $\vec K+\gamma \vec L=0$ and show that  this implies \eqref{pk}.}

The space ${\cal H}_\gamma$ has a natural Hilbert space structure, inherited from \eqref{Y}.  Notice, however that it is \emph{not} a subspace of the Hilbert space of square integrable functions $L_2[SL(2,\C)]$. This is because it is formed by a discrete linear combination of functions with a sharp value of $p$, which is a continuous label. In other words, it is like a space of linear combinations of delta functions.\footnote{The relation \eqref{quantumsimpl} is exactly true (not just in the large $j$ limit) if we replace $p=\gamma j$ by  $p=\gamma (j+1)$. This alternative has been considered in the literature, but it seems to lead to problems in relating the dynamics of graphs to that on subgraphs.  Sergei Alexandrov has noticed that $H_\gamma$ is not the only subspace with these properties. It is the first of a family of spaces ${\cal H}_\Gamma^r$, defined by $p=\gamma (j\!+\!r)(j\!+\!r\!+\!1)/j,  \ k=j\!+\!r,\ \  \tilde j=j\!+\!r$, with any integer $r$. The parameter $r$ determines a different ordering of the constraints, and I do not consider it here.  On this, see \cite{Ding:2010fw}.}

Consider a graph $\Gamma$ and a function  $\psi(h_l)$ of $SU(2)$ group elements on its links.  The map  $Y_\gamma$ extends immediately (by tensoring it), and sends $\psi(h_l)$ to a (generalized) function of 
$Y_\gamma\psi(g_l)$ of $SL(2,\C)$ elements. This is not $SL(2,\C)$ invariant at the nodes, but we can make it gauge invariant by integrating over a gauge action of $SL(2,\C)$. That is,
denote 
\be
P_{\scriptscriptstyle SL(2,\C)}\psi(g_{l})=\int_{SL(2,\C)^N} dg'_n \ \psi(g_{s({l})}g_{l} g_{t({l})}^{-1}). 
\label{orthpro2}
\ee
where the prime on $dg_n$ indicates that one of the edge integrals is dropped (it is redundant).
Thus, the linear  map
\be
f_\gamma := P_{\scriptscriptstyle SL(2,\C)}\circ Y_\gamma
\label{fg}
\ee
sends $SU(2)$ spin networks into $SL(2,\C)$ spin networks. In particular, $(f_\gamma \psi)(\id_l)$
is a linear functional on the space of $SU(2)$ spin networks. By linearity
\be
(f_\gamma \psi)(\id_l) = \int dh_l\ \psi(h_l) A_\gamma(h_l)
\label{fg1}
\ee
An explicit calculation (see below) shows that this can be written in the form 
\be  
A_{v}(h_l)=\int_{SL(2,\mathbb{C})} dg'_n\ \prod_l K(h_l,g_{s_l}g^{-1}_{t_l})
\label{va}
\ee
where $l$ are the links and $n$ the nodes of $\Gamma$ and the kernel $K$ is 
\be
K(h,g)=\sum_j\int_{SU(2)} dk\ d^2_j\ \overline{\chi^j(hk)} \ \chi^{\gamma j,j}(kg).
\label{K}
\ee 
We use this below.
   
 \problem{(important) Show that the last two equations give \eqref{fg1}. 
[Track: from the definition of the characters
\be
K(h,g)\!=\!\sum_j\int_{SU(2)}\hspace{-2em} dk\ d^2_j\ \overline{Tr[D^j(h)D^j(k)]} \ 
Tr[D^{\gamma j,j}(k)D^{\gamma j,j}(g)] 
\label{K33}
\ee 
but since $k\in SU(2)$,  
$D^{\gamma j,j}(k)^{j'm}{}_{j''m'}=\delta^{j'}_{j''}\ D^j(k)^{m}{}_{m'}
$,  
so that, using \eqref{wigner}, the integration over $k$ gives 
\be
K(h,g)=\sum_j   d_j\  \ 
\overline{D^j(h)^m{}_{m'}}\, D^{\gamma j,j}(g)^{jm'}{}_{jm},
\ee 
or, using the definition \eqref{Y}  of $Y_\gamma$,
\be
K(h,g)=\sum_j   d_j\ Tr[\overline{D^j(h)} Y^\dagger_\gamma D^{\gamma j,j}(g) Y_\gamma].
\ee 
Inserting this into \eqref{va} gives
\be 
A_{v}(h_l)=\int_{SL(2,\mathbb{C})} dg'_n\ \prod_l 
\sum_j   d_j\ Tr[\overline{D^j(h_l)} Y^\dagger_\gamma D^{\gamma j,j}(g_{s_l}g^{-1}_{t_l}) Y_\gamma].
\label{vaaaa}
\ee
Using this, the right hand side of \eqref{fg1} reads
\be
 \int_{SU(2)} \hspace{-2em} dh_l\, \psi(h_l) \!\!
\int_{SL(2,\mathbb{C})} \hspace{-2.2em} dg'_n\ \prod_l 
\sum_j   d_j\, Tr[\overline{D^j(h_l)} Y^\dagger_\gamma D^{\gamma j,j}\!(g_{s_l}g^{-1}_{t_l}\!) Y_\gamma]. 
\label{questaqui}
\ee
On the other hand, using the definitions, we have 
\be
(Y_\gamma \psi)(g_l) \!=\! 
\sum_j   d_j  \int_{SU(2)} \hspace{-2em} dh_l\; \psi(h_l) 
 \prod_l 
 Tr[\overline{D^j(h_l)} Y^\dagger_\gamma D^{\gamma j,j}(g_l) Y_\gamma] 
\ee
and the left hand side of \eqref{fg1} gives
\begin{multline}
(f_\gamma \psi)(g_l) = 
(P_{SL(2,\mathbb{C})}Y_\gamma \psi)(g) = \\   \hspace*{1em} \nonumber =
\int_{SL(2,\mathbb{C})} \hspace{-2em} dg'_n\ \sum_j   d_j\ 
 \int_{SU(2)} \hspace{-1em} dh_l\ \psi(h_l)  
 %\\ && \ \ \nonumber 
 \prod_l 
Tr[\overline{D^j(h_l)} Y^\dagger_\gamma D^{\gamma j,j}(g_{s_l}g_lg^{-1}_{t_l}) Y_\gamma] 
\end{multline}
which is equal to \eqref{questaqui} when $g_l=\id_l$.]\\

Finally, observe that the map $f_\gamma$ defined in \eqref{fg} sends $SU(2)$ intertwiners to $SL(2,C)$ intertwiners. Indeed $Y_\gamma$ sends tensors that transform under $SU(2)$ representations into tensors that transform under $SL(2,C)$ representations, and $P_{\scriptscriptstyle SL(2,\C)}$ projects these tensors on their $SL(2,C)$ invariant subspace.

\label{sect2b}

}  

\subsection{Elementary math: 2-complexes}\label{math22}

A (combinatorial) two-complex $\cal C=({\cal F}, {\cal E}, {\cal V}, \partial)$ is defined by a finite set $\cal F$ of $F$ elements $f$ called ``faces", 
a finite set $\cal E$ of $E$ elements $e$ called ``edges", 
a finite set $\cal V$ of $V$ elements $v$ called ``vertices", and a boundary relation  $\partial$ that associates to each edge an ordered couple of vertices $\partial e=(s_e,t_e))$ and to each face a cyclic sequence of edges.  A cyclic sequences of edges is a sequence of $n_f$ edges or reversed edges $e$ such that $t_{e_n}=s_{e_{n+1}}$ with $n_f+1=1$.  

The boundary $\Gamma=\partial{\cal C}$ of a two-complex $\cal C$ is a (possibly disconnected) graph $\Gamma$, whose links $l$ are edges of $\cal C$ bounding a single face and whose nodes $n$ are vertices of $\cal C$ bounding (links and) a single internal edge. 

For each vertex $v$, call $\Gamma_v$ the graph formed by intersection of the two-complex with a small sphere surrounding it.

The same definitions of automorphisms and partial order I gave for the graphs hold for the two-complexes.  In particular, it makes sense to define the limit
\be
  f_\infty = \lim_{{\cal C}\to\infty} f_{\cal C}
\ee
of a function that depends on a complex.  

A two-complex can be visualized as a set of polygons $f$ meeting along edges $e$ in turn joining at vertices $v$ (see Figure \ref{vrtx2}).

\begin{figure}[h]
\centerline{
\includegraphics[scale=0.2]{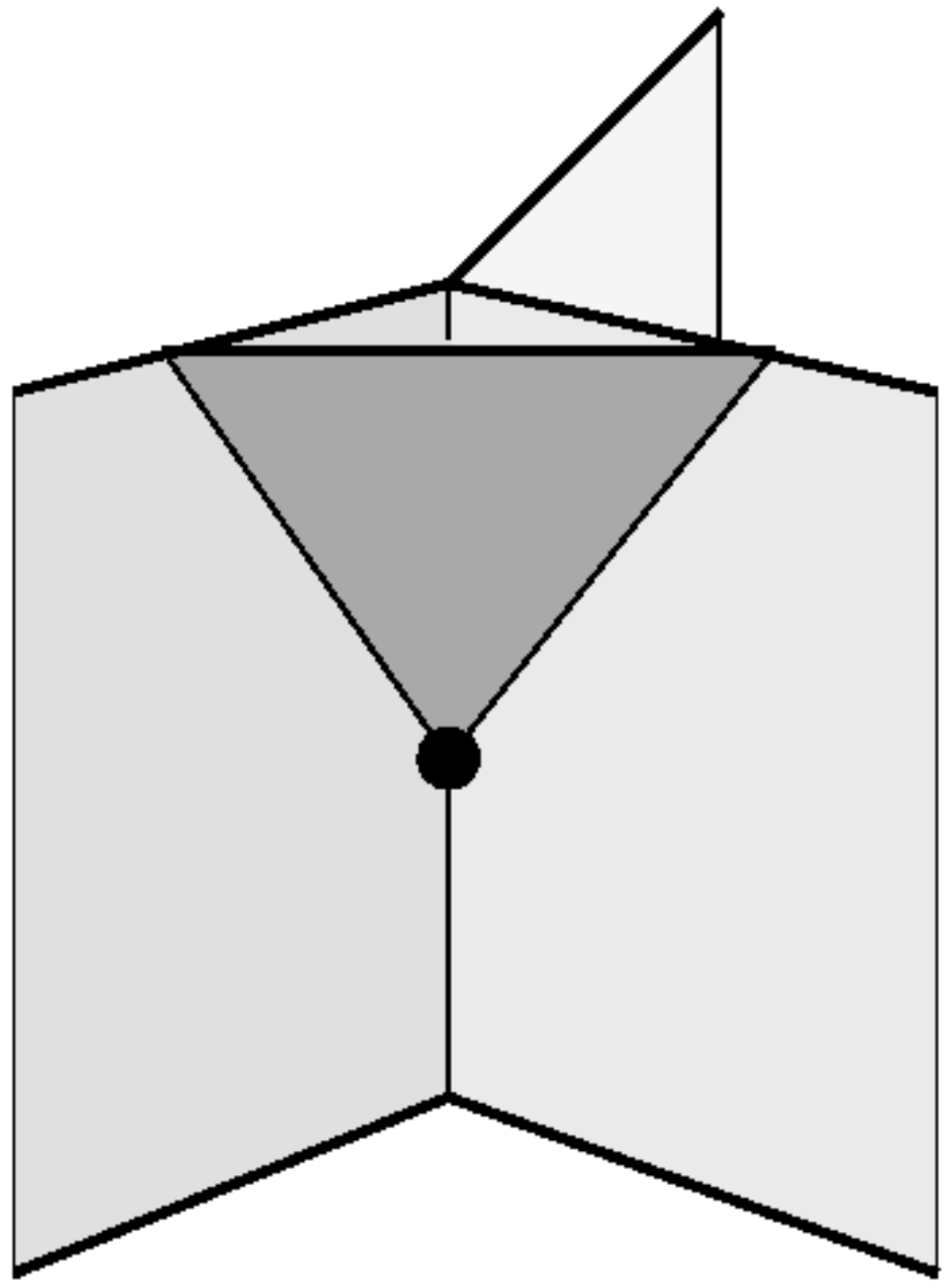}
$\begin{picture}(20,25)
\put(-38,33){\small $v$}
\put(-42,53){\small $f$}
\put(-45,22){\small $e$}
\end{picture}$}
\caption{A two-complex with one internal vertex.}
\label{vrtx2}
\end{figure}

Recall that in 3d, a geometrical picture can be obtained observing that  the dual to a cellular decomposition of space defines a graph.  The same will be true for two-complexes in 4d: the dual of a cellular decomposition of spacetime defines a two-complex.\footnote{The two-complex is the two-skeleton of the dual: its elements of dimension 0, 1 and 2.}   \emph{Vertices} can be thought as dual to 4d cells in spacetime.  \emph{Edges} are dual to 3d cell bounding the 4d cells. Importantly, \emph{faces} are dual to the surfaces to which we have assigned areas in 3d.  

Notice that boundary relations then hold correctly: An edge hitting the boundary of the two-complex is simply a 3d cell that happens to sit on the boundary of a 4d cellular decomposition of spacetime.  A face hitting the boundary of the two-complex is a 2d surface that sits on the boundary.  Thus, for instance a 2d surface in spacetime is represented by a \emph{link} $l$ in the 3d graph, but it is represented by a \emph{face} $f$ in the 4d complex. If the surface is on the boundary, $l$ is bounds $f$. 

We now have all the ingredients for defining the transition amplitudes of quantum gravity.

}

\subsection{Transition amplitudes}

As mentioned in the introduction, the Ponzano-Regge partition function of 3d quantum gravity on a triangulation $\Delta$ is defined by 
\be
   Z_{\cal C}=\sum_{j_f} \prod_f (2j_f+1)\ \prod_v \{6j\}.
   \label{ZPRR}
\ee 
Each tetrahedron $v$ of $\Delta$ has four triangles $e$, each bounded by three Regge bones $f$. Let $j_1,j_2,j_3$ be the spins of the three bones around the triangle $e$. In the dual two-complex $\cal C$, $v$ is a vertex where four edges $e$ meet, and each edge bounds three faces $f$.   The intertwiner space ${\cal K}_e={\cal K}_{j_1,j_2,j_3}$ is one-dimensional. Let $\vol_e$ be the single (normalized) intertwiner in $K_{j_1,j_2,j_3}$.  The Wigner $\{6j\}$ symbol is defined by 
\be
\{6j\} = Tr\left[\otimes_{e\in v} \vol_{e}\right]
\ee
The trace is taken on the repeated tensors indices (two indices in each representation, because each Regge bone $f$ joins two triangles of the tetrahedron $v$.) 

Let me now come to the main point of these lectures: the definition of the partition function of 4d Lorentzian LQG. This is defined by 
\be
   Z_{\cal C}=\sum_{j_f,\vol_e} \ \prod_f (2j_f+1)\ \prod_v A_v(j_f,\vol_e),
   \label{ZEPRLL}
\ee 
where $\cal C$ is a two-complex with faces $f$, edges $e$ and vertices $v$, the intertwiners $\vol_e$ are in the space ${\cal K}_e={\cal K}_{j_{f_1}...j_{f_1}}$ where $f_1,...,f_n$ are the faces meeting at the edge $e$ and
\be
A_v(j_f,\vol_e)=Tr\left[\otimes_{e\in v} (f_\gamma \vol_{e})\right]. 
\label{Ave}
\ee
where $f_\gamma$ is given in \eqref{fg} and $\gamma$ is a dimensionless parameter that characterizes the quantum theory, called Immirzi, or  Barbero-Immirzi, parameter.  This is the definition of the covariant dynamics of LQG. 

Notice that the theory is entirely determined by the imbedding $Y_\gamma$ of $SU(2)$ functions into $SL(2,\C)$ functions, defined in section \ref{math2}, see equation \eqref{Y}. An intuitive track for understanding what is happening is the following. If we erase $f_\gamma$ in \eqref{Ave} we obtain the Ooguri quantization of BF theory \cite{Ooguri:1992eb}. A shown above in \ref{math2}, $f_\gamma$ implements equation \eqref{clsimpl}, which is precisely the relation that transforms BF theory into general relativity, as shown in detail below in \ref{math2}.

The trace in \eqref{Ave} requires a bit of care, due to the infinite dimensionality of the $SL(2,C)$ representations involved. To make this explicit, I write the same partition function in an equivalent form: 
\be
Z_{\cal C}=   \int_{SU(2)} dh_{vf}\ \prod_f\delta(h_f)\ \prod_v A_v(h_{vf}).
\label{int1} 
\ee
Here $h_f=\prod_{v\in f} h_{vf}$ is the oriented product of the group elements around the face $f$ and the vertex amplitude is given by \eqref{va} and \eqref{K}, which I repeat here for completeness:
\be  
A_{v}(h_l)=\int_{SL(2,\mathbb{C})} dg'_n\ \prod_l K(h_l,g_{s_l}g^{-1}_{t_l})
\label{vava}
\ee
\be
K(h,g)=\sum_j\int_{SU(2)} dk\ d^2_j\ \overline{\chi^j(hk)} \ \chi^{\gamma j,j}(kg).
\label{KK}
\ee 
The last three equations define the partition function in a completely explicit manner. 

Using the problem at the end of Section \ref{sect2b}, we see that the vertex amplitude for a spin-network state $\psi$ on $\Gamma_v$, namely around a vertex, can be also written in the compact form
\be
A_v(\psi)= (f_\gamma\psi)(\id).
\label{K3}
\ee

\insertion{\underline{Problem:} \ Show that \eqref{Ave} and \eqref{K3} are equivalent.}

The transition amplitudes are obtained by choosing a two-complex $\cal C$ with a boundary, and are functions of the boundary coloring. In the spinfoam basis \eqref{ZEPRLL}, they read
\be
   W_{\cal C}(j_l,\vol_n)=\sum_{j_f,v_e} \prod_f (2j_f+1)\ \prod_v A_v(j_e,\vol_n),
   \label{ZEPRLLTR}
\ee 
where $l$ and $n$ are the boundary links and nodes.  In the group basis, they read 
\be
W_{\cal C}(h_l)=   \int_{SU(2)} dh_{vf}\ \prod_f\delta(h_f)\ \prod_v A_v(h_{vf}).
\label{int1hl} 
\ee
where $h_l$ is an $SU(2)$ group element for each boundary link. These expressions define truncations of the full transition amplitudes. The full physical transition amplitude is 
\be
W(h_l)= \lim_{{\cal C}\to\infty} W_{\cal C}(h_l). 
\label{limit}
\ee
In a general-covariant quantum theory, the dynamics can be given by associating an amplitude to each boundary state \cite{Oeckl:2003vu,Oeckl:2005bv}. This is determined by the linear functional $W$ on $\cal H$. The modulus square 
\be
P(\psi)=|\langle W|\psi\rangle|^2
\ee
determines (with suitable normalization) the probability associated to the process defined by the boundary state $\psi$.  In Section \ref{Ap} I  show how these can be used to compute the probability of interesting physical processes.

\insertion{In the definition I have given, there is no restriction on the two-complex $\cal C$.  On physical grounds, this may be too general, and it may prove necessary to restrict the class of two complexes to consider \cite{Hellmann:2011jn}. A natural choice is to demand that $\cal C$ comes from a cellular decomposition.}

\subsection{Properties and comments}

The most important property of the vertex amplitude \eq{K3} is that it appears to yield the Einstein equations in the large distance classical limit. There is a number of result in the literature supporting this indication. The relation between the vertex amplitude \eqref{K3} and the action of general relativity has been first studied in the Euclidean context (not discussed here) \cite{Barrett:2009mw,Barrett:2009gg,Barrett:2009cj} and then extended to the relevant Lorentzian domain \cite{Pereira:2010}. For a five-valent vertex (dual to a 4-simplex), it has been shown that the amplitude is essentially given by the exponential of the 4d Regge action in an appropriate semiclassical limit \cite{Bianchi:2010mw}. 
\be
A_v \sim 
% \sum\raisebox{-1mm}{${}_k$}\
  e^{i S_{Regge}}
\label{not}
\ee
To understand this relation, observe that the amplitude is a function of the boundary state, and this can be chosen to be peaked on a given boundary geometry of a flat Regge cell. The corresponding Regge action is then well defined.   

The result can be extended to arbitrary triangulations, showing that the spinfoam sum is dominated by configurations that admit a Regge interpretation and whose amplitude is given by the exponential of the Regge action \cite{Conrady:2008ea,Magliaro:2011qm,Magliaro:2011dz}. 
All these results hold in the semiclassical regime of large quantum numbers.  For small spins (small distances), the theory departs strongly form naive quantum Regge calculus, in particular, the discreteness of the spins implements the intrinsic short-distance cut-off, which is not present in naive quantum Regge calculus. 

\insertion{I do not give the explicit derivation of these results here. The key technique used is to write the amplitude as an integral over group elements and spheres, write the integrand as an exponential (this is done in Appendix \ref{coherentform}), and then notice that for large spins we are in the regime where a saddle point approximation holds. A good detailed technical introduction to these calculation techniques is Roberto Pereira thesis \cite{Pereira:2010}.  See also \cite{Bianchi:2011fk,Krajewski:2011uq,Magliaro:2011dz}.}

The evidence for the emergence of the Regge action is now multifold. Several issues remain open. For instance, the limit has not been studied for two complexes that are not dual to triangulations.  

\insertion{In \cite{Barrett:2009mw} it was shown that $A_v\sim e^{i S_{Regge}}+ e^{-i S_{Regge}}$, and concern has been raised by the appearance of the two terms. This concern is excessive in my opinion, for two reasons.  First, in the holomorphic representation (see below) only one of the terms in survives \cite{Bianchi:2010mw}. This is because of the ubiquitous mechanism of phase cancellations between propagator and boundary state in quantum mechanics. See \cite{Bianchi:2006uf} for a discussion of this mechanism. Therefore the existence of different terms in does not affect the classical limit.  Second, the amplitude of the theory \emph{should} include both terms. This appears clearly in the three dimensional Ponzano Regge theory \cite{Rovelli:1993kc} as well as in low dimensional models \cite{Colosi:2003si}, and is related to the fact that the classical dynamics does not distinguish propagation ``ahead in (proper) time" or ``backward in (proper) time", in a theory where coordinate time is an unphysical parameter.\footnote{It is sometime argued that the presence of the two terms follows from the fact that one has failed to select the ``positive energy" solutions in the quantization.  But such a selection makes sense only in the context of the specific strategy for quantization which consists in considering \emph{complex} solutions of the classical equations and then \emph{discarding} solutions with ``negative energy". This strategy is not available here, because of the absence of a preferred time and energy. Other quantization strategies  are available: we quantize the \emph{real} solution space and keep \emph{all} solutions.  The physical scalar product is determined by all \emph{real} solutions with the proper symplectic structure, not by a ``positive energy sector" of the complex solutions. A simple system illustrating the situation is given in \cite{Colosi:2003si}.}}

The emergence of the Einstein theory from a natural group structure based on $SU(2)\subset SL(2,\C)$ is certainly surprising. 
The skepticism prompted by this surprise may be tempered by two considerations. The first is that the same happens in QED.  The simple vertex amplitude
\ba
   \langle A| \psi_{e_1}^A(p_1), \psi_{e_2}^B(p_2), \psi_\mu^\gamma(k) \rangle
   &=& \raisebox{-.5cm}{\includegraphics[scale=0.4]{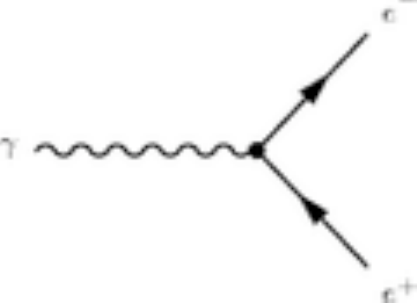}}\\
   &=&
   e\ \gamma_\mu^{AB}\ \delta (p_{1}\!\!+\!p_{2}\!\!+\! k )\nonumber
\ea
codes the full complexity of the interacting Dirac-Maxwell field equations. In other words, QED, with its fantastic phenomenology and its 12 decimal digits accurate predictions, is little more than momentum conservation plus the Dirac matrices $\gamma_\mu^{AB}$, which, like $f_\gamma$, are essentially Clebsch-Gordan coefficients.

The second consideration is that, as I discuss later on, general relativity is $BF$ theory plus the simplicity constraints. $BF$ theory means flat curvature. Hence in a sense GR is flat curvature plus simplicity conditions (see \eq{simplicity} below). The map $f_\gamma$ implements the simplicity conditions, since it maps the states to the space where the simplicity conditions \eqref{quantumsimpl} hold; while the evaluation on $G_l=\id$ codes (local) flatness. \\[-2.mm]

{\footnotesize  The last observation does not imply that the theory describes flat geometries, for the same reason for which Regge calculus describes curved geometries using \emph{flat} 4-simplices. In fact, there is a derivation of the vertex (\ref{K3}) which is precisely based on Regge calculus, and a single vertex is interpreted as a flat 4-simplex \cite{Engle:2007qf,Engle:2007wy}. In this derivation one only considers 4-valent nodes and 5-valent vertices. The resulting expression naturally generalized to an arbitrary number of nodes and vertices, and therefore defines the dynamics in full LQG. The existence of this generalization was  emphasized in \cite{Kaminski:2009fm}.

}\vskip.3cm

It is interesting to observe that the form of the amplitude $W$ is largely determined by general principles:  Feynman's superposition principle, locality, and local Lorentz invariance  \cite{Bianchi:2010Nice}. 

\begin{enumerate}
\item{\em Superposition principle.} Following Feynman, we expect that the amplitude $\bk{W}\psi$ can be expanded in a sum over ``histories" $W(\sigma)$. The integral in \eqref{int1} is like a truncated version of a Feynman path integral, analogous to the integration over the group elements in lattice QCD. The integration variables are precisely the $SU(2)$ group elements that form a basis in the Hilbert space of the theory. As mentioned, a given two-complex can also be viewed as a history of quanta of spaces. The integration in \eqref{int1} is then analog to the momenta integration in Feynman diagrams.

\item{\em Locality.}
We expect the amplitude $W(\sigma)$ of a single history to be built in terms of \emph{products} of elementary amplitudes associated to local \emph{elementary} processes.\footnote{Notice that this is true for the Feynman integral amplitudes, which are exponential of integrals, namely limits of exponentials of sums, which is to say (limits of) products of (exponentials of) terms which are local in spacetime, as well as for the amplitudes of the QED perturbation expansion, which are products of vertex amplitudes and propagators.  In particular, in QED the QED vertex is the elementary dynamical process that gives an amplitude to the boundary Hilbert space of the states of two electrons and one photon.}  This is the case for the integrand of \eqref{int1}, which is a product of local face and vertex amplitudes. The face amplitude $\delta(h_f)$ simply glues the different vertices amplitude.  The non-trivial part of the amplitude is  in the vertex amplitudes $A_v$. 

\item{\em Local Lorentz invariance.} Classical GR has a local Lorentz invariance, and we expect the individual spinfoam vertex to be Lorentz invariant in an appropriate sense. If spinfoams states were $SL(2,\C)$ spin networks $\Psi(g_{l}), g_{l}\in SL(2,\C)$, gauge invariance could be easily implemented by projecting on locally Lorentz invariant states with $P_{SL(2,\C)}$. But the Hilbert space ${\cal H}_\Gamma$ has no hint of $SL(2,\C)$. So, to implement local Lorentz invariance, there should be a map from ${\cal H}_\Gamma$ to a Lorentz covariant language that characterizes the vertex.  How?  Well, I have just constructed such a map in the previous section: it is the map $Y_\gamma$, which depends only on a single parameter $\gamma$.  The vertex amplitude is then simply obtained from $Y_\gamma$ and $P_{SL(2,\C)}$, as expressed by \eqref{K3}.
\end{enumerate}

\begin{figure}[h]
\centerline{\includegraphics[scale=0.28]{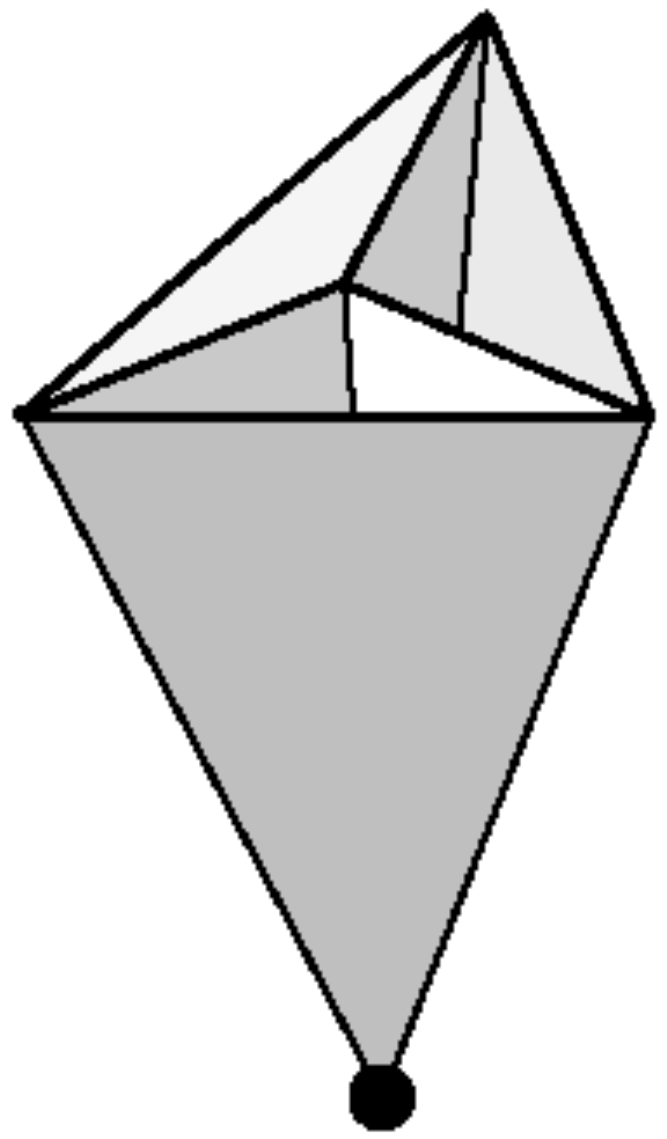}\hspace{-15em}\centerline{\includegraphics[scale=0.2]{1-3vertex.pdf}\hspace{-20em}}}
\caption{Transition from nothing to four quanta of space and from a single quantum of space to three.}
\label{13}
\end{figure}

The vertex amplitude (\ref{K3}) gives the probability amplitude for a single spacetime process, where $n$ grains of space are transformed into one another. It has the same {\em crossing} property as standard QFT vertices. That is, it describes different processes, obtained by splitting differently the boundary nodes into ``in" and ``out" ones. For instance if $n=5$ (this is the case corresponding to a 4-simplex in the triangulation picture), the vertex (\ref{K3})  gives the amplitude for a single grain of space splitting into four grains of space; or for two grains scattering into three, and so on.  See Figure \ref{13}.

More precisely, the vertex $\langle A_v|\psi\rangle$ gives an amplitude associated to the spacetime process defined by a finite region of spacetime, bounded by a 3d region described by the state $\psi$: there is no distinction between ``in" and ``out" states. 

The expressions (\ref{int1}-\ref{K3}) define the quantum field theory of (pure) gravity.\footnote{Recalling the discussion in Section \ref{Hilbert} on the convergence between the QED and the QCD picture, the truncation determined by the choice of a given two simplex can be thought in two equivalent ways. Either as a background independent analog of the truncation provided by a finite 4d lattice in QCD, or as a truncation in the order of the Feynman diagrams, in a background-independent analog to the Feynman perturbative expansion. The two pictures turn out to converge because physically a spacetime lattice is nothing else that a ``history" of space quanta, in the same sense in which a Feynman graph is a ``history" of field quanta.    In the first case, the two-complex can be thought as a discretization of spacetime.  In the second, as a particular Feynman history of quanta of space. 
}  What remains to do is to extract physics from this theory, and show that it gives general relativity in some limit.   Before discussing how to extract physical predictions from the formalism above, however, I add below a brief sections on the TQFT aspect of the construction given.

\insertion{

\subsection{GR as a topological theory, II} \label{ttII} 

In Section \ref{tqft1}, I have mentioned the fact that (a truncation of) the kinematics  GR can be given a topological interpretation: the local degrees of freedom of a 3-metrics on a (topologically trivial) 3-manifold $\cal M$ can be approximated by the global topological degrees of freedom of the topologically non-trivial manifold ${\cal M}^*$ obtained removing the Regge bones from  $\cal M$. 

The same idea works for the dynamics, and can be brought up to four dimensions. In 4d, Regge curvature is concentrated in the triangles of a Regge triangulation, and the relevant topology is now captured by the two complex dual to the triangulation. The faces of the two complex ``wrap around" the Regge triangles and the holonomy around these faces reads out the curvature on the triangles. 

In fact, the entire structure of covariant LQG can be phrased in the language of Topological Quantum Field Theory, as defined by Atiyah \cite{Rovelli:2010vv,Rovelli:2010qx} (for earlier ideas, see \cite{Baez:1997zt,Baez:1999sr}. A related point of view is in  \cite{Crane:2010fk}).  

In Atiyah's scheme, an $(n+1)$-dimensional TQFT is as a functorial association of a finite dimensional Hilbert  space ${\cal H}_\Gamma$ to each closed oriented \emph{$n$-manifold} $\Gamma$, and a vector $W_{\cal C}\in{\cal H}_{\Gamma}$ to each oriented \emph{$(n+1)$-manifold} $\cal C$ having $\Gamma$ as its boundary \cite{Atiyah:1988fk,Atiyah:1990uq}. 

Similarly, quantum gravity in the covariant loop formalism can be defined  \`a la Atiyah as the association \eqref{n-1} of a Hilbert space ${\cal H}_\Gamma$  to each oriented \emph{graph} $\Gamma$, and of a vector $W_{\cal C}\in{\cal H}_\Gamma$, defined by  \eqref{ZEPRLLTR}, to each \emph{two-complex} $\cal C$ having $\Gamma$ as its boundary.  (Generalized)  Atiyah axioms for this association require then 
\\[1mm]   $\bullet$ (multiplicativity) ${\cal H}_{\Gamma_1\cup\Gamma_2}={\cal H}_{\Gamma_1}\otimes{\cal H}_{\Gamma_2}$, 
\\
 $\bullet$  (duality) ${\cal H}_{\overline{\Gamma}}={\cal H}_{\Gamma}^{*}\quad\textrm{and}\quad{W}_{\overline{\cal C}}= W_{\cal C}^{\dagger}$,
 \\
  $\bullet$  (functoriality) $W_{{\cal C}_1\cup_{\Gamma}{\cal C}_2}=\langle W_{\overline{{\cal C}_2}}\vert W_{{\cal C}_1}\rangle_{{\cal H}_{\Gamma}}=\langle W_{\overline{{\cal C}_1}}\vert W_{{\cal C}_2}\rangle_{{\cal H}_{\overline{\Gamma}}}$,
  \\[1mm]
where the overline means reversing the orientation, and ${\cal C}_1\cup_{\Gamma}{\cal C}_2$ is the gluing of two two-complexes via a common boundary connected component $\Gamma$.   See \cite{Rovelli:2010qx} for the details of the construction and a development of this point of view.

}

\section{Extracting physics} \label{Ap}

A theory from which we cannot compute, is not a good theory. This is unfortunately the case of many ``quantum gravity" theories. Therefore this section is, in a sense, the most important of all. 
The difficulty of computing in quantum gravity is very well known, and nearly mythical.  The problem is inherited from general relativity. Remember that Einstein himself got confused on issues such as the interpretation of the Schwarzschild horizon and whether or not gravitational  waves (solution of the linearized Einstein equations) have any physical effect on matter. It took decades to sort out whether or not the gravitational waves are real.\footnote{It was only with Bondi in the 60' that the problem was clarified. As Bondi put it, in principle  ``you can boil a cup of water with gravitational waves".} 

These conceptual intricacies, which  make general relativity difficult, and interesting, are of course amplified in quantum gravity.  But, as they were eventually fully clarified for the classical theory, I expect that clarity should be reached in the quantum theory as well. 

The predictions of the theory are in its transition amplitudes. Given a boundary state, the formalism presented above defines truncated transition amplitudes, namely associates probabilities to boundary states (processes) \cite{Rovelli:2009ee,Rovelli:2004fk,Rovelli:2001bz}.

We are particularly interested in processes involving (background)  semiclassical geometries.  Since the formalism is background independent, the information about the background over which we are computing amplitude must be fed into the calculation. This can only be done with the choice of the boundary state.

Consider a three-dimensional surface $\Sigma$ with the topology of a 3-sphere. Let $(q,k)$ be the three-metric and the extrinsic curvature of $\Sigma$.    The \emph{classical} Einstein equations determine uniquely whether or not $(q,k)$ are physical: that is, whether or not there exist a Ricci-flat spacetime $\cal M$ (a solution of the Einstein equation) which is bounded by $(\Sigma, q, k)$.

\insertion{This is the extension of the Hamilton-function formulation of dynamics to general covariant field theory. Calling $q,p$ the coordinate and momenta at initial time $t$ and $q',p'$ at a final time $t'$, dynamics is fully captured by the conditions the quadruplet $(q,p,q',p')$ must satisfy in order to bound a physical trajectory. For a free particle, for instance, these are $p=p'=m(q'-q)/(t'-t)$. 

These relations can be directly deduced from the Hamilton function $S(q,t,q',t')$, using $p=\partial S/\partial q$ and $p'=\partial S/\partial q'$.   The Hamilton function is the royal tool for quantum gravity.  I recall that in general the Hamilton function is the value of the action on solutions of equations of motion, viewed as a function of the boundary extended configuration variables (that is, configurations variables plus time). For instance, the Hamilton function  of a free particle is $S(q,t,q',t')=m(q'-q)^2/2(t'-t)$.   The Hamilton function is directly related to the quantum mechanics amplitude $W$ by the small $\hbar$ expansion $W\sim e^{\frac{i}{\hbar}S}$.   For a full discussion of this way of writing dynamics in general covariant language, see Chapters 3 and 4 of \cite{Rovelli:2004fk}.} 

The \emph{quantum} theory assigns an amplitude to any semiclassical boundary state peaked on a given boundary geometry $(q, k)$, and for classical GR to be recovered this amplitude must (in the semiclassical regime) be suppressed if $(\Sigma, q, k)$ does not bound a solution of the Einstein equations.\footnote{Cfr: in the non-relativistic theory, if $\psi_{q,p}$ is a coherent state peaked on $q,p$, then $\langle W| \psi_{q',p'}\otimes \overline{\psi_{q,p}}\rangle\equiv \langle \psi_{q',p'}|e^{-iH(t'-t)}| \psi_{q,p}\rangle$ is suppressed unless the conditions mentioned above are satisfied.}

Now, consider a (normalized) \emph{semiclassical} boundary state $\psi_{(q,k)}$ that approximates the  classical geometry $(q,k)$ (these states are discussed in detail in the following section). If $q,k$ is a solution of the Einstein equations, we must expect that, within the given approximation
\be 
P(\psi_{(q,k)})= |\langle W|\psi_{(q,k)}\rangle|^2 \sim 1.
\ee

Next, if we modify the state $\psi_{(q,k)}$ with field operators $E_{n_1},... ,E_{n_N}$, then the amplitude
\be
W_{(q,k)}(n_1,...,n_N) =  \langle W|E_{n_1} ... E_{n_N} | \psi_{(q,k)}\rangle.
\ee
can be interpreted as a scattering amplitude between the $N$ ``particles" (quanta) created by the field operators $E_n$ over the spacetime $\cal M$. (The possibility of using the notion of ``particle" in this context is discussed in detail in \cite{Colosi:2004vw}.) Since we know how to write the gravitational field operator (the triad), we can in principle compute graviton $N$-point functions in this way.   To use this strategy, we must learn how to write semiclassical states in ${\cal H}_\Gamma$; this is the topic of the next section.

\subsection{Coherent states}\label{coherent}

The relation between quantum states and the classical theory is clarified by the construction of coherent states. These are particularly valuable in the present context, where the relation with the classical theory is more indirect than usual. Various classes of coherent states have been studied.  Here I describe the ``holomorphic" coherent states, developed by a number of people  \cite{Hall:2002, Ashtekar:1994nx,Thiemann:2002vj, Bahr:2007xn, Flori:2009rw} and recently discussed in detail by Bianchi-Magliaro-Perini  \cite{Bianchi:2009ky}, as well as the ``semi coherent" states of Livine-Speziale (LS) \cite{Livine:2007vk}.

Holomorphic states are labelled by an element $H_l$  of $SL(2,\C)$ for each link $l$.\footnote{They are a special case of Thiemann's complexifier coherent states  \cite{Thiemann:2002vj, Bahr:2007xn, Flori:2009rw}}. They are defined by 
\be
      \psi_{H_l}(h_l) = \int_{SU(2)^N} dg_n 
       \bigotimes_{l\in \Gamma} K_{t}(g_{s(l)}H_l g^{-1}h_l^{-1}).
\label{states}
\ee
Here $t$ is a positive real number and  $K_t$ is (the analytic continuation to $SL(2,\C)$ of) the heat kernel on $SU(2)$, which can be written explicitly as
\be
K_t(g) = \sum_j (2j+1)  e^{-j(j+1)t}\  {\rm Tr}[D^j(g)]
\label{Kt}
\ee
where $D^j$ is the (Wigner) representation matrix of the representation $j$. 

\insertion{
These states are analogous to the standard wave packets of non-relativistic quantum theory
\be
\psi_{x_op_o}(x)= e^{-\frac{(x-x_o)^2}{2\sigma}-ip_ox},
\ee
which are peaked in position as well as momentum. Notice that $\psi_{x_op_o}(x)$ can be written as a gaussian peaked on a \emph{complex} position
\be
\psi_{x_op_o}(x) \sim e^{-\frac{(x-h)^2}{2\sigma}},
\ee
where
\be
h=x_o+i\sigma p_0.
\ee
The coherent states \eqref{states} have the very same structure: they are given by a gaussian over the group, peaked on a complex extension of the group. Here the complex extension is \slc. To see that \eqref{Kt} is indeed a Guassian, notice that it is the heat kernel on the group: for $t=0$ is reduces to a delta function, while each $j$ mode decays proportionally to the Casimir, which is the Laplacian on the group.
}

The $SL(2,\C)$ labels $H_{l}$ can be given two related interpretations. First, we can decompose each $SL(2,\C)$ label in the form
\be
         H_{l}=e^{2itL_{l}}\ h_{l}
         \label{heu}
\ee 
where $U_{l}\in SU(2)$ and $E_{l} \in su(2)$.  Then it is not hard to show that $U_{l}$  and $E_{l}$ are the expectation values of the operators $U_{l}$ and $L_{l}$ on the state $\psi_{H_{l}}$ 
\be
         \frac{  \bek{\psi_{H_{l}}}{ h_{l}}{\psi_{H_{l}}}  }{  \bk{\psi_{H_{l}}}{\psi_{H_{l}}} }=h_{l}~, \hspace{1em}
         \frac{  \bek{\psi_{H_{l}}}{L_{l}}{\psi_{H_{l}}}  }{  \bk{\psi_{H_{l}}}{\psi_{H_{l}}} }= L_{l}~,
\ee 
and the corresponding spread is small.\footnote{
Restoring physical units, $ \Delta U_{l} \sim \sqrt{t}$ and $\Delta E_{l} \sim 8\pi\gamma\hbar G \sqrt{1/t}. $
If we fix a length scale $L\gg \sqrt{\hbar G}$ and choose $t=\hbar G/L^2\ll 1$, we have then $
        \Delta U_{l} \sim ~ \sqrt{\hbar G}/L$ and $
        \Delta E_{l} \sim  \sqrt{\hbar G}\,L$, which shows that both spreads go to zero with $\hbar$.}
        
Alternatively, we can decompose each $SL(2,\C)$ label in the form
\be
H_{l} = n_{s,{l}} ~ e^{-i(\xi_{l}+i\eta_{l})\frac{\sigma_3}{2}} ~ n^{-1}_{t,{l}}.
\label{fs}
\ee
where $n\in SU(2)$. Let $\vec z=(0,0,1)$ and $\vec n=D^1(n)\vec z$. Freidel and Speziale discuss a compelling geometrical interpretation for the $(\vec n_s, {\vec n}_t, \xi, \eta)$ labels defined on of each link by \eq{fs} \cite{Freidel:2010aq} (see also \cite{Dittrich:2008ar,Oriti:2009wg,Bonzom:2009wm}). For appropriate four-valent states representing a Regge 3-geometry with intrinsic and extrinsic curvature, the vectors ${\vec n}_s, {\vec n}_t$ are the $3d$ normals to the triangles of the tetrahedra bounded by the triangle;  $\eta$ is the area of the triangle divided by $8\pi\gamma G\hbar$; and $\xi$ is a sum of two parts: the extrinsic curvature at the triangle and the 3d rotation due to the spin connection at the triangle.  This last part can be gauged away locally, but cannot be disregarded globally \cite{Magliaro:2010qz}.)
  For general states, the interpretation extends to a simple generalization of Regge geometries, that Freidel and Speziale have baptized ``twisted geometries". \\[-1mm]

{\footnotesize Freidel and Speziale give a slightly different definition of coherent states \cite{Freidel:2010aq}. The two definitions converge for large spins, but differ at low spins. It would be good to clarify their respective properties, in view of the possible applications in scattering theory (see below).

Of great use are also the Livine-Speziale (LS) ``semi-coherent" states. They are defined as follows. The conventional magnetic basis $|j,m\rangle$ with $m=-j,...,j$, in $H_j$ diagonalizes $L^3$. Its highest spin state $|j,j\rangle :=|j,m=j\rangle$ is a semiclassical state peaked around the classical configuration $\vec L=j\vec z$ of the (non commuting) angular momentum operators.  If we rotate this state, we obtain a state peaked around any configuration $\vec L=j \vec n$.  The state 
\be
      |j,n\rangle = D^j(n)|j,j\rangle= \sum_m D^j_{jm}(n)|j,m\rangle,
\label{n}
\ee
is a semiclassical state peaked on $\vec L=j \vec n=j D^1(n)\vec z$.\\[-2.mm]

The states \eq{n} are generally denoted as
\be
      |j,\vec n\rangle := |j,n\rangle. 
      \label{vecn}
\ee
where $\vec n =D^j(n)\vec z$. I find this notation confusing. The problem is of course that there are many different $n$ (many rotations) that yield the same $\vec n$, therefore the state $|j,\vec n\rangle$ is not defined by this equation.  The common solution is to choose a ``phase convention" that fixes a preferred rotation $\hat n$ for each $\vec n$. For instance, one may require that $D^1(\hat n)$ leave $\vec z\times \vec n$ invariant. I would find it clearer, even after such a phase convention has been chosen, to still add a label to the notation \eq{vecn}, say for every rotation $n_\phi$ that leaves $\vec n$ invariant, 
\be
      |j,\vec n, \phi \rangle := |j,n_\phi \hat n \rangle =  e^{ij\phi}\ |j,\hat n\rangle.
\ee\\[-1mm]
The reason is that this phase has a physical interpretation: it codes the extrinsic curvature at the face.

LS states are states in ${\cal H}_n$, where $n$ is $v$-valent (unfortunate notation: here $n$ indicates a node, not an $SU(2)$ element as above),  labelled by a unit vector $\vec n_l$ for each link $l$ in $n$, defined by  
\be
      |j_l,\vec n_l\rangle = \int_{SU(2)} \!dg\ \bigotimes_{l\in n} D^{j_l}(g)|j_l,\vec n_l\rangle.      
\ee
The integration projects the state on ${\cal H}_n$. These states are not fully coherent:  they are eigenstates of the area, and the observable conjugated to the area (which is related to the extrinsic curvature) is fully spread.

Remarkably, in \cite{Bianchi:2009ky} it is shown that for large $\eta_l$ the holomorphic states are essentially LS states which are also wave packets on the spins. That is
\be
\langle j_l, \vec n_l | \psi_{H_l}\rangle \sim \prod_l e^{-\frac{(j_l-j^0_l)^2}{2\sigma_l}}\ e^{i\xi_l j_l}
\ee
where $\vec n$ and $\vec {\tilde n}$ are identified with the $\vec n$ in $s(l)$ and $t(l)$ respectively and where $2j_l+1=\eta_l/t_l$ and $\sigma_l=1/(2t_l)$. Thus, the different coherent states that have been used in the covariant and the canonical literature, and which were long thought to be unrelated, are in fact essentially the same thing.   \\[-1.mm]

}\vspace{.6em}

In summary, the Hilbert space ${\cal H}_\Gamma$ contains an (over-complete) basis of ``wave packets" $\psi_{H_l}=\psi_{\vec n_l, {\vec n}'_l, \xi_l, \eta_l}$, with a nice interpretation as discrete classical geometries with intrinsic and extrinsic curvature. 

\subsection{Holomorphic representation}
The coherent states define a natural holomorphic representation of ${\cal H}_\Gamma$  \cite{Ashtekar:1994nx,Bianchi:2010mw}, which is particular useful in calculations.  In this representation, states are represented by holomorphic functions on $SL(2,\C)^L$
\be
    \psi(H_l)=\bk{\psi_{H_l}}{\psi}.
\ee 

The vertex amplitude takes a more manageable form when written in terms of coherent states. First, it is easy to show that in terms of LS states, it reads
\be
 A_v(j_l,\vec n_l,\vec n'_l) =\! \int\! d\tilde g_n  \,
 \bigotimes_l  \ 
 \langle \vec n_{l} | g_{s(l)} g_{t(l)}^{\scriptscriptstyle -1}|\vec n'_{l}\rangle_{(\gamma j,j)}
\label{lorentzian}
\ee
The scalar product is taken in the irreducible $SL(2,\C)$ representation $H_{(\gamma j,j)}$ and $|\vec n_l\rangle$ is the coherent state $|j,\vec n_l\rangle$ sitting in the lowest spin subspace of this representation.

Second, the form of the vertex in the holomorphic basis defined by the coherent states \eq{states} can be obtained by combining the definition (\ref{vava},\ref{KK}) of the vertex and the definition \eq{states} of the coherent states. A straightforward calculation \cite{Bianchi:2010mw} gives 
 \ba 
A_v(H_{l})&\!\equiv&\! \bk{W_v}{\psi_{H_{l}}}
\label{daniele2}\\
&=&\int_{SL(2,\C)^N}  d\tilde g_n \, ~
\prod_{{l}}     P(H_{l} \, , \,g_{s(l)} g_{t(l)}^{-1} )
\nonumber
\ea
where 
\be
P(H,g)\! =\! \! 
\sum_{j} {\scriptstyle (2j+1)}\,e^{\scriptscriptstyle- j(j+1)t}\ {\rm Tr}\!\!
\left[
D^{\scriptscriptstyle(j)}\!(H)Y_\gamma^\dagger D^{\scriptscriptstyle(\gamma j,j)}(g)Y_\gamma
 \right]\!.
\label{daniele222}
\ee
Here $D^{(j)}$ is the analytic continuation of the Wigner matrix from $SU(2)$ to $SL(2,\C)$ and $Y_\gamma$ is defined in \eq{Y}. Eqs.\,
(\ref{daniele2}, \ref{daniele222}) give the ``holomorphic" form of the vertex amplitude. 

\problem{Show that \eqref{states} with \eqref{K3} gives  \eqref{daniele2} and \eqref{daniele222}.}

\insertion{
\subsection{The euclidean theory}

Before describing how to use the above definition of the dynamics, it is useful to introduce also ``euclidean quantum gravity", which is the model theory obtained from the one above by replacing $SL(2,\C)$ with $SO(4)$. The representations of $SO(4)$
are labelled by two spins $(j^+,j^-)$. The theory is the same as above with the only difference that (\ref{pk}) is replaced by 
\be
          j^\pm=\frac{|1\pm \gamma|}2
\ee
and $f_\gamma$ maps $H_j$ into the lowest spin component of $H_{j^\pm}$ if $\gamma>1$, but to  the highest spin component of $H_{j^\pm}$ if $\gamma<1$ (the case $\gamma=1$ is ill defined.) All the rest goes through as above. The vertex amplitude can be written in the simpler form
\be
 A_v(j_l,\vec n_l,\vec n'_l) = \int\! dg^\pm_n  \,
 \bigotimes_l  \prod_{i=\pm}
 \langle \vec n_{l} |g^i_{s(l)} (g^i_{t(l)})^{\scriptscriptstyle {\rm-}1}|\vec n'_{l}\rangle^{2j^i}
\label{euclidean}
\ee\vskip3mm\noindent
where now the integration is over $SU(2)^N\times SU(2)^N\sim SO(4)^N$ and the scalar product is in the fundamental representation of $SU(2)$.}

\subsection{The two-complex expansion}\label{expa}

\hfill  \begin{minipage}{8cm}\small 
{\em ``One can do an enormous amount by various approximations which are non-rigorous and unproved mathematically. [...] Historically, the rigorous analysis of whether what one says is true or not comes many years later after the discovery of what is true. [...] \\ Calculate without rigor, in an exploratory way; [...] don't be so rigorous or you will not succeed."}\\  Richard Feynman 1957, addressing relativists at the
Chapel Hill Conference on General Relativity \cite{Feynman:kx}.
\end{minipage}\\
\vspace{1em}

There is no physics without approximations.  We need a way to compute  transition amplitudes perturbatively, as we do for instance order by order in QED, or with the use of finite lattices in numerical lattice QCD.   What approximations can be effective in the background-independent context of quantum gravity?  In this section and the next one a discuss this issue in some detail. 

The theory is given by the formal limit of infinite refinement for transition amplitudes defined on finite two-complexes. But we may not need to take the limit  to extract approximate predictions from the theory.\footnote{We do not need to sum up the full infinite series of Feynman graphs  to extract viable approximate predictions from QED.}
The amplitudes defined on a small two-complex can be a good approximation in appropriate regimes of boundary states. Indications in this sense have  appeared in  concrete examples (see below), and can be understood in general terms, as I discuss here.

The discretization of a physical system is an approximation that requires the introduction of a discretization parameter $a$, the lattice spacing. The physical limit is recovered by appropriately sending the number $N$ of lattice sites to infinity  and $a$ to zero.  The lattice spacing $a$ can be absorbed into a redefinition of dynamical variables and coupling constants -- then we recover the continuum limit by taking a coupling constant to a critical value, corresponding to $a\to 0$.  

This behavior of discretized systems is common, but not universal.  In a diff-invariant system (or a system invariant under reparametrization of its evolution parameter), the structure of the continuum limit can be substantially different.  This is because, since coordinates are unphysical, the size of the discretization parameter drops out from the dynamics entirely.   The first consequence is that the continuum limit is obtained solely from taking $N$ to infinity, without any lattice spacing or coupling constant to send to a critical value.  The second consequence is more important. If there is a regime where the system approaches a topological theory, then in this regime the discretization becomes nearly exact, and $N$ behaves an effective expansion parameter. 

\insertion{
A simple example of this scenario is provided by the action of an harmonic oscillator  \cite{Rovelli:2011fk},
\be
  S=\frac{m}{2} \int dt \left(\left(\frac{dq}{dt}\right)^2-\omega^2 q^2 \right).
\ee
This can be discretized as 
\be
  S_N=\frac{m}{2} \sum_n a \left(\left(\frac{q_{n+1}-q_{n}}{a}\right)^2-\omega^2 q_n^2 \right).
\label{discrete1}
\ee
With rescaled dimensionless variables $Q_n=\sqrt{\frac{m}{a \hbar} }\,q_n$ and $\Omega=a\omega$, 
the dimensionless action reads
\be
 \frac{S_N}\hbar=\frac{1}2 \sum_{n=1}^N \left((Q_{n+1}-Q_{n})^2-\Omega^2 Q_n^2 \right) 
\ee
and the continuum limit is obtained by sending $N\to\infty$ \underline{and} $\Omega\to 0$.  But consider instead the parametrized version of the same system
\be
  S=\frac{m}{2} \int d\tau \left(\frac{\dot q^2}{\dot t}-\omega^2 \dot t\, q^2 \right)
\ee
Its discretization gives 
\begin{eqnarray}
  S_N&\!=&\!\frac{m}{2} \sum_{n=1}^N a \left(\frac{(\frac{q_{n+1}-q_{n}}a)^2}{\frac{t_{n+1}-t_{n}}a}-\omega^2 \frac{t_{n+1}-t_{n}}a\, q_n^2 \right)
  \nonumber \\
 &\!=&\!\frac{m}{2} \sum_{n=1}^N \frac{({q_{n+1}\!-\!q_{n}})^2}{{t_{n+1}\!-\!t_{n}}}-\omega^2 (t_{n+1}\!-\!t_{n}) q_n^2 .
\label{ho}
\end{eqnarray}
which independent from $a$. The continuum limit is given by $N\to\infty$ \underline{at fixed} $\omega$. We can define rescaled dimensionless variables without using the lattice spacing, as $Q_n=\sqrt{\frac{m\omega}{\hbar} }q_n$ and $T_n=\omega t_n$. These are defined in the 
natural units given by the quantum dynamics itself (in general relativity these natural units are provided by the Planck length). This yields the dimensionless action 
\be
  \frac{S_N}{\hbar}=\frac{1}{2}\sum_{n=1}^N\ \frac{({Q_{n+1}-Q_{n}})^2}{{T_{n+1}-T_{n}}}-(T_{n+1}-T_{n})Q_n^2 .
\ee
It is then shown in \cite{Rovelli:2011fk} that when the boundary values are in the regime
\be
       {\omega ({t_f-t_i})}\ll {\left| \frac{q_f-q_i}{q_i}  \right|} ,
\label{flat}
\ee
the exact transition amplitudes are well approximated by the ones computed with a discretized path integral with \emph{small} $N$. In other words, $N$ is a good expansion parameter in this regime. In this regime, indeed, the kinetic term of the action dominates. The corresponding ``topological" theory is the discretized free action \eqref{ho} with $\omega=0$, which is invariant under refinement:  its transition amplitudes are exactly independent from $N$.\footnote{This is because the discretization is  in the ``perfect action" form, in the language of \cite{Bahr:2009qc}. Notice that here the perfect action is directly realized by the Hamilton function.}}

For gravity, this regime may be near flatness, where the Hamilton function of general relativity approaches that of BF theory, which is topological. If the boundary state is sufficiently close to the boundary data of a flat spacetime, the Hamilton function is close to that of BF theory, and a coarse foam amplitude may approximate the exact one. Intuitively, any further refinment takes the (non-suppressed) amplitudes closer to flatness, where the GR amplitude is the same as the BF amplitude. But this is topological and invariant under a further refinement has no effect. This near-topological-invariance in approaching a regime has been denoted ``Ditt-invariance", from the work of Bianca Dittrich, who has first pointed out its importance  \cite{Dittrich:2008pw,Bahr:2009qc,Bahr:2009mc,Bahr:2010cq,Bahr:2011uj}. 

Recall also the theory approaches Regge gravity in the semiclassical limit.  A single vertex corresponds then to a {\em flat} 4-simplex. Cutting the theory to small $N$ defines an approximation valid around flat space,  where relevant wavelengths are not much shorter than the bounded scattering region $\cal R$. 
More precisely, let $\cal C$ be the two-complex dual to a Regge triangulation, fix boundary data for the Regge equations and 
let $L$ be the maximal length of a bulk Regge bone and $L_R$ the minimal curvature length on the classical solution to these equations. Then the small parameter in which the foam expansion is taken is 
\be
\theta\sim  {L}/{L_R}
\ee
which is essentially the largest value of the Regge deficit angle. Intuitively, a triangulation at a scale smaller that the classical curvature scale suffices. 

This picture is far from being rigorously demonstrated.  But it justifies the exploration of the expansion for small foams as a physical expansion.  In other words, it is reasonable to take the \emph{truncated} transition amplitudes \eqref{ZEPRLLTR} as a family of approximations to the exact amplitudes, and study whether the expansion in the complexity of the foam converges rapidly. 

We can break the expansion in two steps: the graph expansion and the vertex expansion. 

\subsubsection{Graph expansion} 

Consider the component ${\cal H}_\Gamma$ of ${\cal H}$.  Notice that because of the equivalence relation defined in Section \ref{Hs}, all the states that have support on graphs smaller than (subgraphs of) $\Gamma$ are already contained in  ${\cal H}_\Gamma$, provided that we include also the $j=0$ representations. Therefore if we truncate the theory to a single Hilbert space  ${\cal H}_\Gamma$ for a given fixed $\Gamma$, what we lose are only states that need a ``larger" graph to be defined. Let us therefore consider the truncation of the theory to a given graph.\footnote{The analog in QFT is to truncate the theory to the sector of Fock space with a 
number of particles less than a finite fixed maximum number.  It is important to stress that  that virtually {\em all} calculations in perturbative QED are performed within this truncation.}

What kind of truncation is this?  It is a truncation of the degrees of freedom of general relativity down to a finite number; which can be interpreted as describing the lowest modes on a mode expansion of the gravitational field on a compact space. Strictly speaking this is neither an ultraviolet nor an infrared truncation, because the whole physical space can still be large or small. What are lost are not wavelengths shorter than a given length, but rather wavelengths  $k$ times shorter than the full size of physical space, for some integer $k$.  

Therefore the truncation defines an approximation viable for gravitational phenomena where the ratio between the largest and the smallest relevant wavelengths in the boundary state is small.\footnote{A similar situation holds in numerical lattice QCD. The number of lattice sites concretely needed for a numerical calculation is determined by the ratio between the smallest and largest wavelenghts involved in the phenomenon studied. For instance, in studying the proton mass it is determined by the ratio between the quarks and the pion Compton wavelengths. [I thank Laurent Lellouch and Alberto Ramon for clarifications on this point.]}

A good indication supporting the viability of this expansion is given by a recent result in the application of the formalism to cosmology: in  it is shown in \cite{Vidotto:2011qa} that  the large distance limit of the transition amplitude in cosmology \emph{does not change} if the boundary graph is refined.

Notice that the graph expansion resolves the apparent problem that the operators of the theory are defined on ${\cal H}_\Gamma$ rather than on $\cal H$, since all calculation in this approximation can be performed on a single graph. 

\subsubsection{Vertex expansion} 

The second part of the expansion in the number $N$ of vertices of $\cal C$.  

Notice the similarity of this expansion with the standard perturbation expansion of QED. In both cases, we describe a quantum field in terms of interactions of a finite number of its ``quanta".  In the case of QED, these are the photons.   In the case of LQG, these are the ``quanta of space", or ``chunks of space", described in Section \ref{interpr}. In the QED case, individual photons can have small or large energy; in the quantum gravity case, the quanta of space can have small or large volume.  

\begin{figure}[h]
\centerline{\includegraphics[scale=0.03]{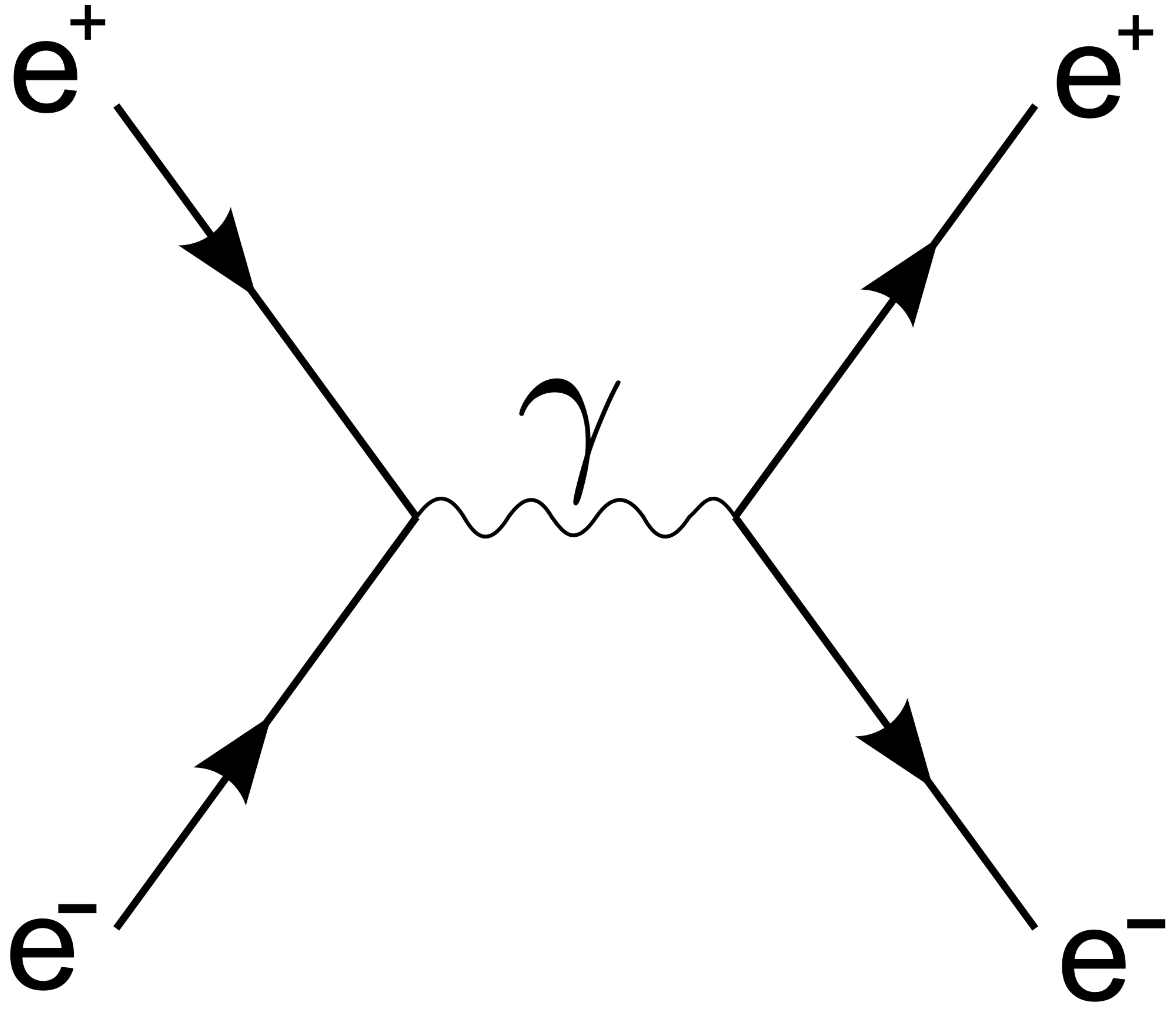}}
\caption{Electromagnetic scattering of two electrons at first relevant order in QED.}
\label{ee}
\end{figure}

In QED, one should be careful not to take the photon picture too literally when looking at the semiclassical limit of the theory. For instance, the Feynman graph for the Coulomb   scattering of two electrons is given in Figure \ref{ee}. But Figure \ref{ee} does \emph{not} provide a viable picture of the continuous electric field in the scattering region, nor of the smoothly curved trajectory of the scattering electrons.   Similarly, if we compute a transition amplitude between geometries at first order in the vertex expansion, we should not mistake the corresponding spinfoam for a faithful geometrical picture of the gravitational field in the corresponding classical spacetime. 

An important issue regards the effect of the radiative corrections on the expansion. The QED perturbative expansion is viable because the effect of all the radiative corrections due to the higher frequency modes can be absorbed into the renormalization of a few parameters.  Does the same happen in LQG?  Preliminary calculations are encouraging: they indicate \emph{finite} radiative corrections of the vertex and corrections \emph{logarithmic} in the cosmological constant (the IR cut-off) for the ``self energy" \cite{Perini:2008pd}\footnote{When \cite{Perini:2008pd} was written, the choice of the face amplitude was still  unclear. Later independent arguments were found selecting the most favorable case ($d_j\sim j$) \cite{Bianchi:2010fj} .}.  But we need to understand this better.

\subsection{Large distance expansion}

An independent approximation can be taken by choosing the boundary to be large.  This means that the boundary state is peaked on a boundary geometry which is large compared with the Planck length. In particular, we can chose holomorphic  boundary states $\psi_{H_l}$ where $\eta_l\gg1$ in each $H_l$.  Recall that  $\eta$ determines the area of the faces in Planck units (see equation \eqref{fs} and the discussion that follows it). Therefore large $\eta$ means that the states we consider describe results of measurements of geometrical quantities that are large compared to the Planck area. 

It is important not to confuse two distinct limits of the theory, and not to confuse two different sources of discreteness.  Let me begin by the limits.  

Each two complex $\cal C$ defines a truncation of the theory. The full theory is recovered by the continuous refinement limit $\cal C\to\infty$ as an expansion in the number of degrees of freedom. This truncation can be done both in the quantum theory and in the classical theory.  In the quantum theory, the truncation is defined by the transition amplitudes $W_{\cal C}$. In the classical theory, the truncation is defined by Regge calculus on a triangulation (whose dual is) $\cal C$.  Each truncation gives an approximation to the full dynamics, which is good as long as $L\ll L_R$. 

A separate, independent limit is the semiclassical limit. This is necessarily a large distance limit, since there is no Planck scale semiclassical regime, because gravitational quantum fluctuations are never smaller than expectation values at the Planck scale. In other words, a necessary condition for the semiclassical limit is Bohr large quantum number conditions, which here becomes large spins $j\to\infty$, therefore large distances compared to the Planck scale.  This is the limit in which $L\gg L_{\rm Planck}$.

The semiclassical limit $j\to\infty$ can be taken at finite $\cal C$. In this limit, the transition amplitudes  $W_{\cal C}$ converge to the dynamics of classical Regge calculus on $\cal C$.  More precisely, they converge to the exponential of the Regge Hamilton function on $\cal C$, as a function of the boundary geometry.   In synthesis:
\begin{table}[htbp]
  \centering 
\multirow{2}{10mm}{\begin{sideways}\parbox{20mm}{$\xrightarrow{ \hspace*{2em}{\rm Continuous\ limit} \hspace*{2em}}$}\end{sideways}}\ \ \ 
\ \ \   \begin{tabular}{@{} ccc @{}}
   \parbox{2cm}{\scriptsize Full quantum \\ gravity ($W$)} \hspace*{1em}& $\xrightarrow{j\to\infty}$ &\hspace*{1em} \parbox{2cm}{\scriptsize General\\ Relativity} \\[4mm]
{\begin{sideways}\parbox{10mm}{ $\xrightarrow{{\cal C}\to\infty}$}\end{sideways}} \hspace*{1em}&  &\hspace*{1em} {\begin{sideways}\parbox{10mm}{ $\xrightarrow{{\cal C}\to\infty}$}\end{sideways}}  \\[2mm]
   \parbox{2cm}{ \scriptsize Truncated \\ amplitudes ($W_{\cal C}$)}\hspace*{1em} & $\xrightarrow{j\to\infty}$ & \hspace*{1em}\parbox{2cm}{\scriptsize Regge calculus\\ on $\cal C$} \\ 
  \end{tabular}\\[.8cm]
\parbox{35mm}{$\xrightarrow{ \hspace*{4em}{\rm Semiclassical\ limit} \hspace*{4em}}$}
  \label{tab:label}
\end{table}

Finally, recall that the truncation introduced by $\cal C$ should not be confused with the quantum discreteness of the geometry.  The quantum discreteness of the geometry is the fact that the geometrical size of the cells of the complex takes discrete values. It disappears in the semiclassical limit, where the theory is studied at distances large than the Planck scale, while it persists in the continuum limit, where an arbitrary large two-complex is considered.  In other words, the refinement of the cellular complex does not make the size of the cells go smoothly to zero, because geometry is physically discrete at the Planck scale.  This is the most characteristic aspect of the quantum gravity.

\subsection{Some concrete calculation}

Using the structure of approximations discussed above, some transition amplitudes have already been computed using the covariant theory. So far, the focus has been limited to the semiclassical limit, in order to show that the formalism is viable, and that converges to the expected classical quantities.  I briefly mention here two series calculations, referring the reader to the literature for all the details.  The first is  the derivation of classical cosmology from the full quantum gravity theory.  The second is the derivation of graviton $n$-point functions.

\subsubsection{Cosmology}

The classical dynamics of homogeneous and isotropic cosmology has been first derived from the full quantum theory in \cite{Bianchi:2010zs}  then with a cosmological constant in \cite{Bianchi:2011ym}, and extended to arbitrary regular boundary graphs in \cite{Vidotto:2011qa}.  The idea of the calculation is simple. Consider a closed universe with scale factor $a(t)$ and derive the Friedman dynamics for $a(t)$ from the full quantum theory.  To this aim, we want to compute the transition amplitude between the initial and final homogeneous isotropic geometry of a 3-sphere. A well known classical solution of the problem, in the presence of a cosmological constant $\Lambda$ and in the regime of large scale factor is given by the deSitter solution.
\be 
                   a(t)=e^{\sqrt{\frac{\Lambda}{3}}t},
\ee
and therefore
\be 
              \dot a = \sqrt{\frac{\Lambda}{3}}\ a.
\label{region}
\ee 
$\dot a$ is the proper time derivative of the scale factor, which is determined by the extrinsic curvature, which is turn is the canonical variable conjugate to $a$.  Consider a semiclassical state $\psi_{a,\dot a} $ of quantum geometry, peaked on an homogeneous geometry with scale factor $a$ and on a certain value $\dot a$.   If the quantum dynamics has the correct classical limit, the quantum amplitude for this state must be suppressed everywhere except in the region \eqref{region}. Showing this is the aim of the calculation.\footnote{Notice that in this case the dynamical equations do not rely final canonical coordinate and momenta with initial ones; rather, they constrain the final coordinate and momenta among themselves. This can also be see from the factorization of the Hamilton function of homogenous isotropic cosmology, which is easily computed inserting the deSitter solution into the general relativity action with boundary terms, and reads
\be 
S(a_i,a_f)=\frac23\sqrt{\frac{\Lambda}{3}}(a_f^3-a_i^3)\;,  
\ee
which is the sum of an initial and final term. Since at first order in $\hbar$ the transition amplitude is the exponent of the Hamilton function, we expect then the transition amplitude to factorize in this approximation.}

Following the general approximation strategy outlined in the previous sections, we truncate the theory down to a simple graph. In particular, we triangulate the 3-sphere with two tetrahedra glued to one another through all their faces.  The dual graph is $\Gamma=\Delta_2^*$,  the ``dipole" graph \cite{Rovelli:2008dx} formed by two nodes connected by four links. 
 \begin{center}
\begin{picture}(20,40)
\put(-36,28) {$\Delta_2^*\  = $}
% cerchi 
%\put(20,30) {\circle{30}}
\put(04,30) {\circle*{3}} 
\put(36,30) {\circle*{3}}  
% linee 
\qbezier(4,30)(20,53)(36,30)
\qbezier(4,30)(20,21)(36,30)
\qbezier(4,30)(20,39)(36,30)
\qbezier(4,30)(20,7)(36,30)
\put(50,27) {\circle*{1}} 
\end{picture}
\end{center} 
\vspace{-1.5em}
We take two such dipole graphs, representing initial and final 3-sphere, and select the simplest non-trivial two-complex bounded by these.  This is given by a single vertex, four edges and eight faces and is represented by \\
\begin{center}
\begin{picture}(20,30)
%\put(-36,28) {$\Delta_2^*\  $}
% cerchi 
%\put(20,30) {\circle{30}}
\put(04,30) {\circle*{3}} 
\put(36,30) {\circle*{3}}  
% linee 
\qbezier(4,30)(20,53)(36,30)
\qbezier(4,30)(20,21)(36,30)
\qbezier(4,30)(20,39)(36,30)
\qbezier(4,30)(20,7)(36,30)
%\put(-36,-32) {$\Delta_2^*\  $}
% cerchi 
%\put(20,-30) {\circle{30}}
\put(04,-30) {\circle*{3}} 
\put(36,-30) {\circle*{3}}  
% linee 
\qbezier(4,-30)(20,-53)(36,-30)
\qbezier(4,-30)(20,-21)(36,-30)
\qbezier(4,-30)(20,-39)(36,-30)
\qbezier(4,-30)(20,-07)(36,-30)
%vertice
%{\color{red}
\linethickness{0.4mm}
\put(20,0) {\circle*{6}} 
\qbezier(4,29)(20,0)(36,-29)
\qbezier(4,-29)(20,0)(36,29)
%}
\end{picture} 
\end{center} 
\vspace{15mm}
This is the two-complex $\cal C$ whose amplitude gives the first approximation to the quantum transition amplitude.\footnote{In \cite{Hellmann:2011jn}, the effect of some ``wild" two-complexes has been considered, leading to to suggestions that some conditions on the regularity of the two-complex are needed.}

We want to compute the transition amplitude between \emph{coherent} homogenous isotropic states, namely semiclassical wave packets peaked on a given coordinate and momentum. These will depend on two variables, essentially $a$ and $\dot a$, at each $\Delta_2^*$. 

To build these semiclassical wave packets, we use the coherent state technology developed in Section \ref{coherent}. To this aim, consider a regular (symmetric) triangulation of a \emph{metric} 3-sphere, formed by two \emph{regular} tetrahedra joined along all their faces. Let $\vec n_{l}, \ {l}=1,...,4$ be four unit vectors in $R^3$ normal to the faces of a regular tetrahedron, that is, such that $\vec n_{l}Ê\cdot  \vec n_{{l}'}=-\frac13$. The isotropic homogeneous states of a 3-sphere in this approximation are then the states 
\be
|z\rangle= |H_{l}(z)\rangle
\ee
where $H_{l}(z)$ are given by 
\be
H_{l}(z)= e^{z \vec n_{l}\cdot \vec \tau}. 
\label{Hl}
\ee
where $z=c+itp$.  Following  Section \ref{coherent}, we recognize $p$ as the area of the triangles of the tetrahedra, namely as a quantity proportional to $a^2$, and $c$ as a quantity proportional to the extrinsic curvature, namely $\dot a$.   

These coherent quantum states represent truncations of the GR degrees of freedom, but include more degrees of freedom than just the scale factor, because the geometry of the triangulation captures more degrees of freedom of the metric than just the scale factor. A more refined graph captures increasingly more degrees of freedom (these have been studied in \cite{Battisti:2009kp}, \cite{Magliaro:2010qz} and  \cite{Vidotto:2011qa}).   The coherent states include the relevant quantum fluctuations of these degrees of freedom, around the classical homogenous isotropic configuration. 

We are interested in the transition amplitude between two such states, that is
\be
W(z,z')= \langle W| (|z\rangle\otimes|z'\rangle).
\ee
Using the explicit form of the transition amplitude in this coherent state basis, given in \eqref{daniele2}, we obtain \cite{Bianchi:2010zs}
\ba
W(z,z')&=&\int_{SL(2,\C)^4} dg_1... dg_4  \\ \nonumber  && \times \prod_{{l}=1,4} \sum_{j_{l}} d_{j_{l}}  e^{-tj_{l}(j_{l}+1)}
P_j(H_l(z),g_1g_2^{-1})  \\  && \times 
 \prod_{{l}=1,4} \sum_{j_{l}} d_{j_{l}}e^{-tj_{l}(j_{l}+1)}
P_j(H_l(z'),g_3g_4^{-1}). \nonumber
\ea
That is 
\be
W(z,z')= W(z)W(z')
\ee
where
\be
W(z)=\int_{{}_{SL(2,\C)}} \hspace{-1.5em}dg\  \prod_{{l}=1,4} \sum_{j_{l}} d_{j_{l}}
e^{-tj_{l}(j_{l}+1)}
P_j(H_l(z),g).
\ee

\problem{Show this.}

The factorization property was expected in this limit, for the reason given in the last footnote. 

We are interested in computing this for large spaces, namely when the imaginary part of $z$ is large.  Consider the form \eqref{Hl} of $H_{l}$ and $D^{(j)}(H_{l})$ in this limit. Orienting the $z$-axis along $n_{l}$, we have 
\be
D^{(j)}(H_{l})=D^{(j)}(e^{z\tau_3})^m{}_n=\delta^m{}_n e^{imz}.
\ee
When the imaginary part of $z$ is large, only the highest magnetic quantum number  $m=j$ survives, therefore in this limit
\be
D^{(j)}(H_{l})=\delta^m{}_j\delta^j{}_n e^{ijz}=e^{ijz}P_{n_{l}}
\ee
where $P_{n_{l}}$ is the projector on the highest magnetic number eigenstate in the direction $n_{l}$. Observe now that the sum over $j$ is (a discretization of) a gaussian integral in $j_l$, peaked on a large value $j^*\sim t p$.  One can show that the rest of the expression contributes only polynomially, giving finally  \cite{Bianchi:2010zs} 
\be
 W(z) \sim z e^{-\frac{z^2}{2t}}.
\ee
Finally, restoring $\hbar\ne 1$ for clarity, the transition amplitude in this approximation is 
\be
 W(z,z') \sim zz' e^{-\frac{z^2+{z'}^2}{2t\hbar}}.
\ee
This amplitude reproduces the correct Friedmann dynamics in the sense that it satisfies a quantum constraint equation which reduces to the (appropriate limit of the) Friedmann hamiltonian in the classical limit \cite{Bianchi:2010zs}. Once appropriately normalized by the norm of the boundary coherent state, it can be shown to be peaked on the classical solutions of the Friedmann dynamics in the large scale limit we are considering. 

\begin{figure}[h]
\includegraphics[scale=0.5]{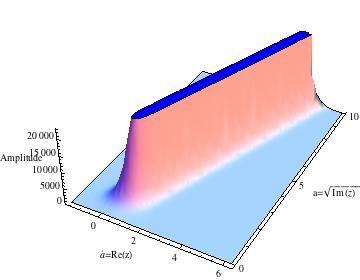}
\caption{Numerical analysis of the transition amplitude \eqref{amplitude}. The only free parameter is $\lambda v_o$, here is set equal to $.3$. The computation has been done truncating the sum over $j$ up to a maximum value $j_{max}=200$. This choice is compatible to maximal scale factor plotted.}
\label{figcosmo}
\end{figure}

The result can be improved by adding a cosmological constant to the theory \cite{Bianchi:2011ym}.  To do this, we use the spin-intertwiner basis version of the amplitude, and modify it as follows
\be
   Z_{\cal C}=\sum_{j_f,\vol_e} \prod_f (2j+1)  \prod_e e^{i\lambda \vol_e} \prod_v A_v(j_f,\vol_e). 
\label{modificazione}
\ee
This resulting amplitude turns out to be 
\be
W(z)=
\sum_{j} { (2j+1)} \,\frac{N_o}{j^3}\ e^{-2t\hbar j(j+1)- i zj-i\lambda \vol_o j^{\frac32}} 
\label{amplitude}
\ee
There are many ways of analyzing this amplitude. The simplest is to plot its modulus square: see figure \ref{figcosmo}. This shows a linear relation between $a$ and $\dot a$, which is readily integrated giving 
\be
a(t)=e^{\sqrt{\frac\Lambda3}t}
\label{desitter}
\ee
where $\Lambda$ is a constant fixed by the parameter $\lambda$ in \eqref{modificazione}. See \cite{Bianchi:2011ym} for more details.  Equation \eqref{desitter} is the DeSitter solution of the Einstein equations.\footnote{For large scale factors, which is the regime in which we are, this also holds also for the spatially compact coordinatization.} 

\subsubsection{$n$-point functions}

Much has changed since the early attempts to find graviton physics in the background independent loop quantum gravity formalism \cite{Iwasaki:1992qy}.  The boundary formalism breakthrough has opened the way to concrete calculations.   The two point function of general relativity over a flat spacetime has been computed in the Euclidean  \cite{Bianchi:2009ri}, as well as in the Lorentzian  \cite{Bianchi:2011fk}  theory, following the earlier attempts in \cite{Rovelli:2005yj,Bianchi:2006uf,Alesci:2007tx,Alesci:2007tg,Alesci:2008ff}, and has been shown to converge to the free graviton propagator of quantum gravity in the large distance limit. 

Higher $n$-point functions have been computed to the same order in \cite{Rovelli:2011kf} (so far only in the Euclidean theory), and they match the corresponding Regge expressions. 

The calculation is to first order in the vertex expansion, on the complete graph with five nodes $\Gamma_5$, and to first order in the large-distance expansion. The boundary state $\psi_L$ is a coherent state determined on the boundary of a 4-simplex. That is, the boundary graph is
\begin{equation}
\Gamma_5 = \hspace{-4em}\setlength{\unitlength}{0.0004in} 
\begin{picture}(5198,1100)(5000,-4330)
\put(8101,-5161){\circle*{68}}
\put(8401,-3961){\circle*{68}}
\put(6601,-3961){\circle*{68}}
\put(7501,-3361){\circle*{68}}
\put(6901,-5161){\circle*{68}}
\put(8101,-5161){\line(1,4){300}}
\put(7501,-3361){\line( 3,-2){900}}
\put(6601,-3961){\line( 3, 2){900}}
\put(8101,-5161){\line(-1, 0){1200}}
\put(6901,-5161){\line(-1, 4){300}}
\put(7876,-4479){\line( 1,-3){225.800}}
\put(7089,-4351){\line(-6, 5){491.312}}
\put(7801,-3961){\line( 1, 0){600}}
\put(6901,-5161){\line( 1, 3){383.100}}
\put(7321,-3871){\line( 1, 3){173.100}}
\put(7501,-3354){\line( 1,-3){325.500}}
\put(7456,-4726){\line(-4,-3){581.920}}
\put(8394,-3961){\line(-5,-4){828.902}}
\put(7569,-4629){\line( 0,-1){  7}}
\put(6601,-3969){\line( 1, 0){1020}}
\put(8109,-5161){\line(-5, 4){867.073}}
\end{picture}\hspace{-3em}
\label{fsimpic}
\end{equation}
\vskip1cm
\noindent and represents a triangulation of a (topological) three sphere immersed in spacetime.  The spinfoam is (vertex and edges drawn)
\begin{equation}
{\cal C} = \hspace{-4em}\setlength{\unitlength}{0.0004in} 
\begin{picture}(5198,1100)(5000,-4330)
\put(8101,-5161){\circle*{68}}
\put(8401,-3961){\circle*{68}}
\put(6601,-3961){\circle*{68}}
\put(7501,-3361){\circle*{68}}
\put(6901,-5161){\circle*{68}}
\put(8101,-5161){\line(1,4){300}}
\put(7501,-3361){\line( 3,-2){900}}
\put(6601,-3961){\line( 3, 2){900}}
\put(8101,-5161){\line(-1, 0){1200}}
\put(6901,-5161){\line(-1, 4){300}}
\put(7876,-4479){\line( 1,-3){225.800}}
\put(7089,-4351){\line(-6, 5){491.312}}
\put(7801,-3961){\line( 1, 0){600}}
\put(6901,-5161){\line( 1, 3){383.100}}
\put(7321,-3871){\line( 1, 3){173.100}}
\put(7501,-3354){\line( 1,-3){325.500}}
\put(7456,-4726){\line(-4,-3){581.920}}
\put(8394,-3961){\line(-5,-4){828.902}}
\put(7569,-4629){\line( 0,-1){  7}}
\put(6601,-3969){\line( 1, 0){1020}}
\put(8109,-5161){\line(-5, 4){867.073}}
\thicklines
\put(7500,-4260){\line(0,1){910}}%%%
\put(7500,-4260){\line(3,1){930}}%%%
\put(7500,-4260){\line(-3,1){930}}%%%
\put(7500,-4260){\line(2,-3){590}}%%%
\put(7500,-4260){\line(-2,-3){590}}%%%
\put(7500,-4260){\line(-2,-3){590}}%%%
\put(7500,-4260){\circle*{170}}%%%
\end{picture}\hspace{-3em}
\label{fsimpic}
\end{equation}
\vskip1.2cm
\noindent  and represents a finite region of spacetime, bounded by $\Gamma_5$.

For the two-point function, the quantity computed is 
\ba
W^{abcd}_{mn} &=&  \langle W| \vec L_{na}\cdot \vec L_{nb}\  \vec L_{mc} \cdot \vec L_{md} | \psi_L\rangle \label{prop}
\\
&& -\langle W| \vec L_{na}\cdot \vec L_{nb} | \psi_L\rangle\langle W\vec L_{mc} \cdot \vec L_{md}^j | \psi_L\rangle.\nonumber
\ea
where $m,n,a,b...=1,...,5$ label the nodes of $\Gamma_5$.  

The boundary state $\psi_L$ is a coherent state peaked on the intrinsic and extrinsic geometry of a four-simplex immersed in flat space.  $L$ is the radius of the four simplex, chosen so that the centers of the two boundary tetrahedra $n$ and $m$ are at a distance $L$. This determines the background space over which the two-point function is computed. 

The operators $\vec L_{na}\cdot \vec L_{nb}$ are Penrose metric operators, on the node $n$, relative to links $(n,a)$ and $(n,b)$. They give angles between the faces of the boundary tetrahedra, and areas of these tetraehdra.  These are directly given by components of the metric tensor (integrated over the faces of the tetrahedra). 

The resulting expression can be compared with the corresponding (connected) quantity 
\ba
W^{abcd}(x_m,x_n) &=&  \langle 0| g^{ab}(x_n)g^{cd}(x_m) | 0\rangle_{\rm c}
\ea
in conventional QFT, where $g^{ab}(x)$ is the gravitational field operator.  Intuitively, consider flat spacetime with the two points $x_n$ and $x_m$. Choose a four-simplex in this flat spacetime, such that the two points $x_n$ and $x_m$ sit at the center of two of the boundary tetrahedra.  Then take the intrinsic and extrinsic geometry of the boundary of this four-simplex to be the classical data determining the semiclassical state $\psi_L$.  This is how the background geometry is fed into the calculation, in the background independent dynamics.

I do not report the calculation here, since is is quite intricate and involves numerous subtleties (gauge fixing, choice of the boundary state, correct identification between smeared quantities, precise definition of the limit...).  The result is that the $n$-point functions computed from the LQG converge to the expected first order (free) one in the large $j$ limit.  
For the propagator, for example, \eqref{prop} scales correctly with the square of the distance and has the full correct tensorial structure of the propagator (in an appropriate gauge).  For a complete discussion, see the original references, and in particular \cite{Rovelli:2005yj,Bianchi:2006uf,Bianchi:2009ri,Bianchi:2011fk}. Progress towards next order is in \cite{Magliaro:2011dz,Magliaro:2011qm,Krajewski:2011uq}.

\section{Derivations}\label{derivation} 
 
I have presented the theory without \emph{deriving} it from classical general relativity. There are a number of distinct derivations that converge to the theory. In this last section, I sketch some basic ideas in these derivations.  A word of caution is however needed. 

Quantum-gravity research has often focused on setting up and following ``quantization paths" from classical general relativity to a quantum theory. These are very useful to provide heuristic indications for constructing the quantum theory, but they are neither sufficient nor necessary for taking us to quantum gravity.  If there was a straightforward quantization route, the quantum theory of gravity would have been found long ago.  Any generalization requires a certain amount of guesswork. The  ``quantization paths"  sketched below must be seen as nothing more than heuristics, which have given suggestions useful for construction of the theory, and shed light on aspects of the definitions.  

The theory itself should not be evaluated on the basis of whether or not quantization procedures have been ``properly followed" in setting it up. It must be judged on the basis of two criteria. The first is whether it provides a coherent scheme  consistent with what we know about Nature, namely with quantum mechanics and, in an appropriate limit, with classical general relativity. The second is to predict new physics that agrees with future empirical observations. This is all we demand of a quantum theory of gravity. 

Since for the moment we do not have so many useful empirical observations, it might sound that the considerations above give us far to much freedom.  How then to choose between different quantum gravity theories, or different ways of constructing the theory? This question is asked often. I think it is a misleading question, for the following reason. At present, we do not have several consistent, complete and predictive theories of quantum gravity. In fact, we are near to have none at all. Most of the quantum gravity approaches lead to very incomplete theories where predictions are impossible. Therefore the scientifically sound problem, today, is whether \emph{any} complete and consistent quantum theory of gravity can be set up at all.  If we can solve  \emph{this} problem, it is already a great success, after decades of search.  The issue of checking whether this is the  \emph{right} theory, namely the theory that agrees with experiments, comes after.  And if history is any guide, solving a problem of this kind has almost always immediately led to the right solution: Maxwell found one of the possible ways of combining electricity and magnetism, and it was the right one, Einstein found one of the possible ways of writing a relativistic field theory of gravity and it was the right one, and so on. The scientifically sound problem today, therefore, is whether \emph{a} complete, consistent, predictive theory of quantum gravity exists.  With this in mind, let us see what is the formal relation between classical general relativity and the quantum theory constructed above.  Accordingly, this section is mostly sketchy, and relies on pointers to existing literature.

\subsection{Dynamics} \label{derivationdynamics}

General relativity can be presented as the field theory for the field $g_{\mu\nu}(x)$ defined by the classical equations of motion that follow for instance from the action 
\be
       S[g]=\int dx\ \sqrt{g}\ (R-2\Lambda).
       \label{actiongr}
\ee
In  this section I neglect the cosmological constant $\Lambda$.  The metric field $g_{\mu\nu}(x)$ cannot be the fundamental field, because it does not allow fermion coupling.  A better presentation of the gravitational field, compatible with the physical existence of fermions, is the tetrad formulation, where the gravitational field is represented by the field $e(x)= e_\mu(x) dx^\mu$, where $e_\mu(x)= (e_\mu^I(x), I=0,1,2,3)$ is a vector in Minkowski space. The relation with the metric is well known: $g_{\mu\nu}=\eta_{IJ}e_\mu^Ie_\nu^J$. It convenient to treat the theory in the so called first-order formalism, in which the connection is treated as an independent variable.  Therefore we introduce a (a priori) independent $SL(2,\C)$ connection field $\omega(x)= \omega_\mu(x) dx^\mu$, where $\omega_\mu(x)= (\omega_\mu^{IJ}(x))$ is an antisymmetric Minkowski tensor, namely an element in the adjoint representation of $SL(2,\C)$. Another element in the adjoint representation is the Plebanski two form
\be
    \Sigma\equiv e\wedge e,
\ee 
which is important in what follows. Rewriting \eqref{actiongr} in terms of these quantities gives 
\be
S[e,\omega]=\int (e\wedge e)^*\wedge F[\omega]\label{acct}
\ee
where $F$ is the curvature of $\omega$, the star indicates the Hodge dual in the Minkowski indices, that is $(e\wedge e)^*_{IJ}\equiv\frac12\ \epsilon_{IJKL}\ e^K\wedge e^L,
$ and a trace in the adjoint representation is understood (that is $\Sigma F\equiv \Sigma_{IJ} F^{IJ}$)

\problem{Derive the action \eqref{acct} from  \eqref{actiongr}.}

Now, recall that in QCD we add to the classical action $\int F^*\wedge F$ a parity violating $\theta$-term $\theta\int F\wedge F$  that does not affect the equations of motion, but which has an effect in the quantum theory. The same can be done in general relativity, giving the Holst action
\be
S[e,\omega]=\int [(e\wedge e)^*+\frac1\gamma (e\wedge e)]\wedge F[\omega].
\label{accct}
\ee
This action is equivalent to \eqref{actiongr} in the sense that all solutions of the Einstein equations of motion are also solutions of this action.  Therefore this action is  just as empirically valid as standard general relativity. 

\problem{Write the second term in \eqref{accct} in metric variables (assuming the connection is determined by the metric): What does it give? The result can be guessed from its parity character.}
We are interested in the quantum states of this theory. Quantum states live at fixed time, or, in a more covariant language, on a 3d surface bounding a spacetime region.  Let us therefore consider a region of spacetime with a boundary $\Sigma$, and let's study the canonical formalism associated to this boundary. The momentum conjugate to $\omega|_\Sigma$, namely to the restriction of $\omega$ to $\Sigma$, is immediately read out of the action:
\be
\pi=\left.\left((e\wedge e)^*+\frac1\gamma(e\wedge e)\right)\right|_\Sigma.
\ee
$\pi$ lives in the adjoint representation of $SL(2,\mathbb{C})$. In the quantization of any theory with an internal gauge, the momentum conjugate to the connection is the generator of the local gauge transformations, thus we identify $\pi$ with the \slc generator in the quantum theory. 

 It is convenient to partially gauge fix the internal $SL(2,\C)$ symmetry. Choose (continuously) a scalar field $n=(n_I)$ with (timelike) values in Minkowski space at every point of the boundary, and gauge fix $e$ by requiring that $ne|_\Sigma=n_Ie^I_idx^i=0$, where $x^i$ are coordinates on the boundary.  Define its electric and magnetic components with respect to the gauge defined by $n$, that is
\be
	\vec K:=n\pi,\hspace{1em}\vec L:=-n\pi^*.
\ee
where the arrow reminds us that these are in fact 3d quantities like the electric and magnetic field. 
In the time gauge, using $ne=0$, this gives immediately 
\be
\vec K=n(e\wedge e)^*|_\Sigma,\hspace{1em}\vec L=-\frac1\gamma n(e\wedge e)^*|_\Sigma,
\ee
that is 
\be
          \vec K+\gamma \vec L = 0,
          \label{simplicity}
\ee
which is called the ``linear simplicity constraint". 
 $\vec L$, formed by the space-space components of $\pi$, generates rotations, while $\vec K$, formed by the time-space components of $\pi$, generates boosts.  Therefore the states of the quantum theory will have to satisfy the constraint equation \eqref{simplicity}. Compare this equation with equation  \eqref{quantumsimpl}, and recall that the space  ${\cal H}_\gamma$ is the subspace of the space of the \slc representations where this equation holds. It is clear that the restriction of the \slc representations defined by ${\cal H}_\gamma$ is precisely an implementation of the general relativity constraint  \eqref{simplicity}.

Now, consider a theory defined by the action
\be
S[B,\omega]=\int B\wedge F[\omega].
\ee
where $B$ is an arbitrary two-form with values in the adjoint representation. This is the same theory as general relativity but \emph{without} the condition that $B$ has the form $B=(e\wedge e)^*+\frac1\gamma (e\wedge e)$ for some $e$, or, equivalently, without the condition that in the time gauge \emph{simplicity} holds.  This theory, which is called BF theory, is well understood both in the classical and quantum domains. The equations of motion give $F=0$, therefore the connection is flat.  Therefore the dynamics of general relativity can be thought as the combination of two ingredients: an \slc theory of a flat connection, but with the additional constraint \eqref{simplicity}. Compare now these observations with the definition of the vertex amplitude \eqref{K3} that defines the LQG dynamics.  The vertex is obtained by mapping $SU(2)$ spin networks into \slc spin networks with the map $Y_\gamma$, and then evaluating these networks at the identity. The first step maps  $SU(2)$ spin networks precisely in the subspace of  \slc spin networks where the simplicity constraint holds. The evaluation at the identity is like the request that the connection is flat. In fact, it can be easily shown that if we replace  $Y_\Gamma$ with the identity, we obtain one of the well known quantizations of $BF$ theory.  This is discussed in many review papers and I will not insist on this here. See for instance \cite{Perez:2004hj} and \cite{Rovelli:2004fk}. 

The considerations above do not represent a derivation of the LQG dynamics from classical general relativity. But they show that the basic ingredients on which the LQG dynamics is defined are precisely the basic ingredients of the general relativity dynamics, when expressed in the tetrad-connection form. 

\subsection{Kinematics}

The consideration above illustrate the formal relation between the \emph{dynamics} of LQG and that of classical general relativity.  Let me now step back and discuss the logic behind the \emph{kinematics} of LQG.  Consider again the three-dimensional spacelike slice $\Sigma$ of spacetime (that is, ``space") and consider the restriction of $\omega$ to this slice.  Fix the time gauge. The ``Ashtekar-Barbero" connection is the field
\be
A\equiv n(\omega^*+\gamma\omega)|_\Sigma.
\ee
It transforms as an $SU(2)$ connection under the transformations generated by $\vec L$ 
and it is a simple calculation to show that $\vec L$ and $A$ are conjugate variables.   Given a two-dimensional surface $l$ in space and a one-dimensional line $\sigma$ , let 
\be
\vec L_l= \int_l \vec L 
\label{LL}
\ee
be the flux of $\vec L$ across the surface and 
\be
U_\sigma= {\cal P}\ e^{\int_\sigma A}
\ee
the parallel transport operator for $A$ along a curve $\sigma$, which is an element of $SU(2)$.  The two quantities $\vec L_l$ and $U_\sigma$ are the basis of the canonical loop quantization of general relativity \cite{Ashtekar:2004eh,ThiemannBook,Rovelli:2004fk}.  Their Poisson algebra can be represented by operators acting on a space $\cal S$ of functionals $\psi[A]$ of the connection. The space $\cal S$ is formed by (limits of sums of products of) functionals that depend on the value of $A$ on graphs.

\problem{(Very important) Compute the Poisson bracket $\{\vec L_l,U_\sigma\}$. Assume for simplicity that $\sigma$ crosses $l$ at a single point, which splits $\sigma$ into two lines $\sigma_1$ and $\sigma_1$. Show that $\{\vec L_l,U_\sigma\}$ is not distributional (why?) and is proportional to $U_{\sigma_1}\vec\tau U_{\sigma_2}$. Hint: the Poisson brackets have a 3d delta function, the flux a 2d integral and the holonomy a 1d integral.  Write everything explicitly in coordinates and observe that the final result is independent from coordinates and parametrization.}

The key gauge invariance is 3d coordinate transformations, which plays three major roles.  First, it is the main hypothesis for a class of theorems stating that the resulting representation is essentially unique \cite{Lewandowski:2005jk,Fleischhack:2006zs}. Second, it ``washes away" the location of the graph $\Gamma$ in $\Sigma$, so that all the Hilbert subspaces associated to distinct but topologically equivalent graphs in $\Sigma$ end up identified \cite{Rovelli:1987df,Rovelli:1989za}.  Depending on the particular class of coordinate transformations one allows in the classical theory, one ends up with the different versions of the Hilbert space mentioned above. Third, this gauge invariance resolves the difficulties that have plagued the previous attempts to use a basis of loop states in continuous gauge theories. 

The other gauge invariance of the canonical theory is formed by the local $SU(2)$ transformations, which gives rise to \eq{gauge}.

The natural definition of the dynamics in the hamiltonian framework is in terms of a hamiltonian constraint operator \cite{ThiemannBook}. 
A spinfoam expansion for the transition amplitudes can in principle be derived from the canonical formalism \cite{Reisenberger:1996pu}, but for the moment we do not know how to derive explicitly the amplitude given above from a well defined Hamiltonian constraint.  \\ 

\insertion{
The purely canonical formulation of the dynamics defined by the Hamiltonian operator is now being developed by an active research program \cite{Giesel:2007wi}, which mostly uses the 
idea of gauge-fixing diffeormorphisms with matter fields.}

\subsection{Covariant lattice quantization} 

A different possibility to build the quantum theory is to discretize general relativity  on a 4d lattice with a boundary, and study the resulting Hilbert space of the lattice theory. This is close in spirit to lattice gauge theory. The difference is diffeomorphism invariance: in general relativity the lattice is a ``coordinate" lattice, and coordinates are gauges. Thus for instance there is no analog of the QCD lattice spacing $a$. More precisely, the physical dimensions (lengths, areas, volumes) of the cells of the lattice are not fixed, as in lattice gauge theory, but are determined by the discretized field variables themselves. 

The (double covering of the) local gauge group of the covariant theory is $SL(2,\C)$ and the boundary space that one obtains on the boundary of the lattice theory is
\ba
{\cal H}^{SL(2,\C)}_\Gamma &=& L_2[SL(2,\C)^L/SL(2,\C)^N].
 \label{hsl}
\ea
where $\Gamma$ is the two-skeleton of the boundary of the lattice. 
The states in this Hilbert space $\psi(g_l), g_l\!\in\! SL(2,\C)$, can be seen as wave functions of the holonomies $H_l \!=\! {\cal P}\exp{\int_{l} \omega}$ of the spin connection $\omega$, along the links $l$.  The corresponding generators $J$ of the Lorentz group must therefore represent the conjugate momentum of $\omega$. 

Notice that  \slc has a natural complex structure and we can define the complex variables $\Pi=K+iL$ and $\overline \Pi=K-iL$.  Then \eqref{simplicity}  can be interpreted as a reality condition.  A technique for implementing reality conditions in the quantum theory is to choose a scalar product appropriately. This is because reality conditions depend on complex conjugation, and this is realized by the adjoint operation in the quantum theory. But the adjoint operation does not depend on the linear structure: it depends on the scalar product. Therefore we can implement the reality conditions by choosing the scalar product appropriately  (see e.g. \cite{Rovelli:1991zi}). Viceversa, we can view the quantum mechanical scalar product as \emph{determined} by the reality conditions. 

For instance, in the usual Schr\"odinger representation the condition that $x$ and $p$ be real translates into the requirement that the linear operators $x$ and $p=-i\frac{d}{dx}$ (which form a linear representation of the poisson algebra of the observables $x$ and $p$) be self adjoint.  The  $L_2$ scalar product is the unique one making them so, and thus satisfying the reality conditions. 

If we apply this idea here, we have to find a scalar product on a space of lineal functionals on \slc, such that \eqref{simplicity} holds.  The solution is clear: this is the scalar product implicitly defined by the map \eqref{Y}. This reduces the space \eqref{hsl} to one isomorphic to \eqref{n-1}. 

\subsection{Polyhedral quantum geometry}  

The idea of polyhedral quantum geometry is to describe ``chunks" of quantum space by quantizing the space $\tilde S$ of the ``shapes" of the geometry of solids figures (tetrahedra, or more general polyhedra) \cite{Barbieri:1997ks,Barrett:1999qw,Barrett:2009cj,Pereira:2010}. This space can be given a rather natural symplectic structure as follows.  Take a flat tetrahedron, for simplicity.   Its shape can be coordinatized by the four normals $\vec L_l, l=1,2,3,4$ to its faces, normalized so that  
 $|\vec L_l|=a_l$ is the area of the face $l$.  A natural $SO(3)$ invariant symplectic structure on $\tilde S$ is $\omega= \sum_l \epsilon_{ijk}\, L_l^i\, dL_l^j\wedge dL_l^k$, or, equivalently, by the Poisson brackets
\be
       \{L_l^i, L_{l'}^j\} =\delta_{ll'}\, \epsilon^{ij}{}_k\  L^k. 
\ee
A quantum representation of this Poisson algebra is precisely defined by the generators of $SU(2)$ on the space ${\cal H}_n$ given in \eq{calHn} (for a 4-valent node $n$). The operator corresponding to the area  $a_l=|\vec L_l|$ is the Casimir of the representation $j_l$, therefore the space  ``quantizes" the space of the shapes of the tetrahedron with areas $j_l(j_l+1)$. Furthermore, the normals of a tetrahedron satisfy 
\be
       \vec C := \sum\!\raisebox{-1mm}{${}_l$} \ \vec L_l= 0. 
\label{C}
\ee
The Hamiltonian flow of $\vec C$, generates the rotations of the tetrahedron in $R^3$. By imposing equation \eq{C} and factoring out the orbits of this flow, the space $\tilde S$ reduces to a space $S$ which is still symplectic.  In the same manner, imposing the operator equation \eq{C} strongly on ${\cal H}_n$ gives the space  ${\cal K}_n$ given in \eq{Hn}.    

The construction generalizes to polyhedra with more than 4 faces. Then the shape of an ensemble of such polyhedra, with the same area and opposing normals on the shared faces\footnote{The area and the normals match, but not the rest of the geometry of the face, in general. Thus, we have ``twisted geometries", in the sense of Freidel and Speziale.}, is quantized precisely by the Hilbert space $\cal H$ defined above. 

What is the relation with gravity? The central physical idea of general relativity is of course the identification of gravitational field and metric geometry.  Consider a polyhedron given on a (say, piecewise linear) manifold. A metric geometry is assigned by giving the value of a metric, or a triad field $e^i=e^i_a dx^a$, namely the gravitational field.  Consider the quantity
\be
 L^i_l =\epsilon_{ijk} \int_l  e^j\wedge e^k.
\ee
Observe that on the one hand this is precisely $L_l$ defined in \eqref{LL}, namely the 
flux of the densitized inverse triad, or the flux of the Ashtekar's Electric field $E^{ia}$ across the face $l$ of the polyhedron:
\be
 E^i_l=\int_l \ n_a E^{ai}\,, 
\ee
where $n_a$ is the normal to the face; on the other hand, in locally flat coordinates it is the normalized normal $\vec n_l$ to the face $l$, multiplied by the area:
\be
 E^i_l 
 =\int_l n_a E^{ai}=\int_l n^i = n_l^i a_l = L_l^i.
\ee
Therefore the quantized normals $\vec L_l$ of simplicial quantum geometry can be interpreted as the quantum operator giving the flux of the Ashtekar electric field, and we recover again the full kinematics of the previous section.

The spinfoam formalism is natural from this point of view, and the first spinfoam amplitude, called the Barrett-Crane amplitude was first formalized in this context \cite{Barrett:1997gw}. The Barrett-Crane model was then improved to give the amplitude defined in these lectures.

\section{Conclusion}

These lectures are far from covering the full spectrum of the research in Loop Quantum Gravity.  I have focused on the covariant formulation of the dynamics, at the expenses of the ongoing research in the canonical language.   A formulation I haven't covered is group field theory \cite{DePietri:1999bx,Oriti06}, which is the language in terms of which current research on the scaling of the theory is formulated \cite{Geloun:2010vj,Krajewski:2010yq,Rivasseau:2010kf}.   I haven't covered the applications of the theory to black hole physics  \cite{Rovelli96,Ashtekar:2000eq,Ashtekar:1999ex,Engle:2010kt}, and to hamiltonian loop quantum cosmology  \cite{Bojowald:2008zzb,Ashtekar:2008zu}, which are by far the most interesting applications of LQG. In particular, loop quantum cosmology is the most likely window for observations.  Also, I have not covered several recent development, such as the manifest Lorentz invariant formulation of the theory  \cite{Rovelli:2010ed}, the coupling to fermions and Yang-Mills fields \cite{Bianchi:2010bn,Han:2011as}, and to a cosmological constant \cite{Fairbairn:2010cp,Han:2010pz}, using a quantum group.  For a wide angle review, complementary to these lectures, on various aspects of the theory, including historical, and a more comprehensive bibliography and tentative overall evaluation, see \cite{Rovelli:2010bf}.

In conclusion, the theory looks simple and beautiful to me, both in its kinematical and its dynamical parts.  Some preliminary physical calculations have been performed and the results are encouraging. The theory is moving ahead fast.  But it  is far from being complete, and we do not yet know if it really works, and there is still very much to do. 

My greatest wish is that one of the students studying these lectures will be able to solve the last problem:

\problem{Show that theory is wrong, correct it. Or: show that the theory is right, find observable consequences.}

\vskip .5cm

\rightline{Verona, July 22nd, 2011}

\vskip 1cm

\centerline{-----}

\vskip 1cm

I thank for corrections and suggestions: Jacek Puchta and Milka Kubalova. A warm thank in particular to Leonard Cottrell.

\appendix

\section{Open problems}\label{problems}

The theory is far from being complete. Below is a list of some of the open problems that require further investigation. 

A great pleasure for me in updating these notes has been to realize that several of the problems posed in the first version of these notes are since been solved.  This witnesses to the state of rapid growth in which the theory is.  For completeness, I leave the old problems here, indicating that the progress made, in square parentheses.

\begin{enumerate}
\item Compute the propagator (\ref{prop}) in the Lorentzian theory, extending the euclidean result of  \cite{Bianchi:2009ri}. [Done in \cite{Bianchi:2011fk}.]
\item Compute the three point function and compare it with the vertex amplitude of conventional perturbative quantum gravity on Minkowski space. [Done in \cite{Rovelli:2011kf}.]
\item Compute the next vertex order of the two point function, for $N=2$.
\item Compute the next graph order of the two point function, for $\Gamma>\Gamma_5$.
\item Understand the normalization factors in these terms, and their relative weight. Find out under which conditions the expansion is viable.
\item Study the radiative corrections in (\ref{int1}) and their possible (infrared) divergences, following the preliminary investigations in \cite{Perini:2008pd}. The potential divergences are associated to ``bubbles" (nontrivial elements of the second homotopy class) in the two-complex. Classify them and study how do deal with these. [Progress in the Group Field Theory version of the formalism \cite{Krajewski:2010yq,Geloun:2010vj}.   Because of Ditt-invariance \cite{Rovelli:2011fk}, this problem might be related to the analysis of the divergences of BF theory, on which there has been substantial progress in \cite{Bonzom:2011br,Bonzom:2010ar,Bonzom:2010zh}. See also the old results on the Barrett Crane model \cite{Crane:2001as,Crane:2001qk}.]
\item Use the analysis of the these radiative corrections to study the scaling of the theory. 
\item In particular, how does $G$ scale?
\item Study the quantum corrections to the tree-level $n$-point functions of classical general relativity. Can any of these be connected to potentially observable phenomena?
\item Is there any reason for a breaking or a deformation of local Lorentz invariance, that could lead to observable phenomena such as $\gamma$ ray bursts energy-dependent time of arrival delays, in this theory? [Observation is proceeding fast on this issue. See \cite{Liberati:2009pf,Jacobson:2005bg,Laurent:2011he}].
\item Compute the cosmological transition amplitude in the Lorentzian theory, extending the euclidean result of  \cite{Bianchi:2010zs}. [Done in \cite{Vidotto:2011qa}.]    Compare with canonical Loop Quantum Cosmology \cite{Ashtekar:2008zu,Bojowald:2006da}. 
\item The possibility of introducing a spinfoam-like expansion starting from Loop Quantum Cosmology has been considered by Ashtekar, Campiglia and Henderson  \cite{Ashtekar:2009dn,Ashtekar:2010ve,Rovelli:2009tp,Henderson:2010qd,Calcagni:2010ad}. Can the convergence between the two approaches be completed? 
\item Find a simple group field theory \cite{Oriti:2009wn} whose expansion gives (\ref{int1}). [Much progress in \cite{Geloun:2010vj,Krajewski:2010yq}.]
\item Find the relation between this formalism and the way dynamics can be treated in the canonical theory.  Formally, if $H$ is the Hamiltonian constraint, we expect something like the main equation
\be
               HW=0
\ee
or $WP=0$ where the operator $P$ is given by $\langle W|\overline\psi\otimes\phi\rangle=\langle \psi | P|\phi\rangle$, since $P$ is formally a projector on the solutions of the Wheeler de Witt equation 
\be
H\psi =0.
\ee
Can we construct the Hamiltonian operator in canonical LQG such that this is realized? 
\item Is the node expansion related to the amount of boundary data available? How? 
\item Where is the cosmological constant in the theory? It is tempting to simply replace \eq{K3} with a corresponding quantum group expression 
\be
   \langle W_v| \psi\rangle =   Ev_q(f\psi).
\label{vertex2}
\ee
where $Ev_q$ is the quantum evaluation in $SL(2,\C)_q$. Does this give a viable theory? Does this give a finite theory? [Solved in  \cite{Han:2010pz,Fairbairn:2010cp,Han:2011vn}.]

\item How to couple fermions and YM fields to this formulation?  The kinematics described above generalizes very easily to include fermions (at the nodes) and Yang Mills fields (on the links). Can we use the simple group theoretical argument that has selected the gravitational vertex also for coupling these matter fields? [A solution of this problem has appeared: \cite{Bianchi:2010bn,Han:2011as}.]

\item Is the scenario sketched in Section \ref{expa} truly realized in the theory? Do the amplitudes converge fast with the refinement, in suitable regime?   Do radiative corrections interfere with this convergence? 

\item Are there interesting variants to the amplitude? Alternative choices for the face amplitude have been considered in the literature. In the Euclidean case, where $SL(2,\C)$ is replaced by $SO(4)$, there is a natural alternative which is the dimension of the $SO(4)$ irreducible into which the representation $j$ is mapped by $Y_\gamma$.  This choice appears to be incompatible with the natural composition properties of the spin foam amplitude \cite{Bianchi:2010fj}.  A simple modification of the theory is to multiply the vertex by a constant $\lambda$. This comes naturally if one derives the transition amplitudes from a group field theory \cite{Oriti:2009wn}: then $\lambda$ is the coupling constant in front of the group-field-theory interaction term.  The physical interpretation of the constant $\lambda$ is debated \cite{Oriti:2009wn,Ashtekar:2010ve}.  Modifications of the amplitudes have recently been explored in \cite{Bahr:2010bs}.

\end{enumerate}

\section{Alternative expressions for the amplitude}\label{Alternative}

I list here various form of the transition amplitudes that have been used in the literature and can be useful.

\subsection{Single equation}

The definition  (\ref{int1},\ref{vava},\ref{KK}) of the amplitude can be written compactly in a single equation in the form 
\begin{eqnarray}
&&W_{\cal C}(h_l)= \int_{(SL2C)^{\scriptscriptstyle 2(E-L)-V}}dg'_{ve}\int_{(SU2)^{ \scriptscriptstyle  {\cal V}-L}}dh_{e\!f}\; \sum_{{j_{\!{}_f}}}
\nonumber\\
\label{int11}
&&\hspace{1em}  N_{\{j_f\}}^{-1}  \prod_{f} d_{j_{\!{}_f}}\;
\chi^{\scriptscriptstyle\gamma {j_{\!{}_f}}\!,{j_{\!{}_f}}}\!\Big(\!\prod_{e\in\partial f}g_{e\!f}^{\epsilon_{l\!f}}\!\Big) \prod_{e\in\partial f}\chi^{j_{\!{}_f}}\!(h_{e\!f})
\end{eqnarray}
Here $\epsilon_{ef}=\pm 1$ if the edge $e$ appears as $e$ or as $e^{-1}$ in the sequence and  
\be
{\cal V} =\sum_f\ n_f. 
\ee
See \cite{Rovelli:2010vv} for details and the next subsection for more on the definition of each term. 

\subsection{Feynman rules}

A more detailed description of this expression is given by the following Feynman rules. 
$W_{\cal C}(h_l)$ is defined as the integral obtained associating:
\begin{enumerate}%\addtolength{\itemsep}{-2mm}
\item Two group integrations  to each internal edge (or one to each adjacent couple  \{internal edge, vertex\})\vspace{-5mm}
\be
\begin{picture}(35,25)
\thicklines
\put(-7,-11){\tiny $g'$}
\put(22,16){\tiny $g$}
\put(0,-6){\line(1,1){20}}
\put(10,-1){\tiny $e$}
\end{picture}\longmapsto\ \ 
\int_{SL2C}dg_{e s_e}\int_{SL2C}dg_{et_e}
\label{un}
\ee
\item A group integration to each couple of adjacent \{face, internal edge\} 
\vspace{-2mm}
\be
\begin{picture}(35,25)
\put(0,-6.2){\line(-1,0){10}}
\put(20.2,14){\line(-1,1){10}}
\thicklines
\put(0,-6){\line(1,1){20}}
\put(7,-3){\tiny $e$}
\put(-8,8){\tiny $f$}
\put(2,9){\tiny $h_{e\!f}$}
\end{picture}\longmapsto\ \ 
\int_{SU2}dh_{e\!f}\; \chi^{j_f}(h_{e\!f})
\label{su2int}
\ee
\item A sum  to each face $f$
\vspace{-2mm}
\be
\hspace{3em}
\begin{picture}(25,25)
\thicklines
\put(-10.1,-6){\line(0,1){20}}
\put(-0.1,-6){\line(-1,0){10.3}}
\put(0,24){\line(1,0){10}}
\put(-10,14){\line(1,1){10}}
\put(20,14){\line(-1,1){10}}
\put(0,-5.9){\line(1,1){20}}
\put(11,0){\tiny $h_{e\!f}$}
\put(-1,8){\tiny $f$}
\put(0,-11){\tiny $g'$}
\put(22,11){\tiny $g$}
\end{picture}
\longmapsto\ \sum_{j_{\!f}}d_{j_{\!f}}\,\chi^{\scriptscriptstyle\gamma j_{\!f},j_{\!f}}\!\Big(\!\prod_{e\in\partial f}g_{e\!f}^{\epsilon_{l\!f}}\!\Big).
\label{sei}
\ee
where $
g_{ef}:= g_{es_e} h_{e\!f}g^{-1}_{et_e}
$ for internal edges, and $g_{e\!f}$ = $h_l\!\in\! SU2$ for boundary edges. $\gamma$ is a fixed real parameter, the Barbero-Immirzi parameter. 
\item At each vertex, one of the integrals $\int_{SL2C} dg_{ev}$ in \eqref{un} (which is redundant) is dropped. (This is the meaning of the prime on $dg_{ev}$)
\item For each coloring $j_f$, divide the local amplitude by the combinatorial factor $N_{j_j}$. This factor has be taken to be unity in the text.  If we choose it to be the number of automorphisms of $\cal C$ that preserves $j_f$, we have an interesting consequence:

The \emph{limit} \eqref{limit} can be equivalently \cite{Rovelli:2010qx} expressed as an infinite \emph{sum} over transitions
\be
W(h_l)= \sum_{\cal C} W^*_{\cal C}(h_l). 
\ee
where $W^*_{\cal C}(h_l)$ is defined by the same expression as $W_{\cal C}(h_l)$, but with the sum over spins going from $\frac12$ to $\infty$ rather than from $0$ to $\infty$.  That is, dropping trivial representations. That is, the full theory can be equivalently recovered by taking the ``infinite refinement limit" or by ``summing" over two-complexes.

\end{enumerate}

\subsection{Using $Y$ explicitly}\label{oo}

The kernel \eqref{K} can be written in the form 
\be
K(h,g)\! =\! \! 
\sum_{j} {\scriptstyle (2j+1)}\, {\rm Tr}\!\!
\left[
D^{\scriptscriptstyle(j)}\!(h)Y_\gamma^\dagger D^{\scriptscriptstyle(\gamma j,j)}(g)Y_\gamma
 \right]\!.
\label{daniele2c}
\ee
That it, using the explicit form of the $Y_\gamma$ map,
\be
K(h,g)\! =\! \! 
\sum_{jmn} {\scriptstyle (2j+1)}\,
D^{\scriptscriptstyle(j)}\!(h)_m{}^n  
D^{\scriptscriptstyle(\gamma j,j)}(g)_{jn}{}^{jm}.
\label{daniele222c}
\ee
We use for this the simplified notation 
\be
K(h,g)\! =\! \! 
\sum_{j} d_j\  {\rm Tr}_j\!\!
\left[Y_\gamma^\dagger gY_\gamma h
 \right]\!.
\label{daniele2c}
\ee

\subsection{Spin-intertwiner basis}

In the spin intertwiner basis, the amplitude reads 
\be
W(j_l,v_n)=\sum_{j_f,v_e}\ \prod_f d_{j_f}\ \prod_v \ A_v(j_f,v_e)
\label{spinint}
\ee 
which is the well known form of the spinfoam state sums. 
This form of the amplitude can be obtained from  \eqref{int1} as follows. First, define the vertex amplitude in the spin network basis by
\be
A_v(h_{vf})\equiv \sum_{j_{vf},v_{ve}}  \psi_{j_{vf},v_{ve}}(h_{vf})  A_v(j_{vf},v_{ve})
\label{spinnnetwk}
\ee 
where $\psi_{j_{l},v_{n}}(h_{l})=\langle h_{l} |j_{l},v_{n} \rangle $ is  the standard kernel of the change of basis from holonomies to group elements, namely simply the contraction of the intertwiners according to the pattern defined by the boundary graph of $v$.   Using this and expanding the delta function in \eqref{int1}, gives (I consider the no-boundary case for simplicity)
\ba
Z&=&   \int dh_{vf}\ \prod_f \ \sum_{j_f} d_{j_f} tr[D^j(h_{v_1f})...D^j(h_{v_nf})]\nonumber \\ 
&& \ \prod_v  \sum_{j_{vf},v_{ve}}  \psi_{j_{vf},v_{ve}}(h_{vf})  A_v(j_{vf},v_{ve}).
\label{spinint6}
\ea 
The integrals can be now performing explictely using \eqref{wigner}.  The result is 
\ba
 \int dh_{vf}&& \!\!\!\! \prod_f \  d_{j_f} tr[D^j(h_{v_1f})...D^j(h_{v_1f})]
\ \prod_v    \psi_{j_{vf},v_{ve}}(h_{vf})    \nonumber \\ &&= \otimes_{j_{vf} } \delta_{j_{vf},j_f}
\otimes_e \langle  v_{es_e} |  v_{et_e} \rangle. 
\ea 
Using this, we have immediately \eqref{spinint}.

\subsection{Coherent states form}\label{coherentform}

The sum over intertwiners can be traded for an integral over coherent states, since these form a basis in intertwiner space. Thus we can write
\be
Z=\sum_{j_f}\int d\vec n_{ef} \prod_f d_{j_f}\ \prod_v \ A_v(j_f,\vec n_{ef})
\label{spinint5}
\ee 
Using \eqref{lorentzian}, we get 
\be
Z=\sum_{j_f}\int\! d\tilde g_{ve} \int dn_{ef} \prod_f d_{j_f}\ 
\prod_{v}  \ 
 \langle -\vec n_{ef} |Y^\dagger g_{e} g_{e'}^{\scriptscriptstyle -1}Y|\vec n_{e'f}\rangle_{j}
\label{lorentzian3}
\ee
where $e$ and $e'$ are the two edges bounding $f$ and $v$. This can be written in the form of a path integral 
\be
Z=\sum_{j_f}\int\! d\tilde g_{ve} \int dn_{ef} \prod_f d_{j_f}\ e^S
\ee
by defining the action
\be
S=\sum_{fv} \ln  \langle -\vec n_{ef} |Y^\dagger g_{e} g_{e'}^{\scriptscriptstyle -1}Y|\vec n_{e'f}\rangle_{j}
\ee
This is the form of the amplitude used to study its asymptotic expansion.

In the euclidean theory,  $Y$ maps the $SU(2)$ representations $j$ of  into the $SO(4)$ representation $(j^+,j^-)$. The matrix elements of $Y$ are the standard Clebsch-Gordan coefficients. Since the coherent states factorize under the Clebsch-Gordan decomposition, we obtain $S=S^++S^-$ with \begin{equation}
S^{\pm}=\sum_{vf}2j^\pm_{f}\ln\bra{-\vec{n}_{ef}}(g_a^{\pm})^{-1}g_b^{\pm}\ket{\vec{n}_{e'f}}_\frac12.
\end{equation}
 
The euclidean vertex amplitude in the spin network basis reads  
\begin{eqnarray}
A(j_{e},v_n)&=&\sum_{v_n^+ v_n^-}
15j\!\left({\scriptstyle \frac{(1+\gamma)j_{l}}{2}};v_n^+\right)
15j\!\left({\scriptstyle\frac{|1-\gamma|j_{l}}{2}};v_n^-\right)\nonumber\\
&& \bigotimes_a f^{i_n}_{i_n^+ i_n^-}(j_{l})
\end{eqnarray}
where the $15j$ are the standard $SU(2)$ Wigner  symbols, and the  ``fusion coefficients" are
\be
f^{v}_{v^+ v^-}:=v^{m_1...m_n}\ 
v^+_{q_1^+... q_n^+}\ v^-_{q_1^-...q_n^-}\ 
\bigotimes_{i} c^{q_i^+q_i^-}_{\ m_i}.
\ee
$c^{q_i^+q_i^-}_{\ m_i}.$ being the Clebsch-Gordan coefficients. 

The vertex amplitude was first constructed in this language. 

In the lorentzian theory:
\begin{eqnarray}
A(j_{l},i_n)&=&\sum_{k_n}\int dp_n (k_n^2+p_n^2)
\left(\bigotimes_a\ 
f^{i_n}_{k_np_n}(j_{l})\right)\nonumber \\
&&
 15j_{SL(2,\mathbb{C})}\left((j_{l},j_{l}\gamma);(k_n,p_n)\right)\
\end{eqnarray}
where we are now using the 15j of $SL(2,\mathbb{C})$ and
\be
f^{v}_{kp}:=v^{m_1...m_4}\ \bar{C}^{kp}_{(j_1,m_1)...(j_4,m_4)},
\ee
where $j_1...j_n$ are the representations meeting at the node.

\subsection{Perez representation}

Finally, Alejandro Perez is writing an introductory review to the spinfoam formalism \cite{perez}. Perez introduces a nice graphical formalism for the theory. For completeness I write here the definition of the partition function for the euclidean theory
\be
Z_{\cal C}=\sum_{j_f v_e} \ \prod_f (2j_f+1) \ \prod_v \raisebox{-1.6cm}{\includegraphics[scale=0.6]{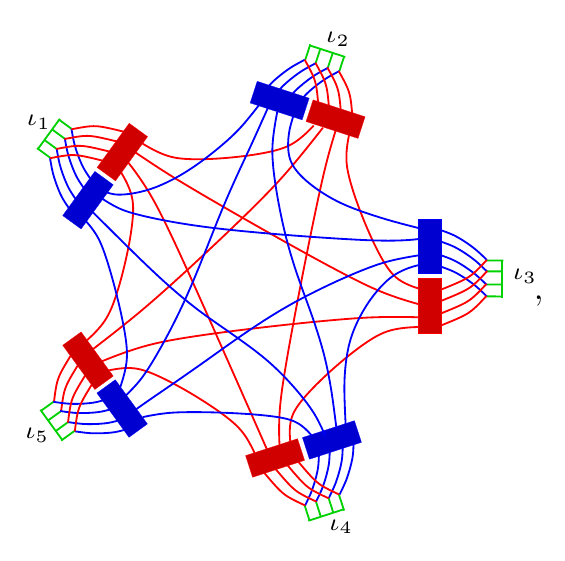}}
\ee
and for the lorentzian theory
\be
Z_{\cal C}=\sum_{j_f v_e} \ \prod_f (2j_f+1) \ \prod_v \raisebox{-1.6cm}{\includegraphics[scale=0.6]{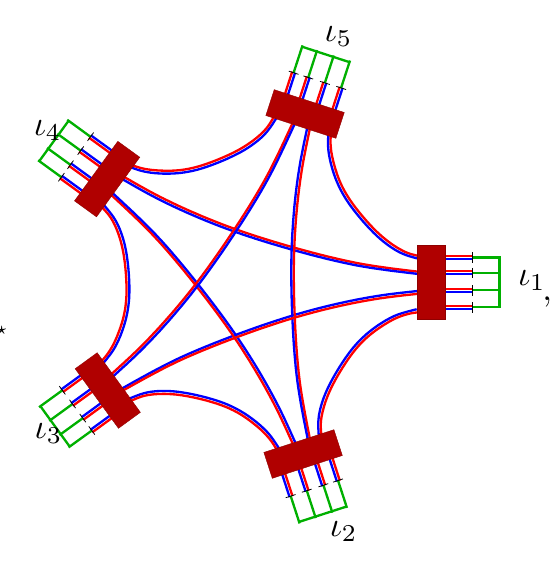}}
\ee
in this notation.  I refer to Perez review for the definition of the notation and the relation with the formalism used here. 
\vspace{2cm}

\bibliographystyle{apsrev4-1}

\bibliography{BiblioCarlo}

\end{document}